\documentclass[preprint]{aastex631}
\usepackage{xspace}
\usepackage{graphicx}
\usepackage{natbib}
\usepackage{url}
\usepackage{float}
\raggedright
\usepackage{epsfig}%\documentclass[manuscript]{aastex63}
\usepackage{amsmath}
\newcommand{\hiisub}{\textnormal{H\,\textsc{ii}}}
\newcommand{\degree}{\ensuremath{\,^\circ}}
\newcommand{\degper}{\rlap.{^{\circ}}}
\newcommand{\arcmper}{\rlap.{^{\prime}}}

\newcommand{\mhz}{\ensuremath{\,{\rm MHz}}}
\newcommand{\ghz}{\ensuremath{\,{\rm GHz}}}
\newcommand{\kel}{\ensuremath{\,{\rm K}}}
\newcommand{\K}{\ensuremath{\,{\rm K}}}
\newcommand{\mk}{\ensuremath{\,{\rm mK}}}

\newcommand{\cm}{\ensuremath{\,{\rm cm}}}

\newcommand{\percc}{\ensuremath{\,{\rm cm^{-3}}}}

\newcommand{\kpc}{\ensuremath{\,{\rm kpc}}}
\newcommand{\pc}{\ensuremath{\,{\rm pc}}}
\newcommand{\kms}{\ensuremath{\,{\rm km\, s^{-1}}}}

\newcommand{\microns}{\ensuremath{\, \mu {\rm m}}}
\newcommand{\ev}{\,eV}
\newcommand{\jy}{\,Jy}

\newcommand{\microK}{\ensuremath{\rm \,\mu K}}

\newcommand{\te}   {\ensuremath{T_{\rm e}}}

\newcommand{\tb}   {\ensuremath{T_{\rm B}}}

\newcommand{\tpk}  {\ensuremath{T_{\rm pk}}}
\newcommand{\tla}  {\ensuremath{T_{\rm L}^{\rm A}}}
\newcommand{\tlb}  {\ensuremath{T_{\rm L}^{\rm B}}}
\newcommand{\tsys} {\ensuremath{T_{\rm sys}}}
\newcommand{\Ne}   {\ensuremath{n_{\rm e}}}
\newcommand{\nerms}{\ensuremath{n_{\rm e}^{\rm rms}}}

\newcommand{\rgal} {\ensuremath{R_{\rm gal}}}

\newcommand{\hna}     {\ensuremath{{\rm \langle Hn\alpha \rangle}}}
\newcommand{\hnb}     {\ensuremath{{\rm \langle Hn\beta  \rangle}}}
\newcommand{\hng}     {\ensuremath{{\rm \langle Hn\gamma \rangle}}}
\newcommand{\hrrl}[1]{H#1}

\newcommand{\halpha}   {\hrrl{91}\ensuremath{\alpha}}

\newcommand{\expo}[1]{\ensuremath{10^{#1}}}

\newcommand{\hi}{H\,{\sc i}}

\newcommand{\hii}{H\,{\sc ii}}

\newcommand{\nii}{N\,{\sc ii}}

\newcommand{\hplus}{H\ensuremath{^{+}}}
\newcommand{\htwo}{H\ensuremath{_{2}}}

\newcommand{\threec}[1]{3C\thinspace #1}

\newcommand{\ngc}[1]{NGC\thinspace #1}

\newcommand{\gsim}{\ensuremath{\gtrsim}}
\newcommand{\lsim}{\ensuremath{\lesssim}}
\newcommand{\urltilda}{\kern -.15em\lower .7ex\hbox{\~{}}\kern .04em}
\newcommand{\Ro}  {\ensuremath{R_{\rm 0}\xspace}}    % Sun-GC distance  
\newcommand{\Rmin}{\ensuremath{R_{\rm min}\xspace}}  % los minimum Rgal
\newcommand{\Dtp} {\ensuremath{d_{\rm TP}\xspace}}   % tangent pt dist from sun
\newcommand{\Dlos}{\ensuremath{d_{\rm LOS}\xspace}}  % los dis to solar
\newcommand{\Dcloud}{\ensuremath{d_{\rm cloud}\xspace}} % cloud diameter kpc
\newcommand{\dsun}{\ensuremath{d_\odot\xspace}}  % distance  from Sun
\newcommand{\gl}   {\ensuremath{\ell}\xspace}
\newcommand{\gb}   {\ensuremath{{\it b}}\xspace}

\newcommand{\vlsr} {\ensuremath{V_{\rm LSR}}\xspace}

\newcommand{\tbv}  {\ensuremath{T_{\rm B}}(\vlsr)}
\newcommand{\vt}   {\ensuremath{V_{\rm T}}\xspace}
\newcommand{\lb}   {\ensuremath{(\gl,\gb)}}
\newcommand{\lv}   {\ensuremath{(\gl,v)}\xspace}

\newcommand{\wrrl} {\ensuremath{W_{\rm RRL}\xspace}}
\newcommand{\area} {\ensuremath{\mk\,\kms}\xspace}
\newcommand{\emeas}{\ensuremath{{\rm \,cm^{-6}\,pc}\xspace}}
\newcommand{\nT}   {\ensuremath{{\rm \,cm^{-3}\,K}\xspace}}
\newcommand{\fwhm} {\ensuremath{\Delta\,V\xspace}}
\newcommand{\dvobs} {\ensuremath{\Delta\,V_{\rm obs}\xspace}}
\newcommand{\dvlos} {\ensuremath{\Delta\,V_{\rm los}\xspace}}
\newcommand{\dvth} {\ensuremath{\Delta\,V_{\rm th}\xspace}}
\newcommand{\dvnt} {\ensuremath{\Delta\,V_{\rm nt}\xspace}}

\newcommand{\dnn}  {\ensuremath{\Delta\,n\xspace}}
\newcommand{\cor}  {\ensuremath{{\rm ^{13}CO}\xspace}}

\shorttitle{Milky Way Warm Ionized Medium}
\shortauthors{Bania, et al.}
\begin{document}

\setlength{\parindent}{15pt} % It works putting it here 

\title{
  The Most Sensitive Radio Recombination Line Measurements Ever Made of
  the Galactic Warm Ionized Medium
}

\correspondingauthor{T.~M. Bania}
\email{bania@bu.edu}

\author[0000-0003-4866-460X]{T. M. Bania}
\affiliation{Institute for Astrophysical Research, Astronomy
  Department, Boston University,\\
  725 Commonwealth Ave., Boston, MA 02215, USA}

\author[0000-0002-2465-7803]{Dana S. Balser}
\affiliation{National Radio Astronomy Observatory, 520 Edgemont
  Rd., Charlottesville, VA 22903, USA}

\author[0000-0003-0640-7787]{Trey V. Wenger}
\affiliation{NSF Astronomy \& Astrophysics Postdoctoral Fellow,\\
  Department of Astronomy, University of Wisconsin--Madison,\\
  Madison, WI, 53706, USA}

\author[0000-0002-9449-2485]{Spencer J. Ireland}
\affiliation{Institute for Astrophysical Research, Astronomy
  Department, Boston University,\\
  725 Commonwealth Ave., Boston, MA 02215, USA}
\affiliation{Department of Physics, New Mexico Tech, Socorro,
  NM 87801, USA}
  
\author[0000-0001-8800-1793]{L. D. Anderson}
\affiliation{Department of Physics and Astronomy, West Virginia
  University,\\
  Morgantown, WV 26506, USA}
\affiliation{Center for Gravitational Waves and Cosmology,
  West Virginia University,\\
  Chestnut Ridge Research Building, Morgantown, WV 26505, USA}
\affiliation{Green Bank Observatory, P.O. Box 2, Green Bank, WV 24944, USA}

\author[0000-0001-8061-216X]{Matteo Luisi}
\affiliation{Department of Physics, Westminster College,
  New Wilmington, PA 16172, USA}
\affiliation{Center for Gravitational Waves and Cosmology,
  West Virginia University,\\
  Chestnut Ridge Research Building, Morgantown, WV 26505, USA}

\begin{abstract}

Diffuse ionized gas pervades the disk of the Milky Way.  We detect
extremely faint emission from this Galactic Warm Ionized Medium (WIM)
using the Green Bank Telescope to make radio recombination line (RRL)
observations toward two Milky Way sight lines:
G20, $(\ell,{\it b})$=$(20\degree,0\degree)$,
and G45, $(\ell,{\it b})$=$(45\degree,0\degree)$.  
We stack 18 consecutive Hn$\alpha$ transitions between 4.3--7.1\,GHz to
derive ${\rm \langle Hn\alpha \rangle}$ spectra that are sensitive to
RRL emission from plasmas with emission measures EM\,$\gsim$\,10\,
${\rm \,cm^{-6}\,pc}$. 
Each sight line has two Gaussian shaped spectral components with
emission measures that range between $\sim$100 and $\sim$300 \emeas.
Because there is no detectable RRL emission at negative LSR velocities
the emitting plasma must be located interior to the Solar orbit. 
The G20 and G45 emission measures imply RMS densities of 0.15 and
0.18$\,{\rm cm^{-3}}$, respectively, if these sight lines are filled
with homogeneous plasma.
The observed ${\rm \langle Hn\beta \rangle}$/${\rm \langle
Hn\alpha\rangle}$ line ratios are consistent with LTE excitation for the
strongest components.
The high velocity component of G20 has a narrow line width, 13.5\kms,
that sets an upper limit of $\lsim$\,4,000\,K for the plasma electron
temperature. This is inconsistent with the ansatz of a canonically
pervasive, low density, $\sim$\,10,000 K WIM plasma.

\end{abstract}

\vspace{-1cm}


\section{The Milky Way Warm Ionized Medium}\label{sec:intro}

\citet{1963WIM} were the first to suggest the existence of a layer of
ionized gas along the Galactic plane with an electron density of $\sim
0.1$\percc\ and an electron temperature of $\sim \expo{4}$\kel.  This
was based on their Hobart, Tasmania discovery of free-free absorption at
frequencies less than 10\mhz\ against the synchrotron background
observed at very low radio frequencies \citep{ellis62, reber56}.  They
estimated that the ionizing radiation necessary to produce the observed
free-free absorption was consistent with flux from OB stars.  The
dispersion of radio waves from pulsars \citep{tanenbaum68} and the
observation of optical emission lines \citep{reynolds73} from the
interstellar medium (ISM) confirmed the existence of a warm ionized
medium (WIM) in the Milky Way.  Detection of H$\alpha$ emission from the
diffuse ionized gas (DIG) in the edge-on spiral galaxy \ngc{891} showed
that other galaxies had similar warm plasmas \citep{dettmar90, rand90}.
Here, we follow the convention of using the term ``WIM'' to refer to the
warm ionized medium in the Milky Way and ``DIG'' to refer to the diffuse
ionized gas in external galaxies \citep{2009WIMreview}.

The WIM has been extensively studied with optical emission lines.
H$\alpha$ maps of the Galaxy show emission from almost every direction
\citep[e.g.,][]{dennison98, gaustad01, haffner03} with a volume filling
factor larger than 0.2 \citep{reynolds91}.  Sensitive observations of
two emission lines from nitrogen indicate that the WIM is about
2,000\kel\ warmer than \hii\ regions \citep{reynolds01}.  Models of
escaping radiation from \hii\ regions indicate a harder spectrum for
photons capable of ionizing hydrogen but a suppression of He-ionizing
photons \citep{wood04}, consistent with observations of helium in the
WIM \citep[e.g.,][]{reynolds95}.

Although known for 60 years now, a detailed understanding of the WIM's
origin, distribution, and physical properties remains to be crafted.
Because of attenuation by dust, studies of the WIM using H$\alpha$ are
mostly limited to regions near to the Sun or at high Galactic
latitudes. Moreover, the relatively low spatial and spectral resolution
of most H$\alpha$ surveys cannot separate emission from the WIM with
that from discrete, OB-star excited \hii\ regions.  This is particularly
a problem in the inner Galactic disk.

Two recent radio recombination line (RRL) surveys overcome these
limitations. The Green Bank Telescope (GBT) Diffuse Ionized Gas Survey
\citep[GDIGS;][]{2021GDIGS} is a 4-8\ghz\ RRL survey of the Milky Way
disk that probes the distribution and properties of the WIM in the inner
Galaxy ($32\degper3 > \gl > -5\degree, |b| < 0\degper5$).  GDIGS has a
angular resolution of $2\arcmper6$ and a spectral resolution of 0.5\kms.
A complimentary survey was completed with the Five-hundred-meter
Aperture Spherical radio Telescope (FAST) using RRL transitions spanning
frequencies between 1--1.5\ghz\ for the Galactic zone $55\degree > \gl >
33\degree, |b| < 2\degree$\citep{hou22}. FAST has an angular resolution
of 3\arcmin\ and a spectral resolution of 2.2\kms.  Both surveys employ
line-stacking of multiple RRL transitions observed simultaneously to
increase the signal to noise ratio.

Using GDIGS data, \citet{luisi20} created a WIM-only RRL map of the W43
star formation complex that is devoid of emission from \hii\ regions.
Employing an empirical model that only accounts for the \hii\ region
locations, angular sizes, and RRL intensities, they were able to
reproduce the observed WIM emission \citep[also see][]{belfiore22}.
This supports the notion that UV photons leaking from the \hii\ regions
surrounding O and B-type stars are the primary source of the ionization
of the WIM.

The main limitation of these new RRL surveys is sensitivity.  For
example, GDIGS is sensitive to emission measures,
EM $\equiv\ \int\,\Ne^2\, {\rm d}\gl\ \gsim 1,100$\emeas,
whereas most H$\alpha$ surveys are several orders of magnitude more
sensitive.  GDIGS detects RRL emission from most directions but not at
all possible velocities where \hi\ gas exists.  This may just be a
sensitivity issue. For example, the GDIGS spectrum toward G20 shows no
RRL emission (see discussion below).

Here, we use the GBT to search for faint RRL emission from two first
Galactic quadrant sight lines: G20 and G45.  These directions, at \lb=
$(20\degree,0\degree)$, and \lb = $(45\degree,0\degree)$ respectively,
were chosen because they contain no detected discrete \hii\ regions, show
extremely weak radio continuum emission, and have infrared emission
morphologies that do not indicate any incipient star formation. Any RRL
emission detected for these lines of sight would thus stem from the WIM.

\section{Observations and Data Analysis}\label{sec:obs}

Using the GBT, we made the RRL observations during a series of sessions
held between June and August of 2020 (project
GBT/20A--483). Measurements were made using the C-band (4-8\ghz)
receiver and the Versatile GBT Astronomical Spectrometer (VEGAS) back
end. Full details of data acquisition protocols, VEGAS tuning
configuration, and calibration are described in our GDIGS paper
\citep{2021GDIGS}. GDIGS survey data were acquired using On the Fly
Mapping.  Here, however, we employ total power position switching by
observing an off source (Off) position for 6 minutes and then the target
(On) position for 6 minutes, for a total time of 12 minutes per
scan. The Off position lies well outside the Galactic plane because the
WIM is expected to be distributed throughout the Galactic disk with a
large volume filling factor. The Off position is offset from the On by
$(\Delta\gl,\Delta\gb)=(0\degper0,-3\degper5)$. 

We tuned VEGAS to 64 different frequencies with two orthogonal linear
polarizations each. A single 12 min OffOn spectral pair thus
simultaneously produces 128 independent 8192 channel spectra, each
spanning a 23.4375\mhz\ bandwidth.  The VEGAS tuning consisted of an
assortment of hydrogen RRL transitions that lie in the 4-8\ghz\ bandpass
of the C-band receiver.  These included $\alpha(\dnn=1), \beta(\dnn=2),
$ and $\gamma(\dnn=3)$ transitions, where $n$ is the transition
principal quantum number.  For $\alpha$ RRL transitions the frequency
extremes of this tuning are H97$\alpha$ (7.09541\ghz\ rest frequency)
and H115$\alpha$ (4.26814\ghz\ rest frequency).  This large frequency
range produces significant differences in the GBT beam size (the
half-power beam width, HPBW) and spectral resolution. The beam size
increases from 107\arcsec\ for H97$\alpha$ to 177\arcsec\ for
H115$\alpha$ and the spectral velocity resolution decreases from
0.12\kms\ per channel for H97$\alpha$ to 0.20\kms\ per channel for
H115$\alpha$.

After acquisition, the telescope data were converted into SDFITS format
and then ported to the {\tt TMBIDL} single dish data analysis software
package \citep{2016tmbidl}.  All subsequent data analysis reported here
is done in the {\tt TMBIDL} environment.

%
% Flux Calibration 
%
The intensity scale is determined by injecting signals from a noise
diode into the signal path every other second during observations. The
flux density scale is then established after data acquisition using
observations of the quasar \threec{286} which is a primary flux
calibrator.  For the GBT at C-band frequencies this quasar is a point
source whose intensity has varied by less than 1\% over decades
\citep{1994OttFluxes, 2000PengFluxes}.  Because the GBT gain at C-band
is 2\K/\jy\ \citep{2001GBTgain}, the \threec{286} observations allow us
to determine the noise diode fluxes. Overall, this procedure establishes
the intensity scale of our spectra to $\sim$5\%\ accuracy.

To maximize our spectral sensitivity, we derive for each target sight
line a single RRL spectrum for each \dnn\ transition, $\alpha, \beta,$
and $\gamma$.  We do this by averaging together all the observed, usable
RRL transitions for a given \dnn.  We call this procedure ``stacking''
and the result of this averaging is a ``stacked'' spectrum: \hna, \hnb,
and \hng. The RRL transitions we use to craft these stacked spectra are
compiled in Appendix Table~\ref{tab:bands}. Listed there for both
targets are the transition, the rest frequency, and the system
temperatures, \tsys, for both linear polarizations (XX and YY).

Stacking is possible because RRLs involve high-$n$ level transitions
wherein adjacent $n$ levels have nearly the same energy. For example, our
usable $\alpha$ transitions span the range from H97$\alpha$ to
H115$\alpha$. In LTE the expected RRL intensities would differ by a
factor of $\sim$1.66 between these two transitions. The expected line
profiles, and hence the spectral line widths, should, however, be
similar for all these transitions.  Since the primary goal here is to
detect RRL emission, all usable RRL transitions can therefore be
averaged together to increase the signal-to-noise ratio, producing the
\hna, \hnb, and \hng\ stacked spectra.

Stacking is not a simple average of the different transitions weighted
by their individual integration times and system temperatures, although
of course that must be part of the procedure. Stacking must correct for
several other effects: (1) each transition has a unique rest frequency
and spectral velocity resolution; (2) each transition's spectrum has a
slightly different center velocity due to our VEGAS tuning; and (3) only
one of the RRL transitions is Doppler tracked due to the limitations of
the GBT IF system hardware. Using the procedure described below we have
successfully stacked RRL spectra taken with the Arecibo
telescope/``Interim Correlator'' spectrometer \citep{roshi05} as well as
the GBT X-band receiver/ACS spectrometer \citep{balser06a}.

To derive the stacked spectra reported here we:
(1) inspect for each transition all the individual OffOn total power
pairs and eliminate bad data; 
(2) calculate for each transition an average spectrum weighted by
integration time and system temperature;
(3) establish a common spectral velocity resolution for each 
transition by using a $sin(x)/x$ interpolation \citep{roshi05}
referenced to our poorest resolution, H97$\alpha$, spectrum;
(4) align all the transitions in velocity referenced to the Doppler
tracked H103$\alpha$ spectrum;
and 
(5) obtain a stacked spectrum by averaging all the individual transition
spectra processed in this way weighted by integration time and system
temperature.

\begin{deluxetable}{ccccccc}[h]
\tablecolumns{7} \tablewidth{0pt}
\tablecaption{Stacked Spectra Properties\label{tab:targets}}
\tablehead{
      \colhead{Source} & \colhead{\gl} & \colhead{\gb} &
      \colhead{\dnn} & \colhead{\# RRLs} &
      \colhead{$t_{\rm intg}$} & \colhead{RMS}\\
      \colhead{} & \colhead{(deg)} & \colhead{(deg)} &
      \colhead{} & \colhead{} & \colhead{(hrs)} & \colhead{(mK)}
}
\startdata
G20 & 20.0  & 0.0   & 1: $\alpha$   & 18     &     243.1 & 0.312 \\
G20 & \dots & \dots & 2: $\beta$    & 22     &     302.4 & 0.288 \\
G20 & \dots & \dots & 3: $\gamma$   & \phn 7 & \phn 91.0 & 0.483 \\
G45 & 45.0  & 0.0   & 1: $\alpha$   & 18     &     207.0 & 0.343 \\
G45 & \dots & \dots & 2: $\beta$    & 22     &     270.8 & 0.314 \\
G45 & \dots & \dots & 3: $\gamma$   & \phn 7 & \phn 79.6 & 0.518 \\
\enddata
\end{deluxetable}

The properties of the stacked spectra for the G20 and G45 sight lines are
summarized in Table~\ref{tab:targets}. Listed are the source name, the 
Galactic position, \lb, the RRL transition order, \dnn, the number of
transitions used to derive the stacked spectrum, the total integration
time of the stacked spectrum, $t_{\rm intg}$, and the root mean square
(RMS) noise of the stacked spectrum after smoothing to
1\kms\ resolution. The integration times for the \hna \ spectra exceed
$\sim$200 hrs and the spectral RMS noise attained thereby is
$\sim$300\microK. We believe these to be the most sensitive measurements
of WIM RRL emission ever made.

%\clearpage
\vskip 10pt
\section{RRL Emission from WIM Plasma}\label{sec:DIGrrl}

\begin{figure}[h]
\centering
%\vskip -5.75cm
\hskip -0.5cm
\includegraphics[angle=0,scale=0.65]{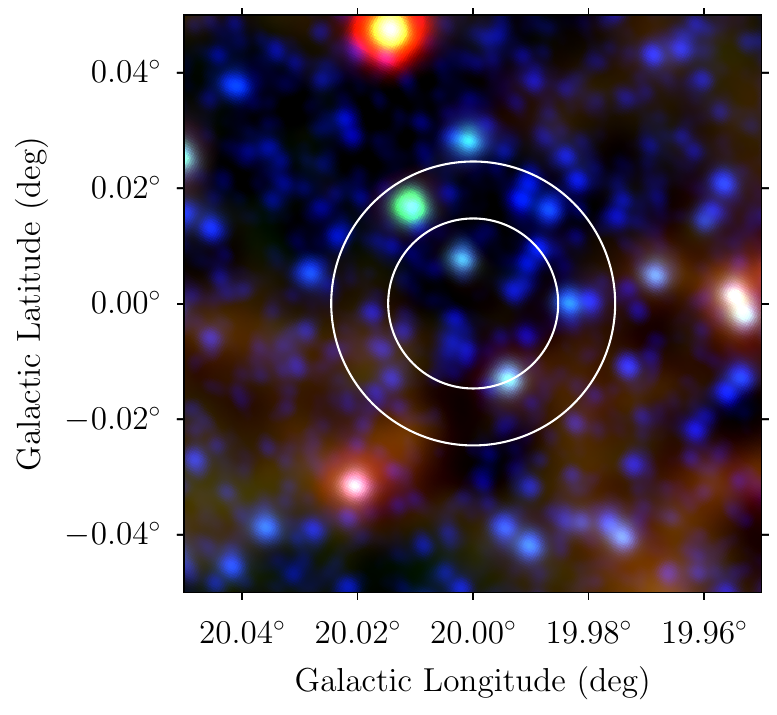}
\includegraphics[angle=0,scale=0.65]{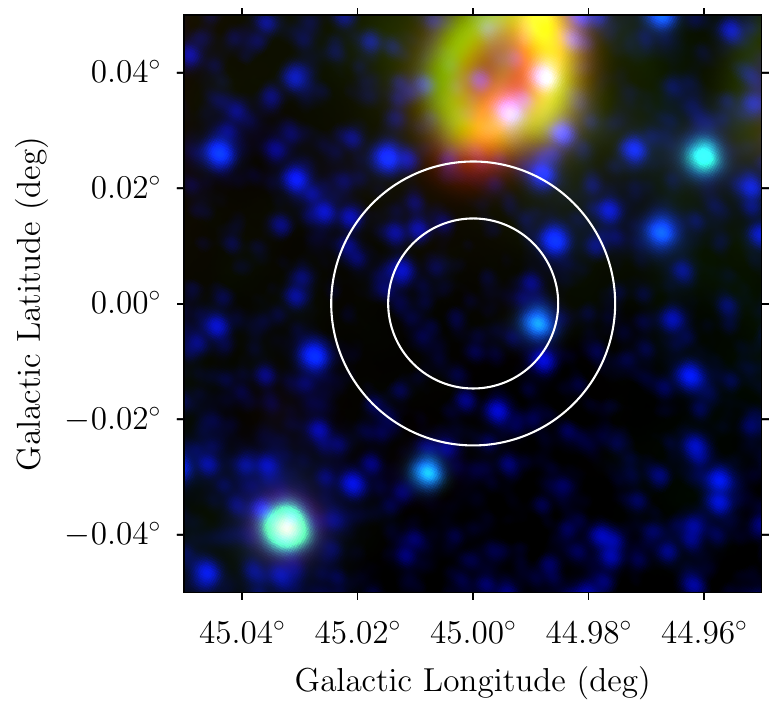}\\
%\vskip -1cm
\caption{
Mid-infrared {\em WISE} images of G20 (left) and G45 (right) where blue is
3.4\microns\ (stars), green 12\microns\ (emission from the
photodissociation region; PDR), and red 22\microns\ (dust). The images
are 6\arcmin\ on a side. The circles show the range of the GBT beam size
(HPBW) for this experiment, from 107\arcsec\ for H97$\alpha$ to
177\arcsec\ for H115$\alpha$.
}
\label{fig:wise}
\end{figure}

\subsection{The G20 and G45 Lines of Sight}\label{sec:los}

We detect weak RRL emission from both first Galactic quadrant sight
lines: G20 and G45.  Images of the mid-infrared emission seen toward
these lines of sight by the Wide-Field Infrared Survey Explorer
\citep[{\em WISE}]{WISE10} satellite are shown in Figure~\ref{fig:wise}.
There is clearly no extended IR emission in these directions consistent
with either protostellar activity or high mass star
formation. Accordingly, the {\em WISE} Catalog of Galactic \hii\ Regions
\citep[][hereafter ``The {\em WISE} Catalog'']{2014WISEcataLOG} lists no
OB-type star excited, discrete \hii\ regions in these fields.  (The
doughnut shaped object seen toward \lb~$\sim$~(44.99,+0.04) is a latent
image from the overexposure of a saturated bright star located beyond
the field of view shown in Figure~\ref{fig:wise}.)

The stacked \hna\ spectra we derive for G20 and G45 are shown in Figure
2. Both sight lines show two components of RRL emission from WIM
plasma. In each direction the higher LSR velocity component has the
stronger peak intensity and narrower line width. The G20 components are
twice as strong as their corresponding G45 components. We fit Gaussian
functions to these components in order to quantify their spectral line
properties. Table 2 compiles the results of these fits. Listed for each
spectral component are the fit values and fit errors for the LSR
velocity, \vlsr , peak antenna temperature at the line center, \tla, the
full width at half maximum, FWHM, line width, \fwhm, the area under the
component, \wrrl, and an estimate of the emission measure, EM. The
errors cited for peak intensity, line area, and emission measure include
the 5\,\% uncertainty in the intensity scale.

How do these spectra compare with the corresponding GDIGS and FAST
measurements?  Figure 3 shows that for G20 our deep integration is much
more sensitive than GDIGS.  GDIGS did not observe the G45 sight line but
FAST did. FAST is more EM sensitive than GDIGS. Inspection of their
Figure 5, however, also reveals no RRL emission. The strongest emission
component that we measure toward either of our targets is therefore
below the GDIGS and FAST sensitivity limits.

\begin{figure}[h]
\centering
\vskip -1cm
\includegraphics[angle=0,scale=0.32]{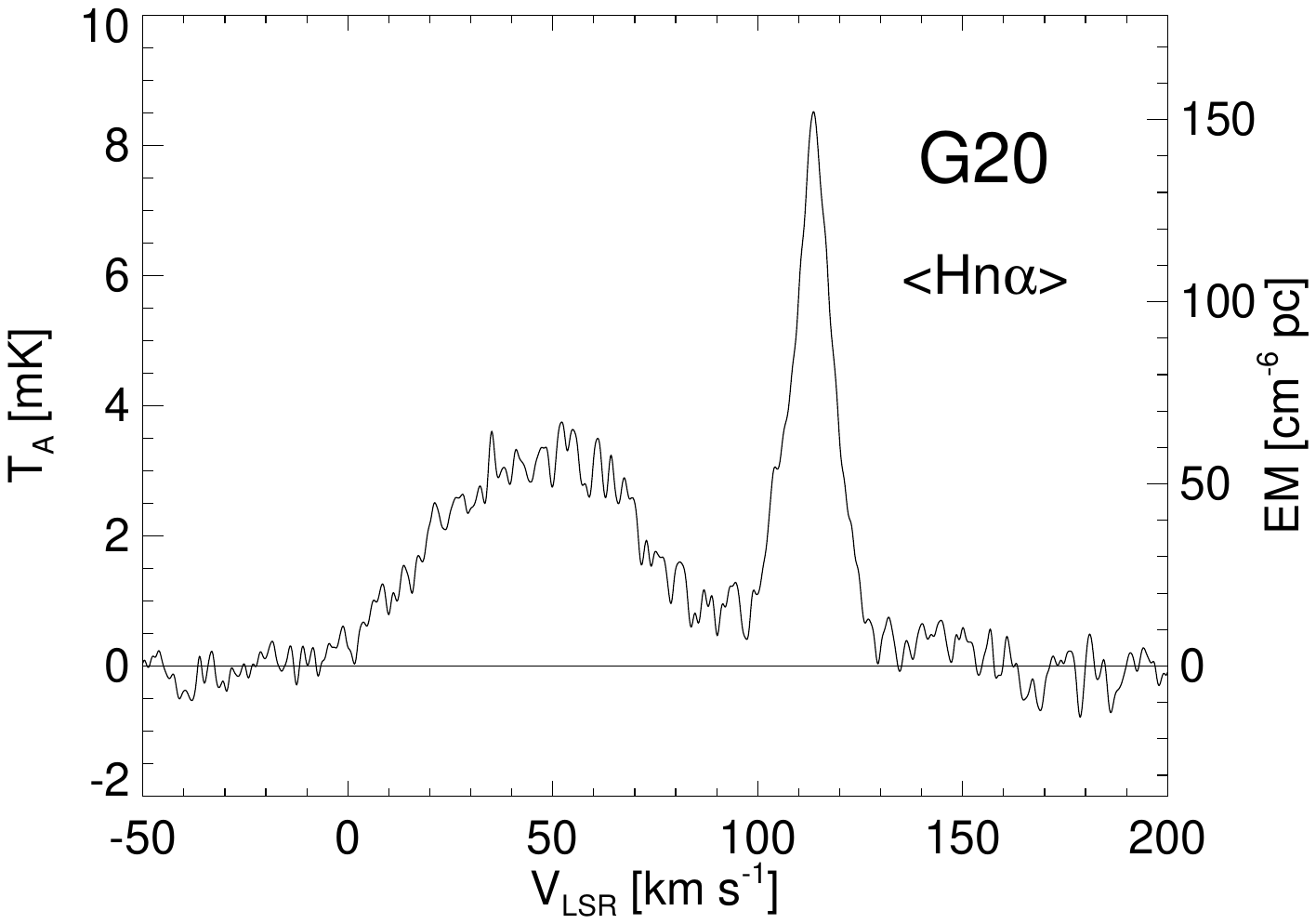}
%\vskip -3cm
\includegraphics[angle=0,scale=0.32]{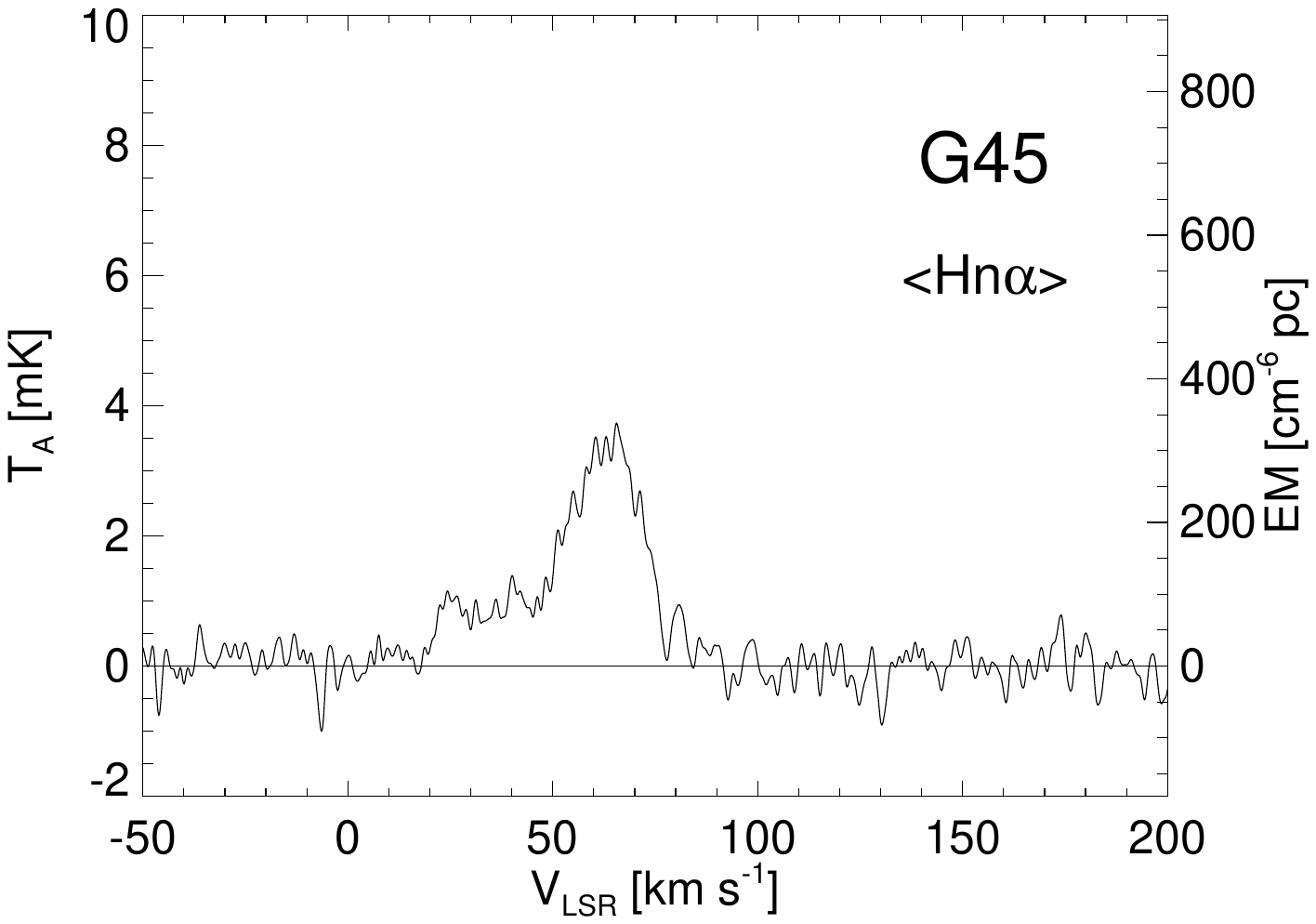}\\
\vskip -0.5cm
\caption{
G20 and G45 stacked spectra for the \hna\ transition.  The spectra were
smoothed to a 1\kms\ velocity resolution and the instrumental spectral
baseline was subtracted. The emission measure, EM, axes  
stem from assumptions described in the text.
}
\label{fig:wim}
\end{figure}

\begin{figure}[h]
\centering
\vskip -1cm
\includegraphics[angle=0,scale=0.32]{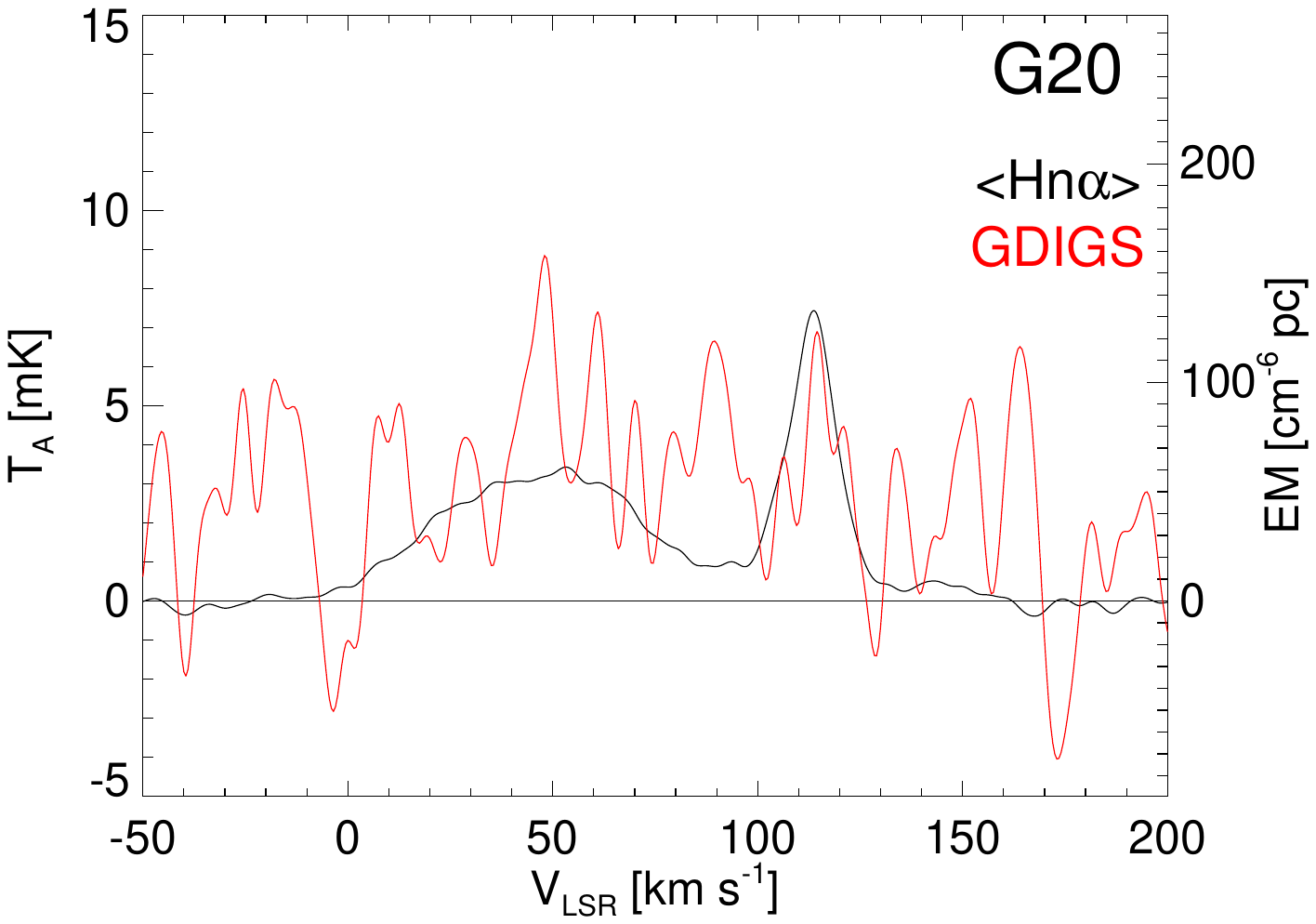}
%\vskip -0.5cm
\caption{
G20 \hna\ spectrum (black) compared with the GDIGS spectral map pixel for
the same direction (red). The \hna\ data were smoothed to
5\kms\ resolution. The $\sim$115\kms\ \hna\ component is below the GDIGS
sensitivity limit. 
}
\label{fig:GDIGS}
\end{figure}

%\clearpage
\subsection{RRL Stacked Spectra}
%\vspace{-0.5cm}
%

For both G20 and G45 RRL emission can be seen at nearly all LSR
velocities between 0\kms\ and the terminal velocity produced by Galactic
rotation. The lack of any detected RRL emission at LSR velocities
below 0\kms\ means that in these directions any RRL emission from WIM
plasma located beyond the Solar orbit about the Galactic Center must be
below the sensitivity limits of our GBT observations.

\begin{deluxetable}{lcccccccccc}
  \tabletypesize{\scriptsize}
\tablecaption{\hna\ Emission Properties\label{tab:WIMfits}}
\tablehead{    
  \colhead{$Source$}     &
  \colhead{\vlsr}        & \colhead{$\sigma_{\rm LSR}$}   &
  \colhead{\tla}         & \colhead{$\sigma_{\rm L}$}    &
  \colhead{\fwhm}        & \colhead{$\sigma_{\Delta V}$}  &
  \colhead{\wrrl    }    & \colhead{$\sigma_{\rm W}$}    &
  \colhead{$EM$}         & \colhead{$\sigma_{\rm EM}$}\\
  \colhead{}             &
  \colhead{(\kms)}       & \colhead{(\kms)}   &
  \colhead{(\mk)}        & \colhead{(\mk)}    &
  \colhead{(\kms)}       & \colhead{(\kms)}   &
  \colhead{(\area)}      & \colhead{(\area)}  &
  \colhead{(\emeas)}     & \colhead{(\emeas)}   
}
\startdata
G20 &\phn 47.68 & 0.22 & 3.32 & 0.33 & 60.1 & 0.60 &    212.2 &    21.2& $<$245.5 & 24.5 \\
%G20 &    113.35 & 0.05 & 7.43 & 0.36 & 13.5 & 0.12 &    107.0&\phn 5.3& 489.2 & 24.3 \\
G20 &    113.35 & 0.05 & 7.43 & 0.36 & 13.5 & 0.12 &    107.0 &\phn 5.3& $<$123.8 & \phn 6.2 \\
G45 &\phn 34.54 & 0.87 & 0.94 & 0.37 & 28.8 & 2.63 &\phn 28.8 &    11.7& $<$112.4 &  45.7 \\
%G45 &\phn 63.29 & 0.20 & 3.44 & 0.39 & 20.3 & 0.42 &\phn 74.4&\phn 8.6&340.0 & 39.2 \\
G45 &\phn 63.29 & 0.20 & 3.44 & 0.39 & 20.3 & 0.42 &\phn 74.4 &\phn 8.6& $<$290.3 &  33.5 \\
\enddata
\end{deluxetable}

We use the observed RRL emission components to estimate the emission
measure.
Assuming that the WIM plasma is extended and always fills the GBT beam,
the brightness temperature at the line center of optically thin
transitions in LTE, \tlb, is [\citet{2002RRLsGordon}, \citet[equation
    14.28 in the 5th edition]{2009TOOLSwilson}, \citet[equation
    16]{2019TeWenger}, and \citet[equation A4]{2021GDIGS}]:

\begin{equation}
  \frac{T_{\rm L}^{\rm B}}{\rm K} = 3013\left(\frac{T_e}{\rm K}\right)^{-1.5}\left(\frac{\rm EM}{\text{cm$^{-6}$ pc}}\right)\left(\frac{\nu_0}{\rm GHz}\right)^{-1}\left(\frac{\Delta V}{\text{km s$^{-1}$}}\right)^{-1}\left(\frac{\Delta n}{n}\right)f_{n+\Delta n, n}
\label{eqn:EM}
\end{equation}

\noindent where
\te\ is the plasma electron temperature,
%$EM$ is the emission measure,
$\nu_0$ is the rest frequency, 
%\fwhm, is the FWHM width of the RRL, 
%$n$ is is the principal quantum number of the transition,
%$\Delta n$ is the change in principle quantum number, 
and
${f_{n+\Delta n,n}}$ is the oscillator strength of the transition between 
$n+\Delta n$ and $n$.
We consider only ${\rm Hn\alpha}~(\Delta n = 1)$ transitions and adopt
the average rest frequency and oscillator strength of our
\hna\ stacked spectra.  Using these values, Equation~\ref{eqn:EM} yields
for the emission measure:

\begin{equation}
  \frac{\rm EM}{\text{cm$^{-6}$ pc}} = 4.57\left(\frac{T_e}{10^4\,\text{K}}\right)^{1.5}\left(\frac{W_{\rm RRL}}{\text{mK km s$^{-1}$}}\right)
\label{eqn:EM-W}
\end{equation}

\noindent
Here, \wrrl\, [\mk\kms], is the spectral area of an emission
component. For a Gaussian line shape, \wrrl = 1.064\,\tlb\,\fwhm.

The RRL intensity values cited in all tables, however, list antenna
temperature, ${T_{\rm L}^{\rm A}}$, the directly measured quantity,
rather than the brightness temperature, \tlb.  
Converting between antenna temperature to brightness temperature depends
on the coupling between the characteristic angular size, $\Theta_{\rm
  WIM}$, of the RRL emitting plasma with the GBT beam.  The stacked
\hna\ spectra stem from frequencies spanning the 4--8\ghz\ instantaneous
bandwidth of the C band receiver.  For this experiment, the GBT HPBW
beam size, $\Theta_{\rm BEAM}$, thus ranges between 177\arcsec\ for
H115$\alpha$ to 107\arcsec\ for H97$\alpha$.  Here, we assume that the
RRL emitting plasma is spatially extended and fills the GBT C
band beam(s): $\Theta_{\rm WIM} \gg \Theta_{\rm BEAM}$ always. For this
case the brightness temperature is $T_{\rm \,B}=T_{\rm \,A}/\eta_{\rm
  \,B}$ where the beam efficiency of the GBT at 5.7578 GHz is $\eta_{\rm
  \,B}= 0.93$ \citep{2001GBTgain,2010GBTetaBeam,2016GBTetaBeam}.

Assuming an electron temperature and FWHM line width, the observed line
antenna temperature, \tla, gives an EM of:

\begin{equation}
  \frac{\rm EM}{\text{cm$^{-6}$ pc}} = 130.71\left(\frac{T_e}{10^4\,\text{K}}\right)^{1.5}\left(\frac{\Delta V}{25\,\text{km s$^{-1}$}}\right)\left(\frac{T_{\rm L}^{\rm A}}{\rm mK}\right)
\label{eqn:EMTA}
\end{equation}

\noindent Here, we use the line peak antenna temperature, \tla, for ease
of interpreting the intensities shown in the spectra and listed in the
tables. The needed conversion to brightness temperature has been
absorbed into Equation~\ref{eqn:EMTA}'s numerical constant.
The EM axes in Figures~\ref{fig:wim} and \ref{fig:GDIGS} and the
values cited in Table~\ref{tab:WIMfits} stem from
Equation~\ref{eqn:EMTA} using the observed \tla\ and \fwhm\ together
with an assumed value for the plasma \te.

Although the electron temperature can be derived from the RRL
line-to-continuum ratio, we did not measure the radio continuum for G20
and G45 and so cannot estimate \te\ from our GBT observations.  Studies
of the WIM/DIG in the Milky Way and other galaxies using ${\rm
  H_\alpha}$, [\ion{N}{2}], [\ion{S}{2}], and [\ion{O}{3}] find that the
electron temperature ranges between 6,000 and
11,000\K\ \citep{2009WIMreview}.
Here, we use the observed line widths, \fwhm, to constrain \te.
Interpreting the observed line width as being due solely to thermal
broadening, \fwhm\ = \dvth, sets a firm upper limit on the plasma
electron temperature:

\begin{equation}
  \frac{T_e}{\rm K} \leq 21.85\left(\frac{\fwhm}{\kms}\right)^2
\label{eqn:telimit}
\end{equation}

\noindent where $k$ is Boltzmann's constant and $m_{\rm H}$ is the mass
of hydrogen.  The high LSR velocity RRL components seen toward G20 and
G45 have the smallest line width for each sight line. These line widths
give \te\ upper limits of $\lsim$\,4,000\,\K\ and
$\lsim$\,9,000\,\K\ for G20 and G45, respectively. These are the
\te\ values used for the EM axes in Figures~\ref{fig:wim} and
\ref{fig:GDIGS} and for the $EM$ values cited in
Table~\ref{tab:WIMfits}.

The \hna\ WIM spectra we derive for G20 and G45 are very deep
integrations. The four \hna\ emission components that we detect toward
G20 and G45 yield emission measures ranging between $\sim$\,100 and
$\sim$\,300\emeas\ (via Equation~\ref{eqn:EM-W} and 
Table~\ref{tab:WIMfits}). All these EM values stem from smoothing the
native VEGAS velocity resolution per channel to 1 \kms. 
The RMS noise cited in Table~\ref{tab:targets} is for this velocity
resolution. The Figure~\ref{fig:wim} \hna\ spectra are thus 
sensitive to emission measures $EM \gsim$\,20\emeas\ (the
3\,$\sigma$ limit via Equation~\ref{eqn:EMTA}).
Below, when we consider LTE excitation or compare RRL models to our
\hna\ spectra, we smooth the per channel velocity resolution to
5\kms. At this resolution these spectra have a 3$\sigma$ sensitivity
EM$\gsim$\,10\emeas.  Nonetheless, even then the spectra show no hint of
additional emission components toward G20 and G45, especially at negative
LSR velocities.

\subsection{Properties of the Stacked Spectra}

Before conducting further analyses of these stacked spectra, we explore
how the RRL emission component line parameters vary as a function of
frequency.  After all, the stacked spectra are averages of many RRL
transitions spanning a significant range of rest frequencies.
To do this, we create ``triad'' spectra by stacking three consecutive
Hn$\alpha$ RRL transitions together and smoothing the resulting spectrum
to 5\kms\ velocity resolution.  Because our stacked \hna\ spectra are
comprised of 18 Hn$\alpha$ RRL transitions, there are six of these triad
spectra for each sight line and they have rest frequencies spanning the
entire bandwidth of the C-band receiver.  These triad spectra have the
sensitivity to enable Gaussian fits for three of the four RRL emission
components present in our sight lines.  Only the weaker, lower velocity
component of G45 cannot be analyzed this way.

%\vspace{-1.0cm}
\begin{figure}[h!]
\centering
\vskip -1.0cm
\includegraphics[angle=0,scale=0.32]{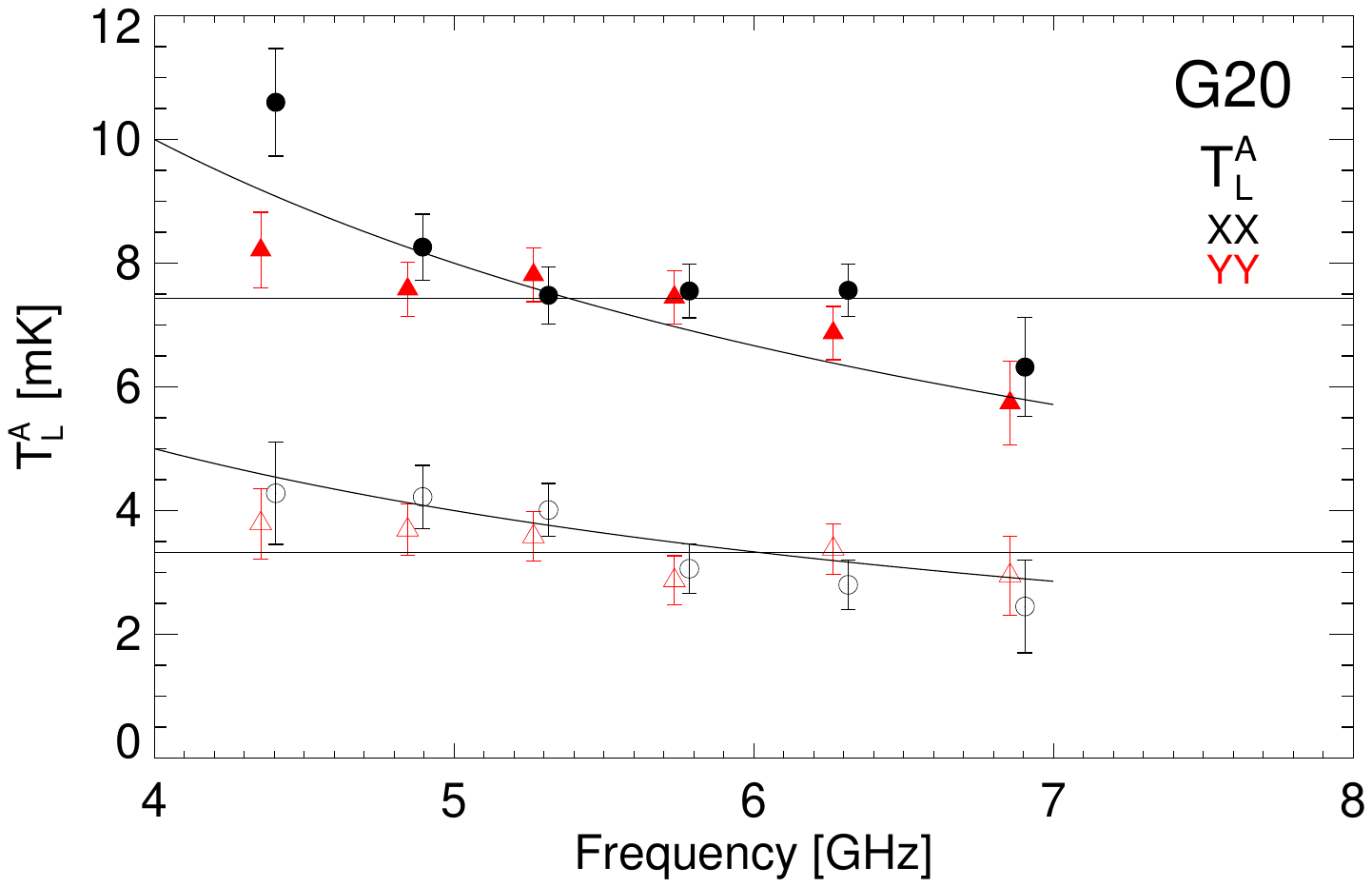}
%\vskip -3cm
\includegraphics[angle=0,scale=0.32]{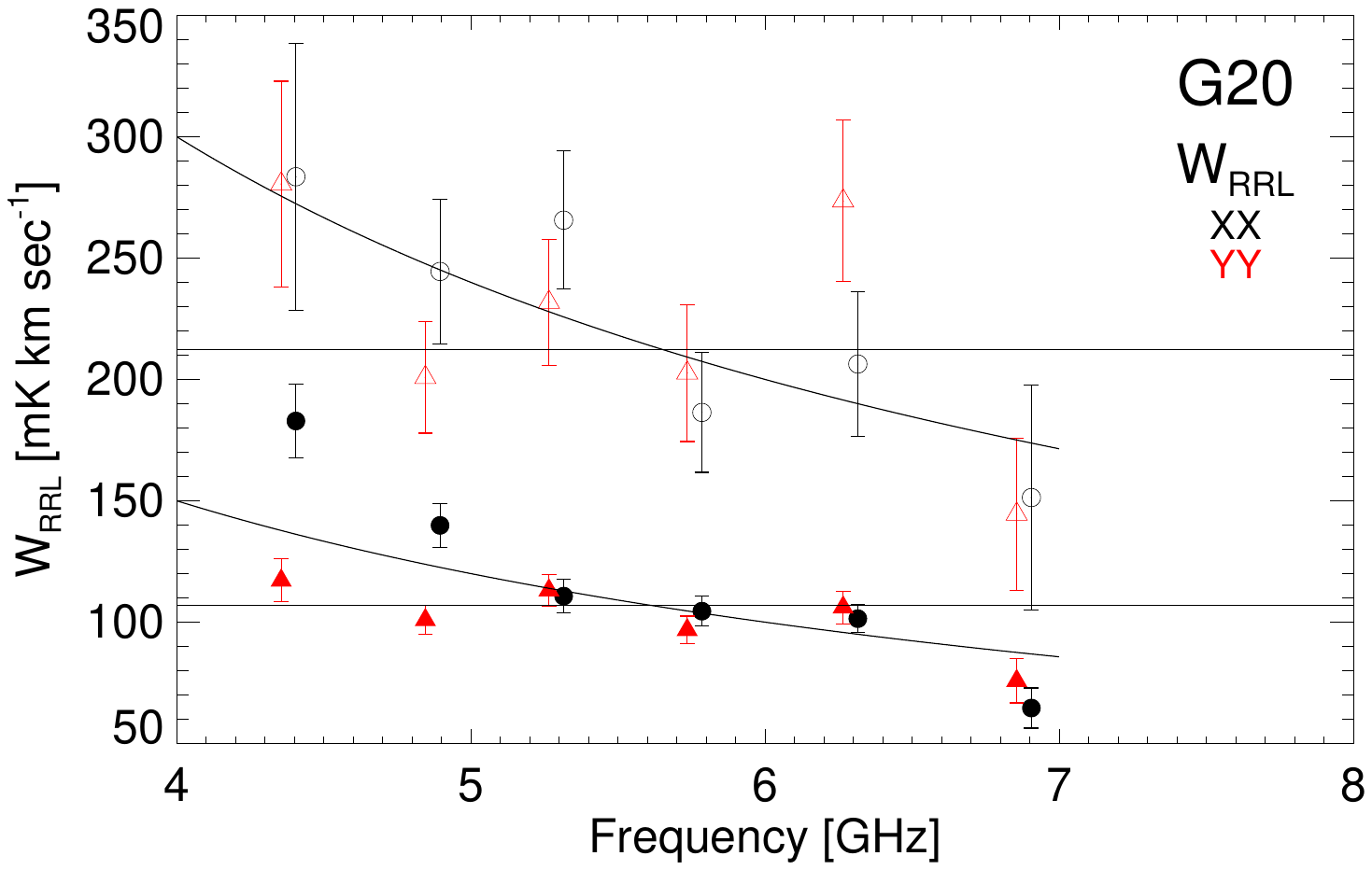}\\
\vskip -1.5cm
\includegraphics[angle=0,scale=0.32]{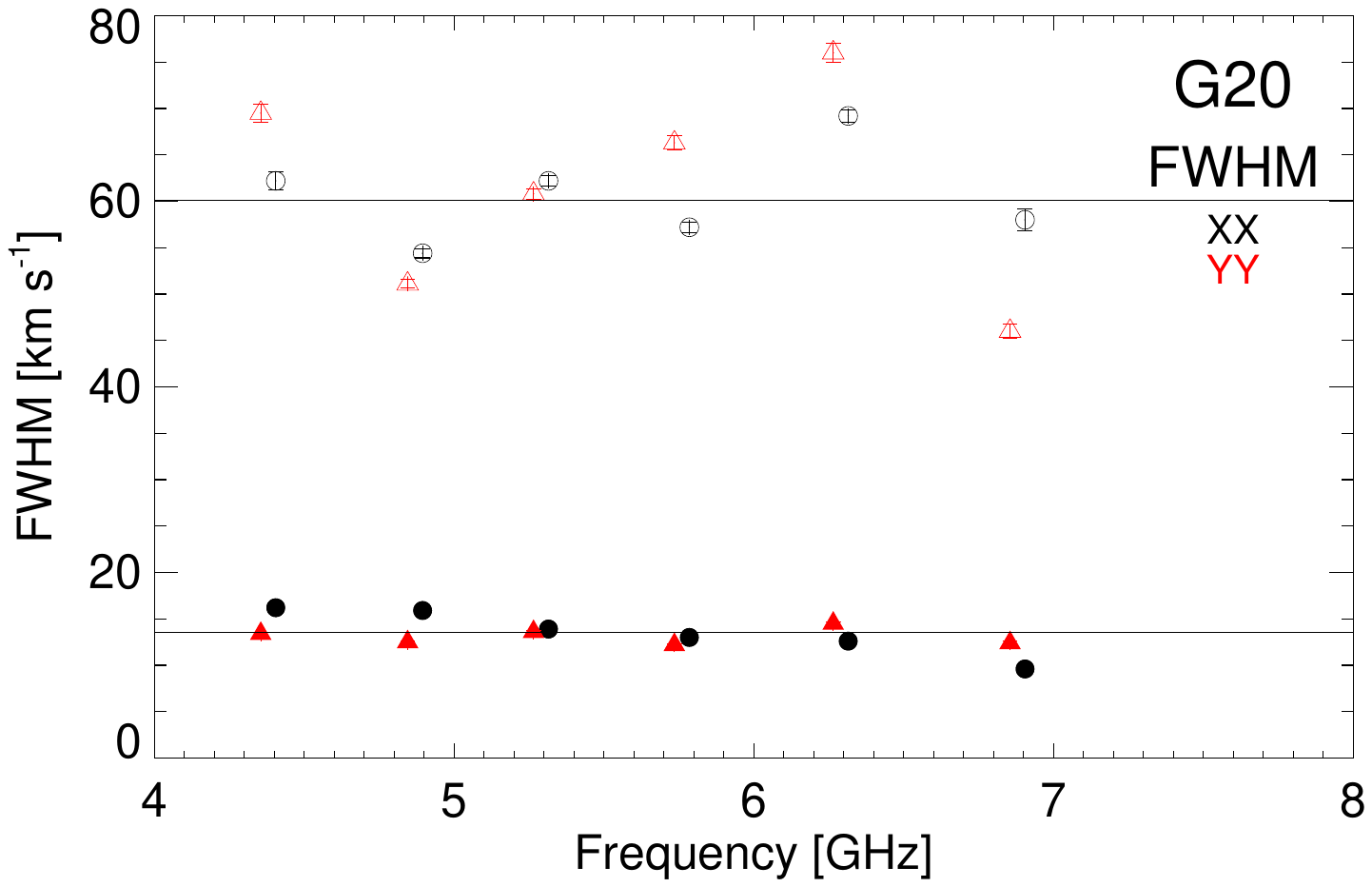}
%\vskip -3cm
\includegraphics[angle=0,scale=0.32]{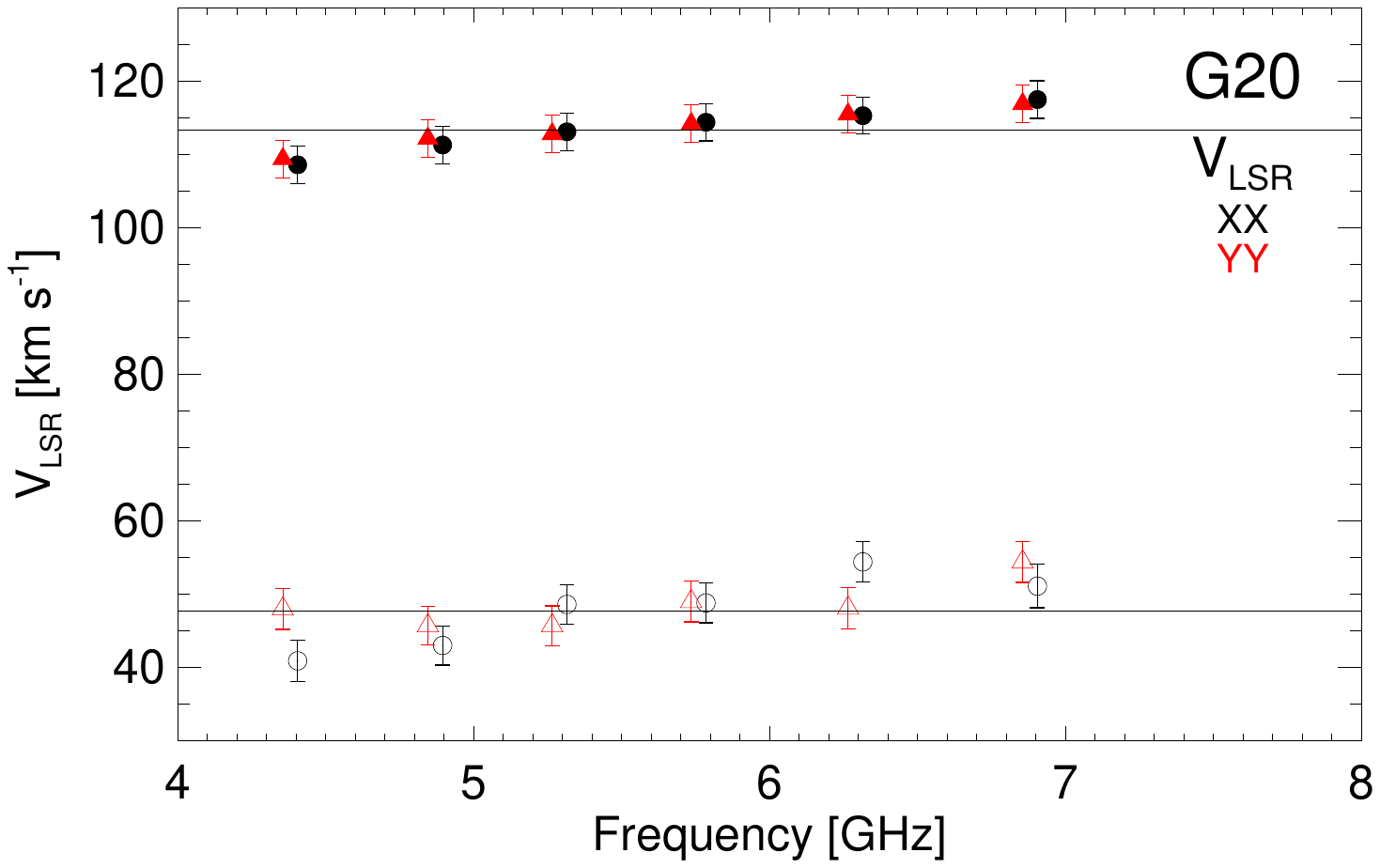}\\
\vskip -0.50cm
\caption{
G20 sight line RRL emission properties for spectral averages of three
consecutive alpha transitions plotted as a function of frequency.
Filled symbols denote the high velocity, stronger component; open
symbols show the low velocity, weaker component.  Horizontal lines flag
the component parameters derived from Gaussian fits to the \hna\ 
spectrum. The curves shown for \tpk\ and \wrrl\ are not fits to the
data, rather they are notional curves that show the $\nu^{-1}$ frequency
dependence expected for plasma in LTE and a source that fills the
telescope beam (see text).
}
\label{fig:g20triads}
\end{figure}

\begin{figure}[h!]
\centering
\vskip -1.0cm
\includegraphics[angle=0,scale=0.32]{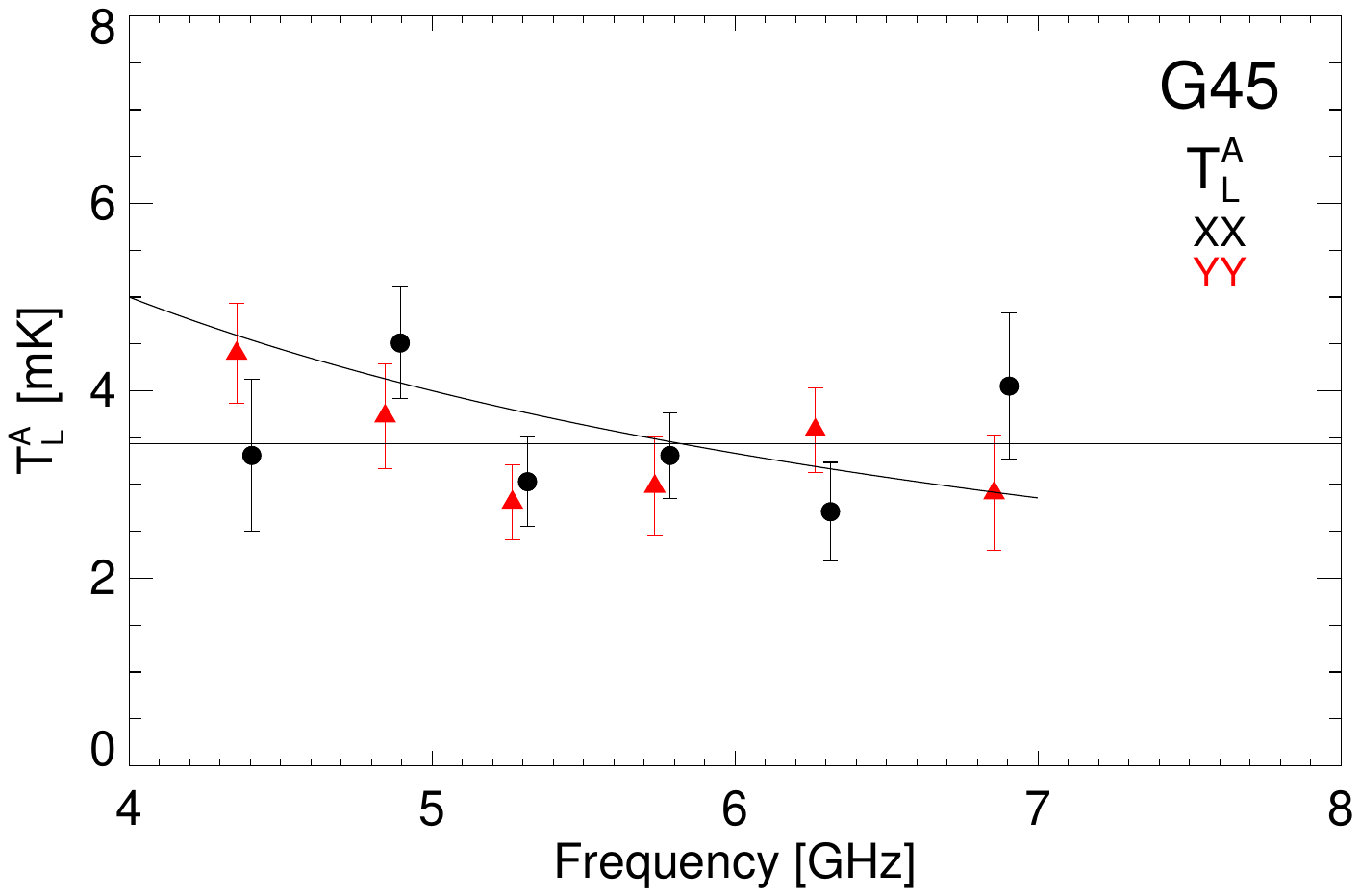}
%\vskip -3cm
\includegraphics[angle=0,scale=0.32]{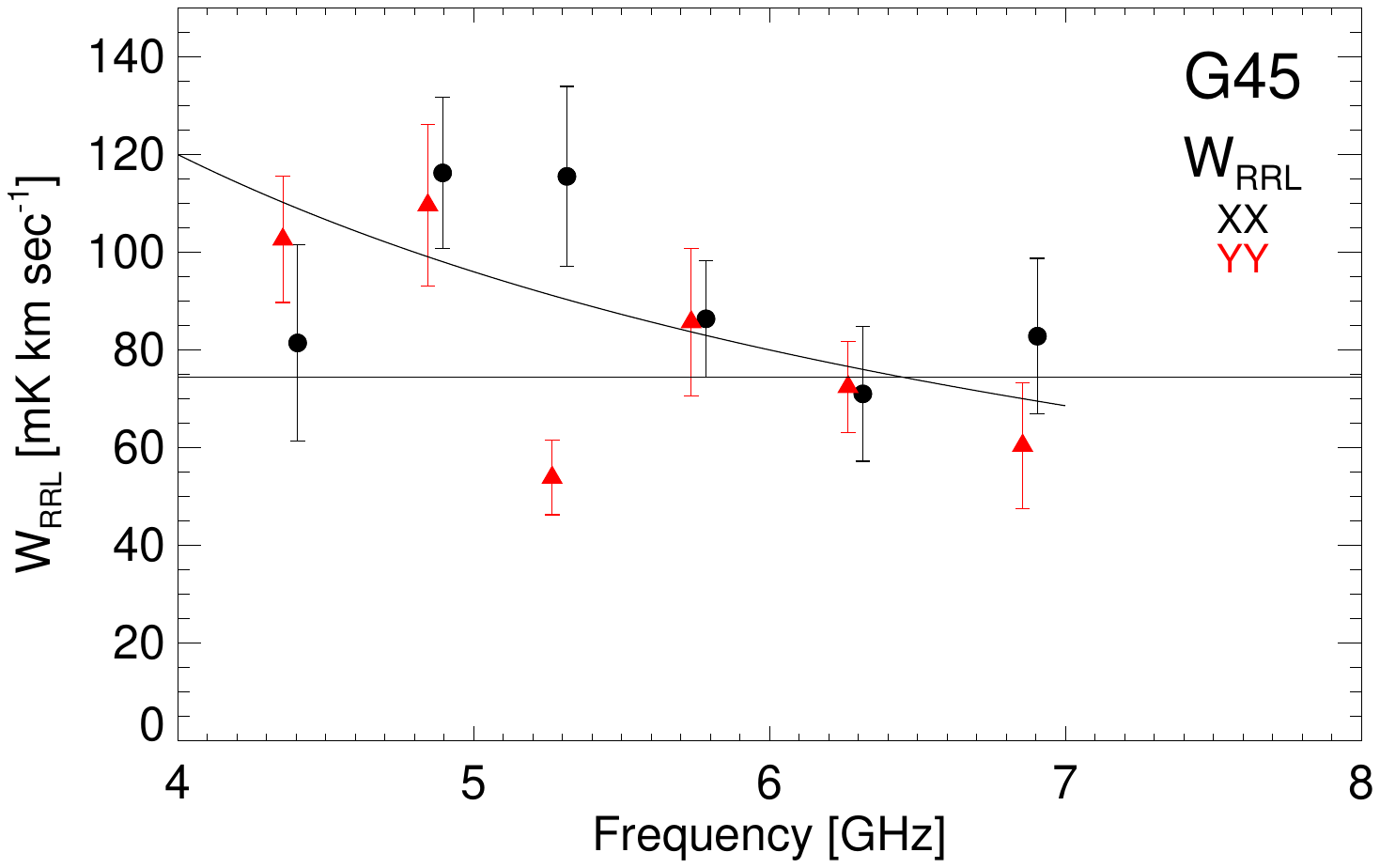}\\
\vskip -1.5cm
\includegraphics[angle=0,scale=0.32]{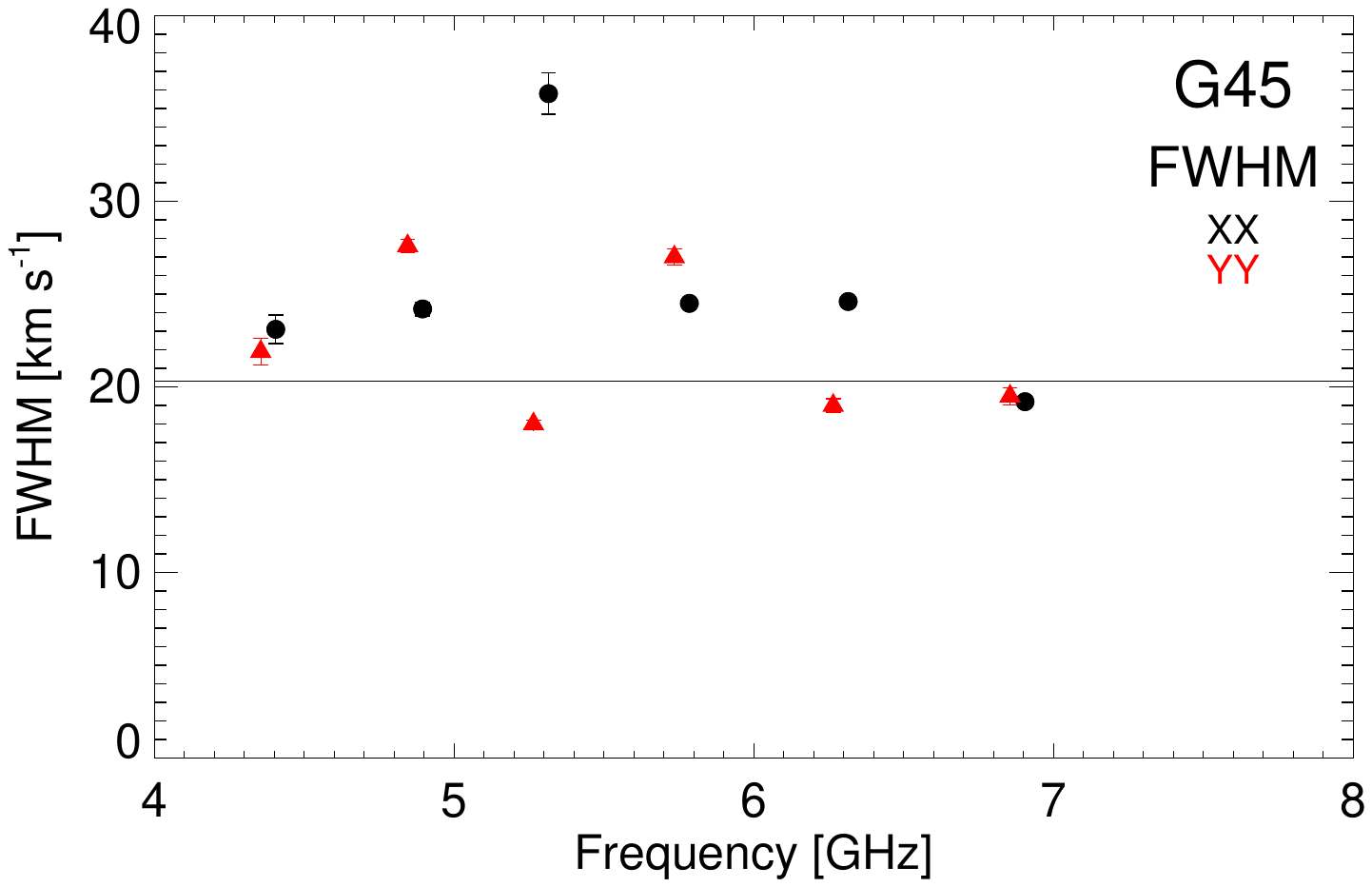}
%\vskip -3cm
\includegraphics[angle=0,scale=0.32]{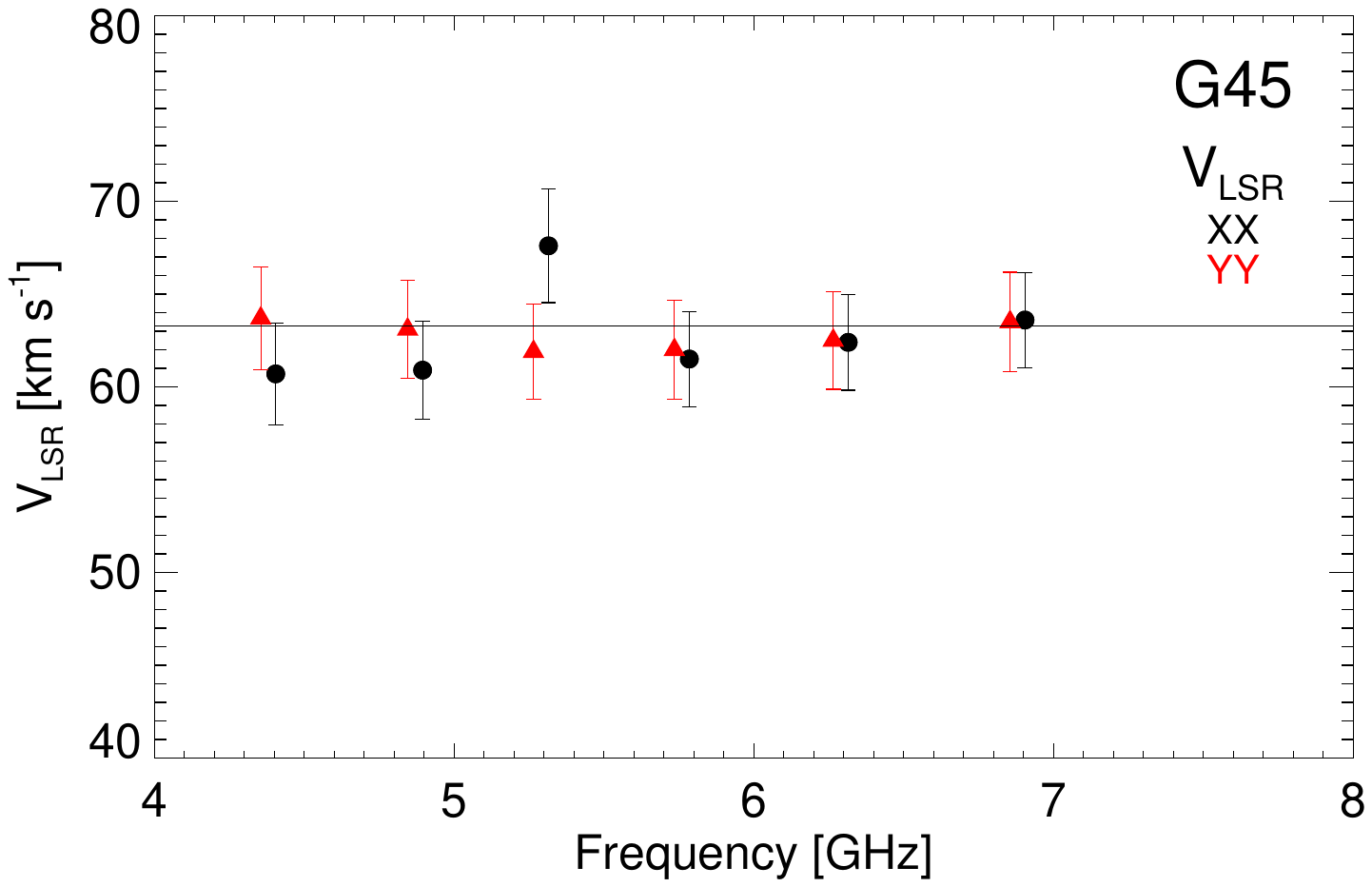}\\
\vskip -0.50cm
\caption{
G45 sight line RRL emission properties for spectral averages of three
consecutive alpha transitions plotted as a function of frequency. Plot
properties are the same as for Figure~\ref{fig:g20triads} except only
one RRL component for this sight line at the higher
$\sim$60\kms\ velocity is strong enough to be analyzed in this manner.
}
\label{fig:g45triads}
\end{figure}

This exploration is summarized in Figure~\ref{fig:g20triads} for G20 and
Figure~\ref{fig:g45triads} for G45. These figures show how the RRL
emission component line parameters vary as a function of frequency.
Filled symbols denote the high velocity, stronger component; open
symbols show the low velocity, weaker component.  For clarity, in these
plots the frequency of these components has been dithered by
$\pm$\,30\mhz. Horizontal lines flag the component parameters derived
from Gaussian fits to the \hna\ spectrum (Table~\ref{tab:WIMfits}).

The triad analysis shown in these figures does not reveal any issues that
might compromise the stacking process. Overall, the triad data are
consistent with the Table~\ref{tab:WIMfits} Gaussian fits to the stacked
\hna\ spectra. Furthermore, none of the RRL emission component FWHM line
widths show any systematic change with frequency. The strongest G20
emission component does have a significant LSR velocity gradient as a
function of frequency. This is probably due to the changing GBT beam
size sampling different volumes of plasma. Neither the weaker G20
component nor the G45 sight line component show any LSR velocity
gradient.

For both sight lines, however, the triad component \tla\ and
\wrrl\ values exhibit significant trends with frequency.
The curves shown for \tpk\ and \wrrl\ are not fits to the
data, rather they are notional curves that show the $\nu^{-1}$ frequency
dependence expected for plasma in LTE and a source that fills the
telescope beam (see Equation~\ref{eqn:EM}).

%\clearpage
\subsection{Are the G20 and G45 Plasmas in LTE?}

\begin{figure}[h]
\centering\vskip -1cm
\includegraphics[angle=0,scale=0.32]{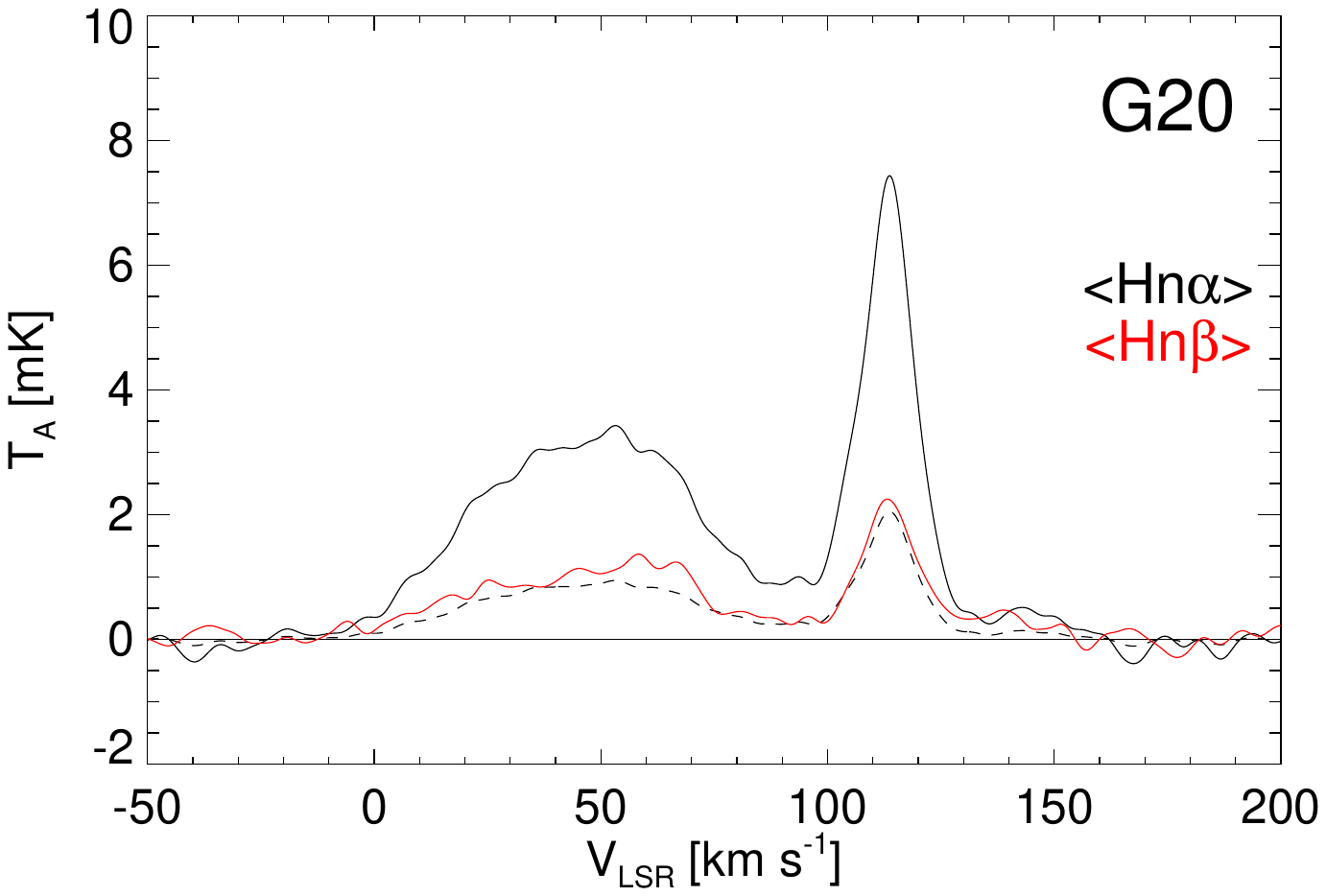}
%\vskip -3cm
\includegraphics[angle=0,scale=0.32]{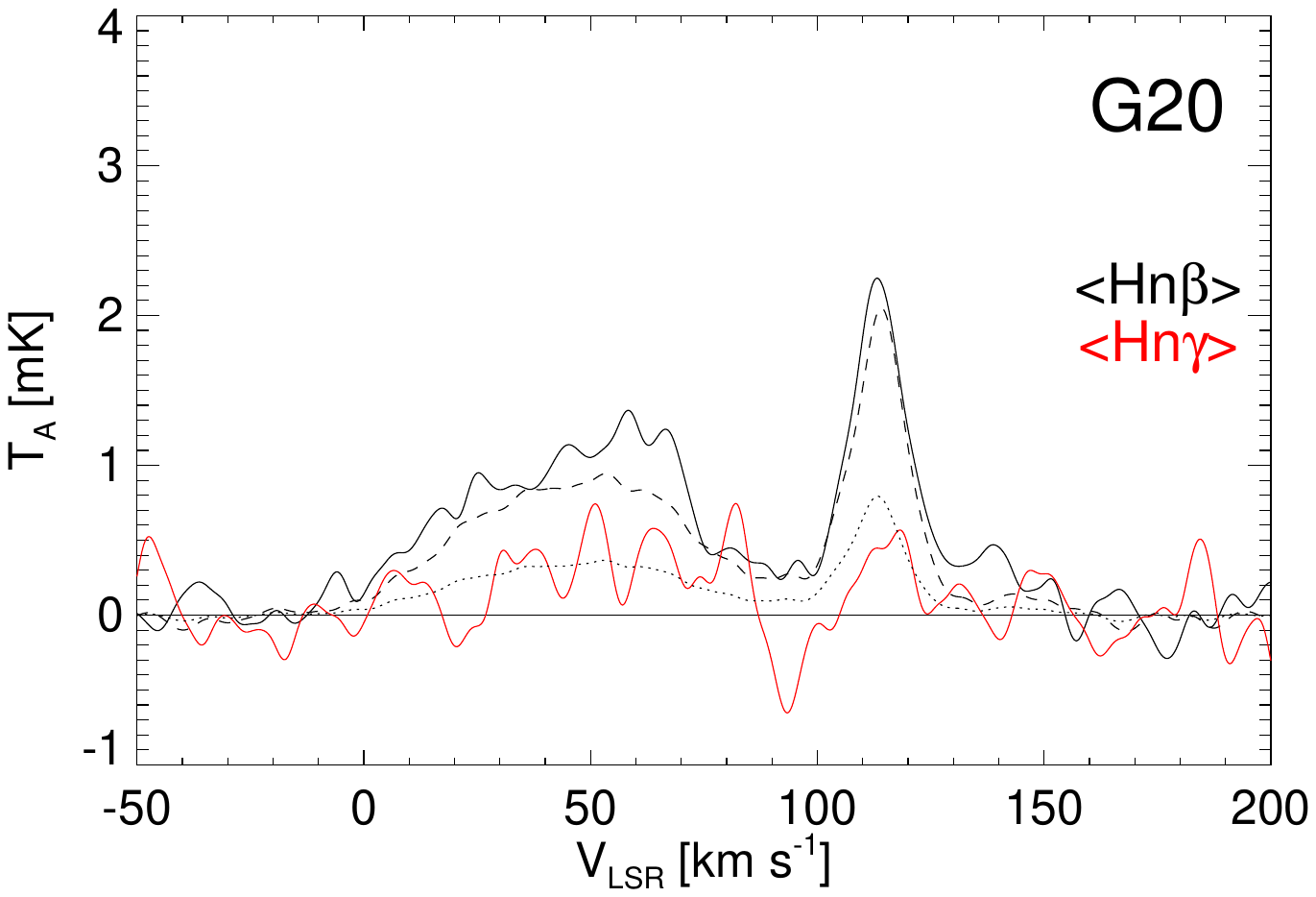}\\
\vskip -1.8cm
\includegraphics[angle=0,scale=0.32]{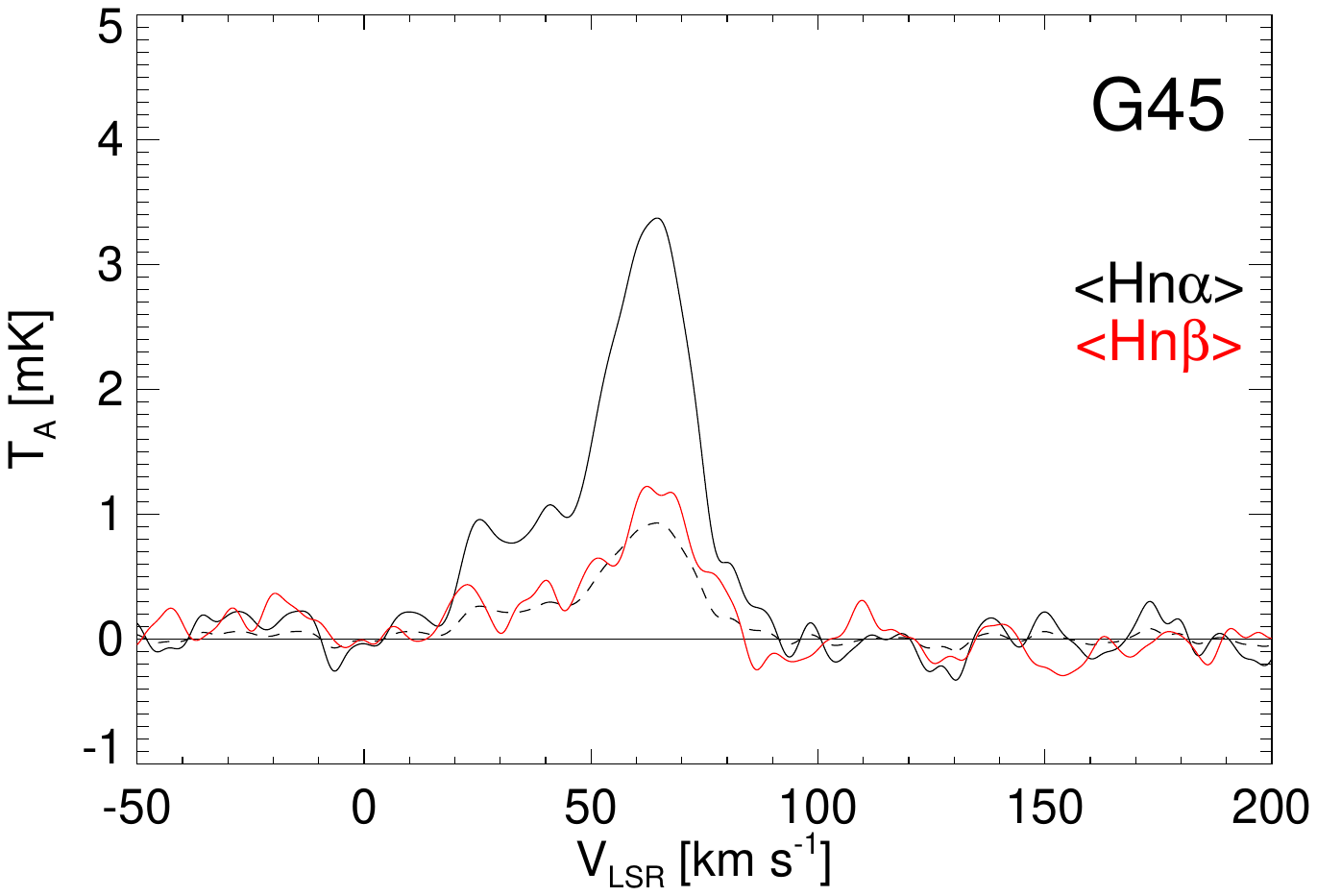}
%\vskip -3cm
\includegraphics[angle=0,scale=0.32]{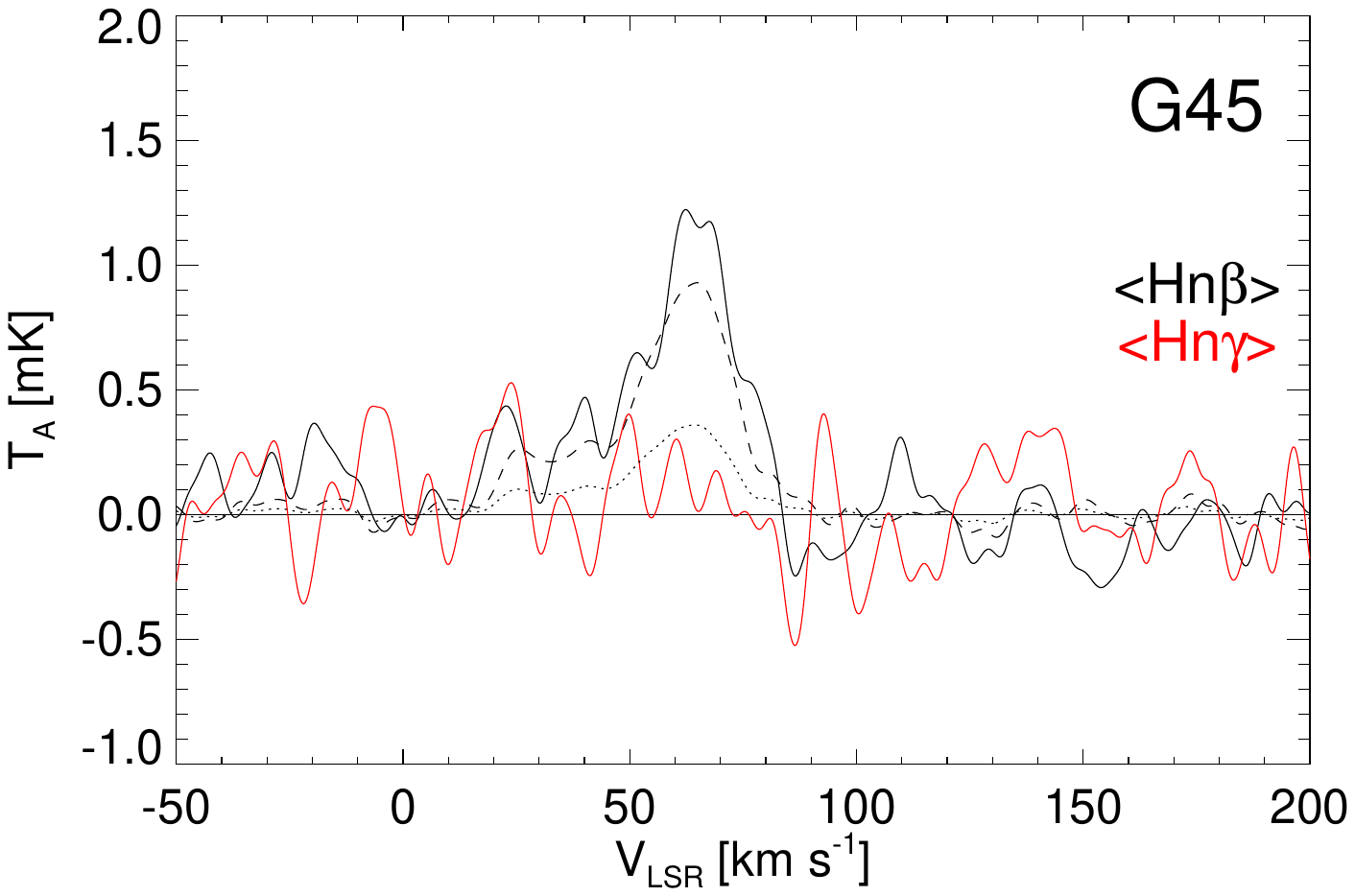}\\
\vskip -0.5cm
\caption{
G20 and G45 stacked spectra for the \hna, \hnb, and \hng\ transitions.
The spectra {\bf have been} smoothed to a 5\kms\ velocity resolution.  {\bf Left
  Panels} show the \hna\ and \hnb\ spectra. The dashed line
spectrum is the expected LTE intensity of the $\beta$ transition.
{\bf  Right Panels} show the \hnb\ and \hng\ spectra. The dashed and
dotted spectra are the expected LTE intensities of the $\beta$ and
$\gamma$ transitions, respectively.
}
\label{fig:LTEspectra}
\end{figure}

Are the \hna, \hnb, and \hng\ stacked spectra consistent with LTE
excitation? Here, we summarize our analysis of this question.  The
quantum mechanical details are provided in Appendix~B where it shown
that for our stacked spectra the expected LTE
$\langle\beta\rangle/\langle\alpha\rangle$ and
$\langle\gamma\rangle/\langle\alpha\rangle$ line intensity ratios should
be 0.27565 and 0.10672, respectively (see Table~\ref{tab:LTEinfo}).
We assess whether LTE excitation holds for our targets in two ways:
(1) we scale the \hna\ spectrum by the Table~\ref{tab:LTEinfo} expected
LTE ratios for $\langle \beta\rangle $ and $\langle \gamma\rangle$ RRLs
and compare this in Figure~\ref{fig:LTEspectra} to our stacked \hnb\ and
\hng\ spectra; and
(2) we fit Gaussian functions to the RRL emission components found in
the \hna, \hnb, and \hng\ spectra and compare these in
Table~\ref{tab:LTEfits} with the expected LTE ratios.
In both cases all the stacked spectra are smoothed to a velocity
resolution of 5\kms\ to improve spectral sensitivity.

The first assessment of LTE is shown in Figure~\ref{fig:LTEspectra} where
the left hand plots compare the \hna\ and \hnb\ spectra for G20 and G45
with the expected LTE \hnb\ spectrum. The right hand plots compare the
\hnb\ and \hng\ spectra with the expected LTE intensities.  Both sight
lines show \hnb\ RRL emission that is slightly stronger than what is
predicted by LTE. Although the G45 $\sim$35\kms\ component matches the
LTE prediction, this \hnb\ component is only a $\sim$\,1\,$\sigma$
signal. Compared with the other transitions the \hng\ spectra are not
very sensitive (see Table~\ref{tab:LTEfits}). Only the
$\sim$\,50\kms\ emission component for G20 has any hint of $\gamma$
emission and this is only a $\gsim\,1.5\,\sigma$ signal.  For all other
\hng\ components we can only provide upper limits for the LTE ratio.

The second evaluation of LTE uses Gaussian fits to the individual
emission components found in the stacked $\alpha$, $\beta$, and $\gamma$
spectra.  To make the LTE assessment that is summarized in
Table~\ref{tab:LTEfits}, we use the observed integrated intensity of
each component, \wrrl.  (The slight differences in the reported fits for
$\alpha$ components that are seen when comparing
Tables~\ref{tab:WIMfits} and \ref{tab:LTEfits} stem from the different
velocity resolutions of the \hna\ spectra.)

The ``LTE?'' column in Table~\ref{tab:LTEfits} gives the ratio between
the observed RRL transition ratios and the expected LTE ratio. In LTE
the value of this ratio of ratios would be unity.  For the strongest WIM
components --- 113.4\kms\ for G20 and 63.3\kms\ for G45 --- this ratio
is $1.11\pm0.11$ and $1.08\pm0.11$, respectively.  We thus find that,
within the errors, our \hnb/\hna\ ratios are consistent with LTE
excitation for both these emission components. The lower velocity
components in these directions, however, have \hnb/\hna\ ratios that are
significantly larger than unity and thus are not consistent with
LTE. Moreover, all ratios involving the \hng\ spectra are compromised by
the much poorer sensitivity of these spectra (see Appendix~B).

\begin{deluxetable}{lccccccccccccc}
\tabletypesize{\scriptsize}
\tablecaption{\hna, \hnb, and \hng\ Emission Properties\label{tab:LTEfits}}
\tablehead{
  \colhead{Source}       &
  \colhead{$\Delta$ n}  &
  \colhead{\vlsr}        & \colhead{$\sigma$}  &
  \colhead{\tla}         & \colhead{$\sigma$}  &
  \colhead{\fwhm}        & \colhead{$\sigma$}  &
  \colhead{$W_{\rm RRL}$}  & \colhead{$\sigma$}  &
  \colhead{$W_{\rm RATIO}$} & \colhead{$\sigma$}  &
  \colhead{LTE?\tablenotemark{a}} &\colhead{$\sigma$} \\
  \colhead{}             &
  \colhead{}             &
  \colhead{(\kms)}       & \colhead{} &
  \colhead{(\mk)}        & \colhead{} &
  \colhead{(\kms)}       & \colhead{} &
  \colhead{(\area)}      & \colhead{} &
  \colhead{($W(\Delta n)\over W(\alpha)$)} &
  \colhead{}             & \colhead{}           
}
\startdata
G20 &1: $\alpha$& \phn 47.8&0.10 & \phs 3.33& 0.17 &59.4 &2.97 &\phd  210.4 &    14.9 &\phs 1.00&\dots   &\phs \dots & \dots \\
    &2: $\beta$ & \phn 48.5&0.21 & \phs 1.17& 0.06 &62.4 &3.12 &\phd\phn 77.8 &\phn 5.5 &\phs 0.37& 0.04 &\phs 1.34  & 0.13  \\
    &3: $\gamma$& \phn 55.3&0.74 & \phs 0.74& 0.37 &42.0 &2.10 &\phd\phn 33.1 &    16.6 &\phs 0.16& 0.08 &\phs 1.48  & 0.75  \\
G20 &1: $\alpha$&     113.4&0.02 & \phs 6.90& 0.35 &14.8 &0.74 &\phd    108.8 &\phn 2.4 &\phs 1.00&\dots &\phs \dots & \dots \\
    &2: $\beta$ &     113.7&0.06 & \phs 2.11& 0.11 &14.8 &0.74 &\phd\phn 33.3 &\phn 2.4 &\phs 0.31& 0.03 &\phs 1.11  & 0.11  \\
    &3: $\gamma$&     115.8&0.33 &   $<$0.54& 0.03 &11.4 &0.57 &\phn   $<$6.5 &\phn\dots&  $<$0.06&\dots &$<$0.56    & \dots \\
G45 &1: $\alpha$& \phn 34.1&0.32 & \phs 0.93& 0.05 &27.5 &1.38 &\phd\phn 27.2 &\phn 1.9 &\phs 1.00&\dots &\phs \dots & \dots \\
    &2: $\beta$ & \phn 37.6&2.34 & \phs 0.32& 0.02 &42.8 &6.41 &\phd\phn 14.6 &\phn 2.3 &\phs 0.54& 0.09 &\phs 1.95  & 0.34  \\
    &3: $\gamma$&     \dots&\dots&   $<$0.50& \dots&\dots&\dots&      $<$14.9 &\phn\dots&  $<$0.55&\dots & $<$5.14   & \dots \\
G45 &1: $\alpha$& \phn 63.3&0.08 & \phs 3.35& 0.17 &21.2 &1.06 &\phd\phn 75.8 &\phn 5.4 &\phs 1.00&\dots &\phs \dots & \dots \\
    &2: $\beta$ & \phn 65.3&0.18 & \phs 1.09& 0.02 &19.4 &1.16 &\phd\phn 22.5 &\phn 1.8 &\phs 0.30& 0.03 &\phs 1.08  & 0.11  \\
    &3: $\gamma$&     \dots&\dots&   $<$0.20& \dots&\dots&\dots&\phn   $<$4.5 &\phn\dots&  $<$0.06&\dots &$<$0.56    & \dots \\
\enddata
\tablenotetext{a}{
  Ratio between observed $W_{\rm RATIO}$ and LTE ratio
  (Table~\ref{tab:LTEinfo}); a plasma in LTE has value unity here.
}
\end{deluxetable}

%\begin{figure}[h]
%\centering
%\vskip -0.5cm
%\includegraphics[angle=0,scale=0.80]{fig7.pdf}
%\vskip -1cm
%\caption{
%Galactic longitude versus LSR velocity, \lv, map for discrete, OB-star
%excited \hii\ regions from the {\em WISE} Catalog compared with the WIM RRL
%emission components observed for our sight lines. Black circles are the
%LSR velocities of discrete \hii\ regions. Red triangles and horizontal
%lines locate the LSR velocity and FWHM of the four RRL emission
%components we see toward G20 and G45.
%}
%\label{fig:LVmap}
%\end{figure}

%\clearpage
\subsection{WIM Ionization from Leakage Radiation?}

WIM plasma halos envelope the PDRs that surround \hii\ regions ionized
by OB-type stars.  Case studies of individual \hii\ regions using RRL
measurements, e.g. RCW\,120 \citep{2015RCW120} and \ngc{7538}
\citep{2016NGC7538}, find that WIM halos are produced by stellar UV
photons leaking through the PDRs and that the fraction of the leaking
radio continuum emission escaping into the WIM is $\sim$25\% and
$\sim$15\%, respectively.
Furthermore, optical H$\alpha$ studies also suggest that if OB stars are
the source of the WIM ionization, then the ISM OB-type star excited 
\hii\ region distribution must allow at least $\sim$15--25\% of the 
H-ionizing Lyman continuum photons emitted by the stars to travel
hundreds of parsecs within the Galactic disk \citep{1991reynoldsWIM}.

\begin{figure}[h!]
\centering
%\vskip -1cm
\includegraphics[angle=0,scale=0.65]{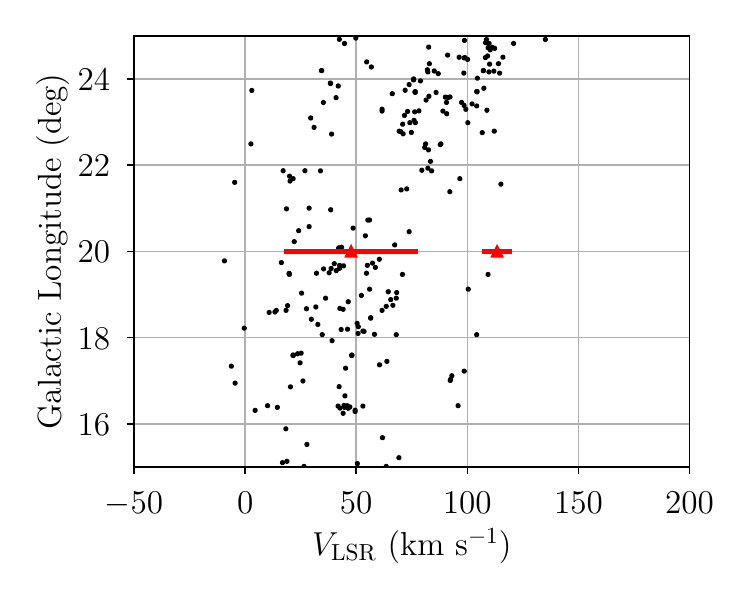}
\includegraphics[angle=0,scale=0.65]{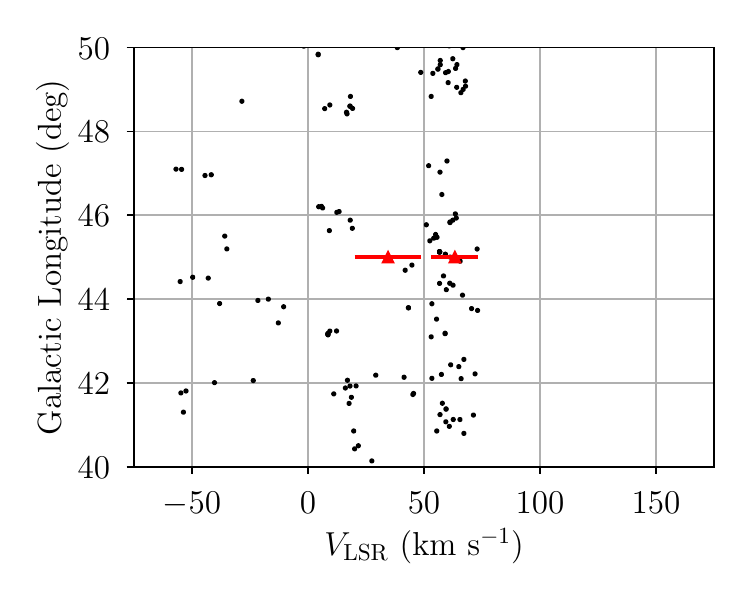}\\
%\vskip -1.5cm
%\includegraphics[angle=0,scale=0.32]{G20_HII.pdf}
%\includegraphics[angle=0,scale=0.32]{G45_HII.pdf}\\
\includegraphics[angle=0,scale=0.55]{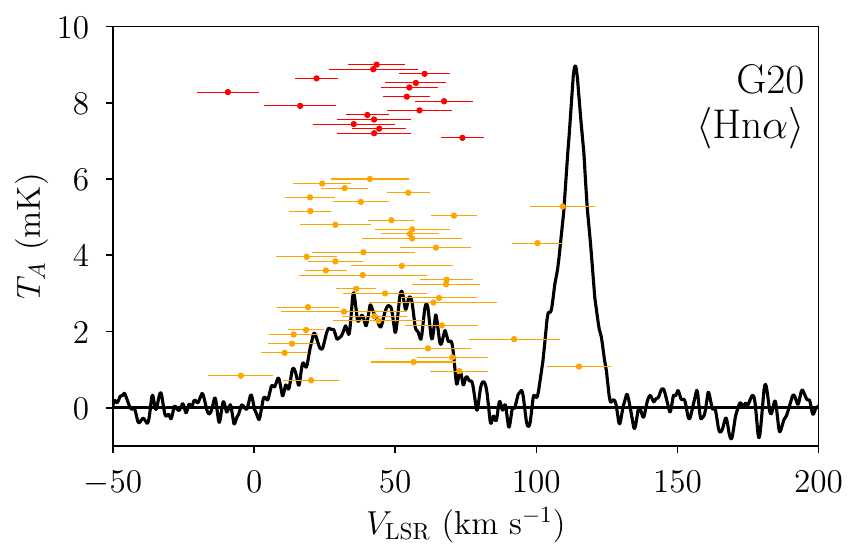}
\includegraphics[angle=0,scale=0.55]{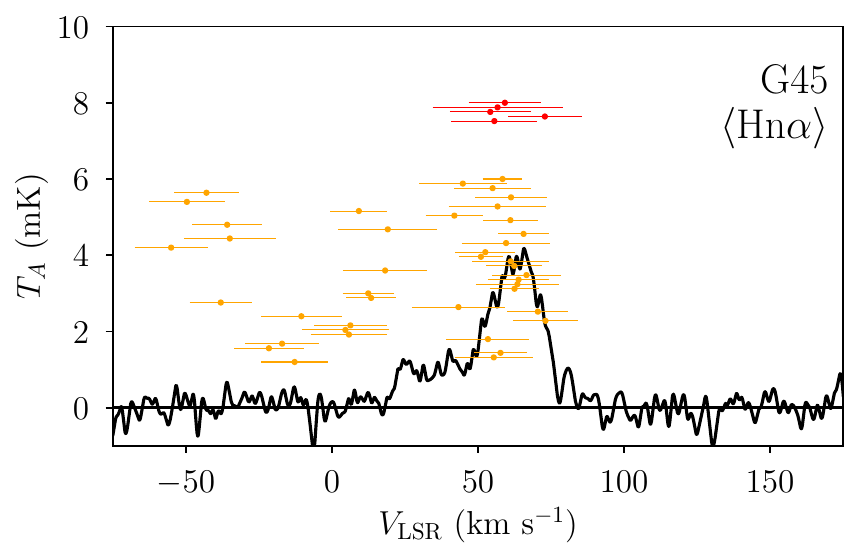}\\
%\vskip -0.5cm
\includegraphics[angle=0,scale=0.55]{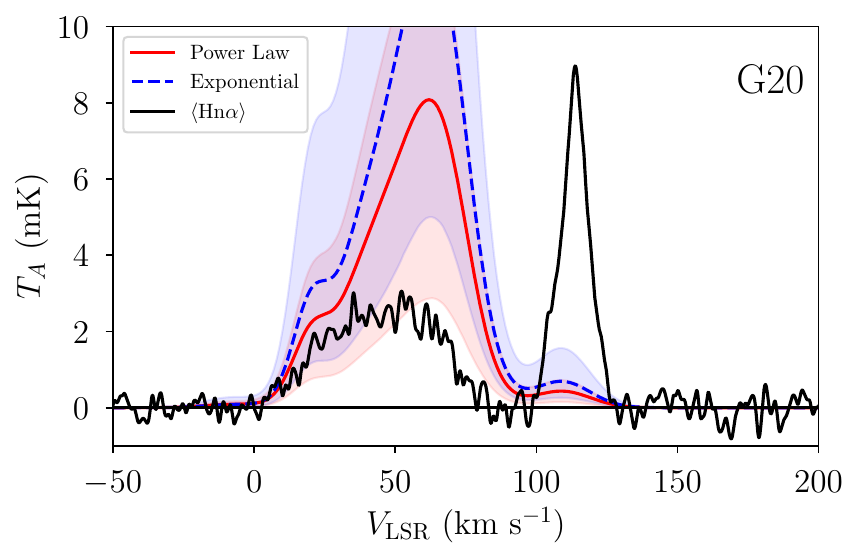}
\includegraphics[angle=0,scale=0.55]{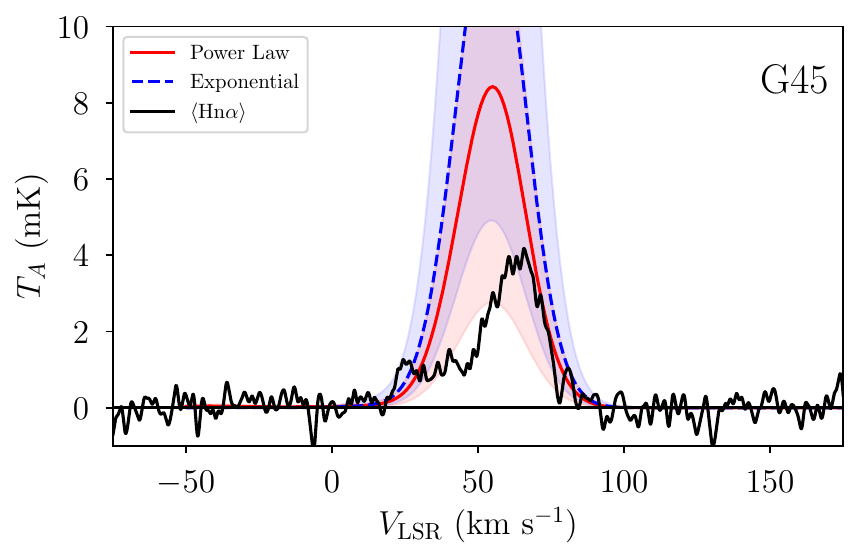}\\
\caption{
Comparison of G20 and G45 WIM emission with that from known
Galactic \hii\ regions.
{\bf Top Plots:}
\lv\ distributions of OB-type star excited \hii\ regions near the G20
and G45 sight lines.
Black circles are the LSR velocities of discrete \hii\ regions.
Red triangles and horizontal lines locate the LSR velocity and FWHM of
the four RRL emission components we see toward G20 and G45. 
{\bf Middle Plots:}
Comparison of the stacked \hna\ WIM spectra with Galactic
\hii\ regions. Shown are the RRL \vlsr\ and FWHM line widths of
\hii\ regions from the {\em WISE} Catalog plotted as a function of
increasing angular separation of the nebular location from the G20 and
G45 sight lines. The nearest \hii\ region has the largest y-axis
value. Red points are sources located within 30\arcmin\ of each sight
line; orange points are those located between 30\arcmin\ and 100\arcmin.
{\bf Bottom Plots:}
Predicted WIM emission due to diffuse ionized gas halos around
\hii\ regions. The solid red and dashed blue curves are the predictions
of the \citet{luisi20} power law and exponential models,
respectively. The shaded regions represent the 95\% confidence
intervals.
}
\label{fig:wimHIIcomposite}
\end{figure}

Here, we assess whether the RRL emitting WIM plasma we find toward G20
and G45 is being ionized by leakage radiation from nearby, discrete
\hii\ regions.  
The Galactic context of our sight lines is provided by the top panels of
Figure~\ref{fig:wimHIIcomposite} which show the locations and LSR
velocities of all OB-type star excited \hii\ regions in the \lv\ zone of
the plots.
Because we chose our targets to be devoid of obvious sources of ionizing
radiation, as expected, neither sight line coincides in \lv-space with
any major concentration of \hii\ regions. This is especially the case
for the $\sim$115\kms\ component of G20 and the $\sim$35\kms\ component
of G45: neither is located on the sky near any significant number of
Galactic \hii\ regions having their LSR velocities.

The middle panels of Figure~\ref{fig:wimHIIcomposite} confirm that there
are few Galactic \hii\ regions that are near both in sky location and
LSR velocity to the G20 $\sim$115\kms\ and G45 $\sim$35\kms\ RRL
emission components.  These plots show the \vlsr\ and FWHM of {\em WISE}
Catalog \hii\ regions that are located on the sky within 100\arcmin\ of
the G20 and G45 sight lines plotted as a function of increasing angular
separation of the nebular location from the G20 and G45 sight lines.

The nearest \hii\ region has the largest y-axis value.  The remaining
\hii\ regions are plotted in separation sequence with the largest
separation having the smallest y-axis value.  (Other than providing a
co-ordinate for plotting the angular separation {\em sequence}, these
y-axis {\em values} have no physical meaning.)  Red points in
Figure~\ref{fig:wimHIIcomposite} are sources located within
30\arcmin\ of each sight line.
There are 17 and 5 {\em WISE} Catalog \hii\ regions located on the
sky within 30\arcmin\ of the G20 and G45 sight lines, respectively. Only
3 nebulae are within 10\arcmin\ of our sight lines (2 for G20 and 1 for
G45) and the closest \hii\ region lies 9\arcmin\ from G45. 

Some properties of these \hii\ regions are compiled in Appendix
Table~\ref{tab:HIIregions}. Listed for each nebula is the angular
separation from the fiducial line of sight (LOS), the IR radius, the
Galactic \lb\ position, the RRL parameters (RRL intensity, LSR
velocity, and FWHM line width), together with the measurement errors.
Again, the G20 $\sim$115\kms\ and G45 $\sim$35\kms\ components show very
little \lv\ correlation with known Galactic \hii\ regions and there are
no \hii\ regions at these velocities within 30\arcmin\ of these sight
lines.

%Can WIM plasma halos around \hii\ regions explain the observed WIM
%emission? 
%
Although there are few \hii\ regions in close \lv proximity to our sight
lines, we may nonetheless be sensitive to extended WIM halos surrounding
these nebulae.  \citet{luisi20} develop empirical plasma halo models for
the \hiisub\ region complex W43. These models reproduce the observed RRL
properties of maps of the WIM surrounding W43.  The models assume that
the WIM RRL intensity depends only on (1) the RRL intensity of nearby
\hii\ regions and (2) the angular distance to those \hii\ regions from
the fiducial LOS relative to the angular size of each nebulae.  For the
W43 complex, they explore both an exponential and a power law WIM
emission distribution model, and they find that the power law model is
better able to reproduce the integrated WIM RRL intensity.

We apply the \citet{luisi20} models to our sight lines in order to
explore the possibility that ionized halos around \hii\ regions can
explain the WIM emission in these directions. The bottom panels of
Figure~\ref{fig:wimHIIcomposite} show the predicted RRL emission due to
WIM halos around all \hii\ regions within 100\arcmin\ of each sight
line. The red solid lines show the power law model prediction,

\begin{equation}
T_{A,\rm model} = k \sum_i T_{i,\hiisub} \left(\frac{r_i}{r_{i,\hiisub}}\right)^m,
\end{equation}

\noindent and the blue dashed lines show the exponential model prediction,

\begin{equation}
T_{A,\rm model} = k \sum_i T_{i,\hiisub}\,{\rm e}^{-m (r_i / r_{i,\hiisub})},
\end{equation}
    
\noindent where $T_{i,\hiisub}$ is the \hii\ region RRL spectrum, $r_{i}$ is the
angular separation between the sight line and the nominal \hii\ region
position, $r_{i,\hiisub}$ is the angular IR radius of the \hii\ region,
and $k$ and $m$ are the free parameters. The \hii\ region RRL spectrum
is evaluated from the RRL parameters in Appendix
Table~\ref{tab:HIIregions}. The sum is taken over all \hii\ regions
within 100\arcmin\ of each sight line for which there is a RRL intensity
measurement listed in the table. We adopt $k = 0.28\pm0.08$ and $m =
-1.85 \pm 0.12$ for the power law model and $k = 0.15\pm0.04$ and $m =
0.33\pm0.05$ for the exponential model as determined by \citet{luisi20}
for the W43 complex.

We note, however, two important differences between \citet{luisi20} and
our analysis: (1) they fit their model to the integrated RRL intensity
rather than the RRL spectra, and (2) they parameterize their model in
terms of the average integrated RRL intensity over the \hii\ region
rather than the ``peak'' RRL spectrum at the nominal \hii\ region
position. For nebulae comparable in size to the telescope beam, the
``peak'' intensity is equal to the average intensity, but for angularly
large nebulae the difference will depend on the emission morphology. The
shaded regions in the bottom panels of Figure~\ref{fig:wimHIIcomposite}
represent the 95\% confidence intervals determined by Monte Carlo
resampling the \hii\ region RRL parameters and the model parameters $k$
and $m$.

These empirical models, fit to the W43 complex, may not be applicable to
every Galactic \hii\ region. By applying such models here, we inherently
assume that the relative distribution of WIM emission around every
\hii\ region is the same as that around nebulae in the W43
complex. \citet{luisi19} find significant variations in the radial
distribution of RRL emission around several \hii\ regions. The magnitude
of this variation is far greater than the statistical uncertainties
shown in the bottom panels of
Figure~\ref{fig:wimHIIcomposite}. Furthermore, the model scaling factor,
$k$, is likely related to the fraction of ionizing photons that leak
into the WIM; thus $k$ probably varies from \hii\ region to \hii\ region
depending on the environment.

It is thus not surprising that the bottom panels of
Figure~\ref{fig:wimHIIcomposite} show clearly that \citet{luisi20}
models with $k$ and $m$ parameter values derived for the W43 complex do
not account for the WIM RRL intensities seen toward G20 and G45. The
model intensities are too high by factors of $\sim$4 and $\sim$2,
respectively, for G20 and G45. These sight lines were purposely chosen
{\it not} to thread through environments even remotely like those found
in massive star forming regions such as W43.  

Finally, it is clear from Figure~\ref{fig:wimHIIcomposite} that the
${\sim}115\kms$ component toward G20 cannot be due to \hii\ region
leakage radiation. No nearby \hii\ region is sufficiently bright or has
a sufficiently narrow line width to explain this feature, as
demonstrated by the \citet{luisi20} models and Appendix
Table~\ref{tab:HIIregions}. We conclude it to be unlikely that this
feature is due to extended ionized gas halos around \hii\ regions.  It
may instead constitute a heretofore unrecognized phase of the WIM (see
Section~\ref{sec:discussG20}).  To make further progress in the
exploration of the nature of plasma halos around \hii\ regions and this
narrow RRL feature we will need sensitive RRL maps of the WIM
surrounding G20 and G45. Such maps will allow us to \textit{fit}
\citet{luisi20} type models to these data.

%\clearpage

\subsection{Galactic ISM Context}

\begin{figure}[h]
\centering
\vskip -1.5cm
\includegraphics[angle=0,scale=0.32]{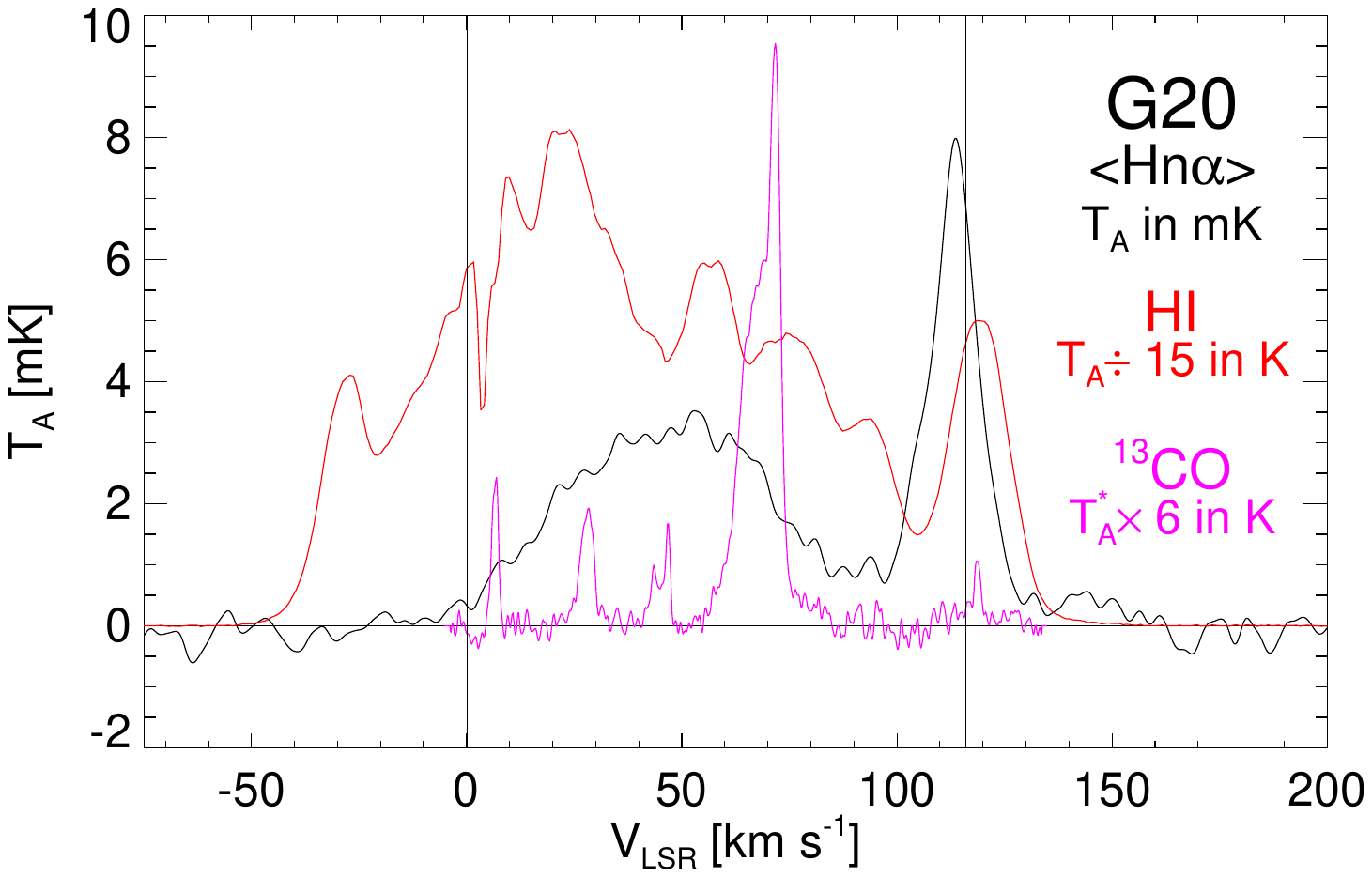}
%\vskip -3cm
\includegraphics[angle=0,scale=0.32]{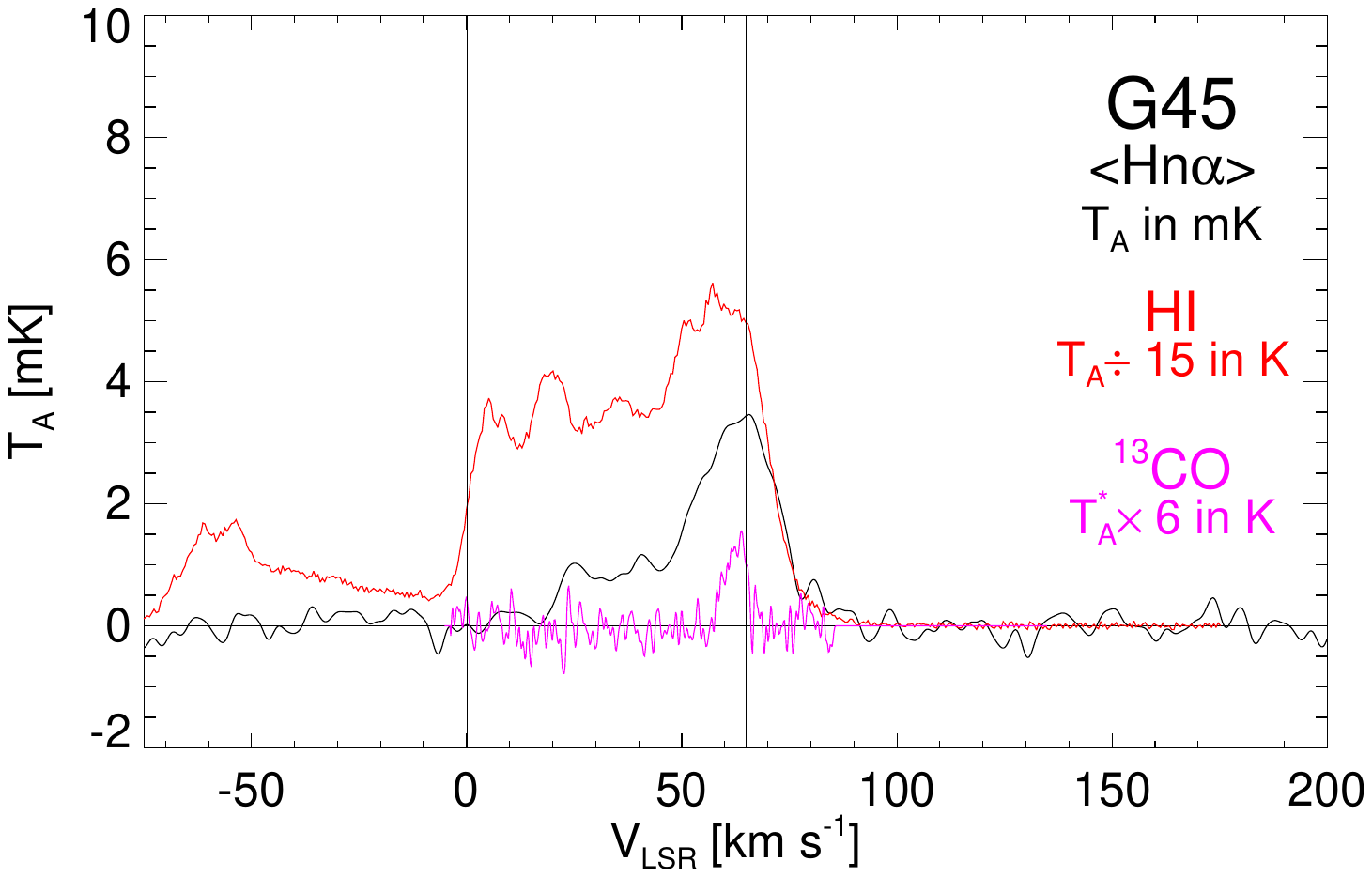}\\
\vskip -0.5cm
\caption{
Spectra toward G20 and G45 showing all phases of gaseous interstellar
hydrogen: \hna\, \hi, and \cor\ (a tracer of \htwo).  For clarity the
\hi\ antenna temperatures in \K\ are divided by a factor of 15 and the
\cor\ $T_{A}^*$ antenna temperatures in \K\ are multiplied by a factor
of 6.  Vertical flags mark LSR velocities 0\kms\ and the terminal
velocity, \vt.
}
\label{fig:all}
\end{figure}

Here, we assess the Galactic context of our G20 and G45 sight lines by
comparing spectra for all phases of ISM hydrogen: \hplus, \hi, and
\htwo.  This is done in Figure~\ref{fig:all} which shows for each sight
line spectra for \hna, 21\cm\ \hi, and \cor(1$\rightarrow$0) which is a
proxy for \htwo.  These spectra were taken by different telescopes over
a large range of frequencies so their beam sizes sample different LOS
volumes.
The \hna\ WIM RRL spectra have a HPBW of $2\arcmper35$.  For G20 the
\hi\ spectrum is from the HI4PI all sky survey
\citep[HPBW=$16\arcmper2$]{2016HI4PI}. The G45 \hi\ spectrum stems from
the Boston University--Arecibo Observatory \hi\ survey
\citep[HPBW=$3\arcmper2$]{1992BUAO}.  The \cor\ spectra come from the
Boston University--Five College Radio Observatory \cor\ Galactic Ring
Survey \citep[HPBW=42\arcsec]{2006GRS}.
Despite their different angular resolutions, comparing these spectra can
nonetheless provide useful insights.  For example, the terminal
velocity, \vt, flagged in Figure~\ref{fig:all} is calculated for each
sight line using the \citet{1985Clemens} rotation curve.  For both sight
lines the \hplus, \hi, and \cor\ spectra all have emission components at
or near the terminal velocity.

Our sight lines probe two important loci in the first quadrant of the
Galaxy. Some key properties of these sight lines, based on an assumed Sun
to Galactic Center distance, $R_0$, of 8.5\kpc, are summarized in
Table~\ref{tab:LOSinfo}. (Other choices for $R_0$ would not alter the
conclusions we reach here anent the spatial distribution of the WIM
plasma.) Listed in Table~\ref{tab:LOSinfo} for each LOS are the minimum
Galactocentric radius, \Rmin=$R_0$\,sin(\gl), tangent point distance
from the Sun, \Dtp=$R_0$\,cos(\gl), and total path length inside the
Solar orbit about the Galactic Center (GC), \Dlos=2$R_0$\,cos(\gl).

\begin{deluxetable}{cccc}[h]
\tabletypesize{\normalsize}
  \tablecolumns{4} \tablewidth{0pt}
\tablecaption{LOS Properties\label{tab:LOSinfo}}
\tablehead{
  \colhead{$LOS$} &
  \colhead{\Rmin} &
  \colhead{\Dtp}  &
  \colhead{\Dlos}\\
  \colhead{} & \colhead{(\kpc)} & \colhead{(\kpc)} & \colhead{(\kpc)}
}
  \vspace{-10pt}
\startdata
G20 & 2.9 & 8.0 & 16.0 \\
G45 & 6.0 & 6.0 & 12.0 \\
\enddata
\tablecomments{$R_{\rm 0}$\,=\,8.5\kpc.}
\end{deluxetable}

Atomic hydrogen gas is ubiquitously distributed throughout the Galactic
ISM.  Because of this, for any line of sight 21\cm\ \hi\ spectra show
emission at all LSR velocities permitted by Galactic rotation.  The
velocity span of these \hi\ spectra defines the maximum velocity spread
permitted by Galactic rotation.
For our first quadrant targets, gas at negative velocities is located in
the Outer Galaxy beyond the Solar orbit about the Galactic Center.  The
Figure~\ref{fig:all} \hi\ spectra have emission at negative LSR
velocities that extends to $\sim-50$\kms\ and $\sim-75$\kms\ for G20 and
G45, respectively.  In contrast, neither of our target sight lines shows
any \hna\ RRL emission at negative LSR velocities at the sensitivity
level achieved here.  G20 shows \hna\ emission at all velocities between
0\kms\ and the terminal velocity whereas G45 only shows RRL emission
between $\sim$+20\kms\ and the terminal velocity.
Assuming perfect circular rotation and no streaming motions, all of the
RRL emitting gas we see toward G20 and G45 must therefore be located in
the Inner Galaxy, inside the Solar orbit. This plasma must reside 
somewhere along the line of sight paths, $d_{\rm LOS}$, summarized in
Table~\ref{tab:LOSinfo}.

Unlike \hi, molecular gas in the Milky Way is found in comparatively
dense, discrete clouds. This means that \htwo/CO spectra will not have
molecular emission spanning all available LSR velocities. As
Figure~\ref{fig:all} clearly shows, the molecular gas seen toward our
target directions is concentrated into a few discrete emission
components. At the sensitivity of the GRS, G20 shows 5 molecular clouds
and G45 has but one. For both sight lines there are \cor\ emission
components that match the LSR velocities of the highest velocity RRL
components in the \hna\ spectra, albeit the $\sim$\,114\kms\ G20 \cor\ 
component is extremely weak.  The GRS spectrometer could not detect
emission at negative LSR velocities so these \cor\ spectra provide no
Outer Galaxy information about molecular clouds toward these sight lines.

Cold \hi\ embedded in a molecular cloud produces the absorption dips
seen in the G20 \hi\ spectrum at \vlsr\ = $\sim5$, $\sim50$, and
$\sim$70\kms. These \hi\ dips are matched by \cor\ emission
components. We suspect that all the \cor\ components produce
\hi\ absorption but cannot definitively prove this due to a combination
of mismatched angular resolution, insufficient spectral sensitivity, and
the complex structure of Galactic \hi\ spectra.

%\clearpage
\vskip 1cm

\section{WIM RRL Emission Models}

Here, using the \hna\ emission seen from our targets, we seek to derive
constraints on the line of sight density and temperature distributions
of the RRL emitting plasmas. Complete details of the modeling are provided in
Appendix~D.
Models for RRL emission from LOS plasmas must specify the 
electron density, \Ne, temperature, \te, and velocity dispersion,
$\sigma$, at every point along the \Dlos.
Each model is comprised of one or more plasma ``clouds'' distributed
along the LOS. All clouds are homogeneous, isothermal plasmas in
LTE. Each cloud's properties are specified at input.  A cloud is defined
by: location along the LOS, \dsun, LOS path length size (aka the cloud
diameter), \Dcloud, electron density, electron temperature, and velocity
dispersion. We use the numerical code described in Appendix~D to compute
synthetic spectra for H109$\alpha$ RRL emission from the model plasmas.
After these model \tbv\ RRL spectra are calculated, we compare them to
\hna\ spectra that are converted to brightness temperature and smoothed
to 5\kms\ resolution.

We emphasize that the models are only intended to provide estimates of
these quantities so we do not attempt to fully explore a large grid of
parameter choices.  Moreover, we do not apply any rigorous numerical
metric to evaluate the ``goodness of fit'' between a model and the
\hna\ spectrum.  We judge a model's fit by eye because we can only set
limits on the electron temperature.

From the observed \vlsr\ span of these spectra we know, assuming
circular rotation, that the RRL emitting plasma must be located within
the Solar orbit. The simplest assumption is that a constant density
isothermal plasma fills the entire \Dlos\ in each sight line (see
Table~\ref{tab:LOSinfo}). For this case, the observed EM provides an
estimate for the rms electron density, \nerms: 

\begin{equation}
  \nerms\ = \left({EM\over d_{\rm los}}\right)^{1/2}
  \label{eqn:Ne}
\end{equation}

\noindent Table~\ref{tab:WIMfits} shows that the total EM from the G20
and G45 emission components is 369.3 and 402.7\,\emeas,
respectively. This gives from Equation~\ref{eqn:Ne} an \nerms\ of 0.15
and 0.18\percc, respectively, for these sight lines. These
\nerms\ estimates, however, do not account for gas clumping: there may
be significant gaps and/or \Ne\ density fluctuations in the plasma
distribution along each LOS.  Many different LOS \Ne\ distributions can
produce identical \nerms\ values.

Due to  the first Galactic quadrant distance ambiguity, we cannot
know {\em a priori} where the RRL emitting plasma is located {\em vis a
  vis} the LOS near/far locations.  A series of models seeking to find
cloud parameters that produce a synthetic spectrum matching the
\hna\ observations confirms that there is no unique solution.
A plethora of models can reproduce the \hna\ spectra. Because of this,
models for our target's emission at LSR velocities having LOS distance
ambiguities provide no meaningful limits for the plasma density and
distribution along the LOS.

\begin{deluxetable}{cccccccccccc}[h]
\tablecolumns{10} \tablewidth{0pt}
\tablecaption{LOS Tangent Point Models 
  \label{tab:finalModels}}
\tablehead{
  \colhead{LOS}      & \colhead{Model}    & \colhead{\rgal}     &
  \colhead{\dsun}    & \colhead{\Dcloud}  & \colhead{\Ne}       &
  \colhead{\te}      & \colhead{$\sigma$} & \colhead{$EM$}      &
  \colhead{$p/k$}\\
  \colhead{}         & \colhead{}         & \colhead{(\kpc)}    &
  \colhead{(\kpc)}   & \colhead{(\kpc)}   & \colhead{(\percc)}  &
  \colhead{(\K)}     & \colhead{(\kms)}   & \colhead{(\emeas)}  &
  \colhead{(\nT)}
}
\startdata
G20&A  &2.9&8.0&\phn 2.0&0.38&4000&5.7&289&    1520\\
   &B  &2.9&8.0&\phn 2.0&0.29&2800&5.7&168&\phn 812\\
   &RMS&2.9&8.0&    16.0&0.15&4000&5.7&360&\phn 600\\
G45&A  &6.0&6.0&\phn 4.0&0.40&9000&8.6&640&    3600\\
   &B  &6.0&6.0&\phn 4.0&0.31&6400&8.6&384&    1984\\
   &RMS&6.0&6.0&    12.0&0.18&9000&8.6&389&    1620\\
\enddata
\end{deluxetable}

\begin{figure}[h]
\centering
\vskip -1.5cm
\includegraphics[angle=0,scale=0.32]{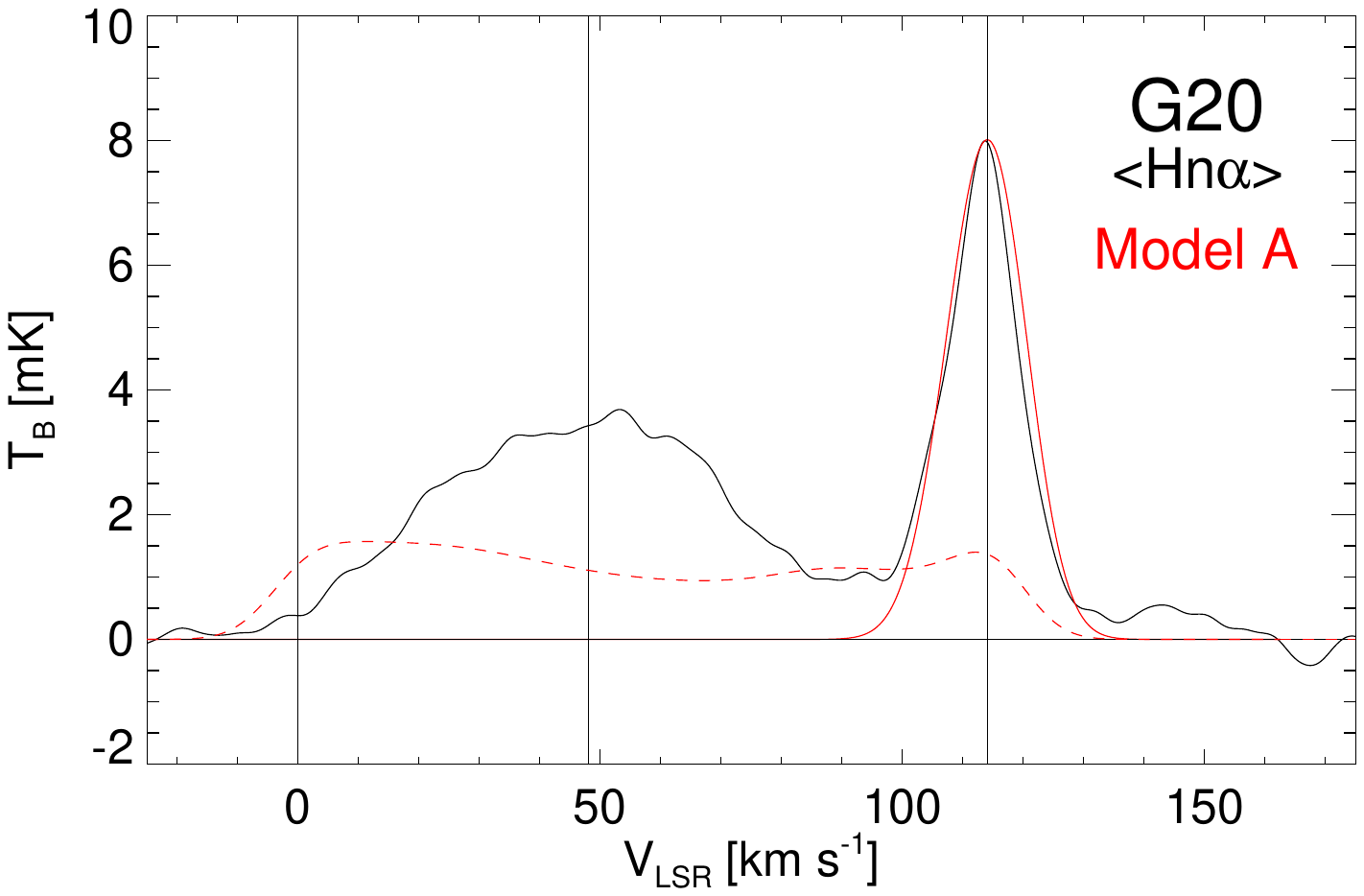}
\includegraphics[angle=0,scale=0.32]{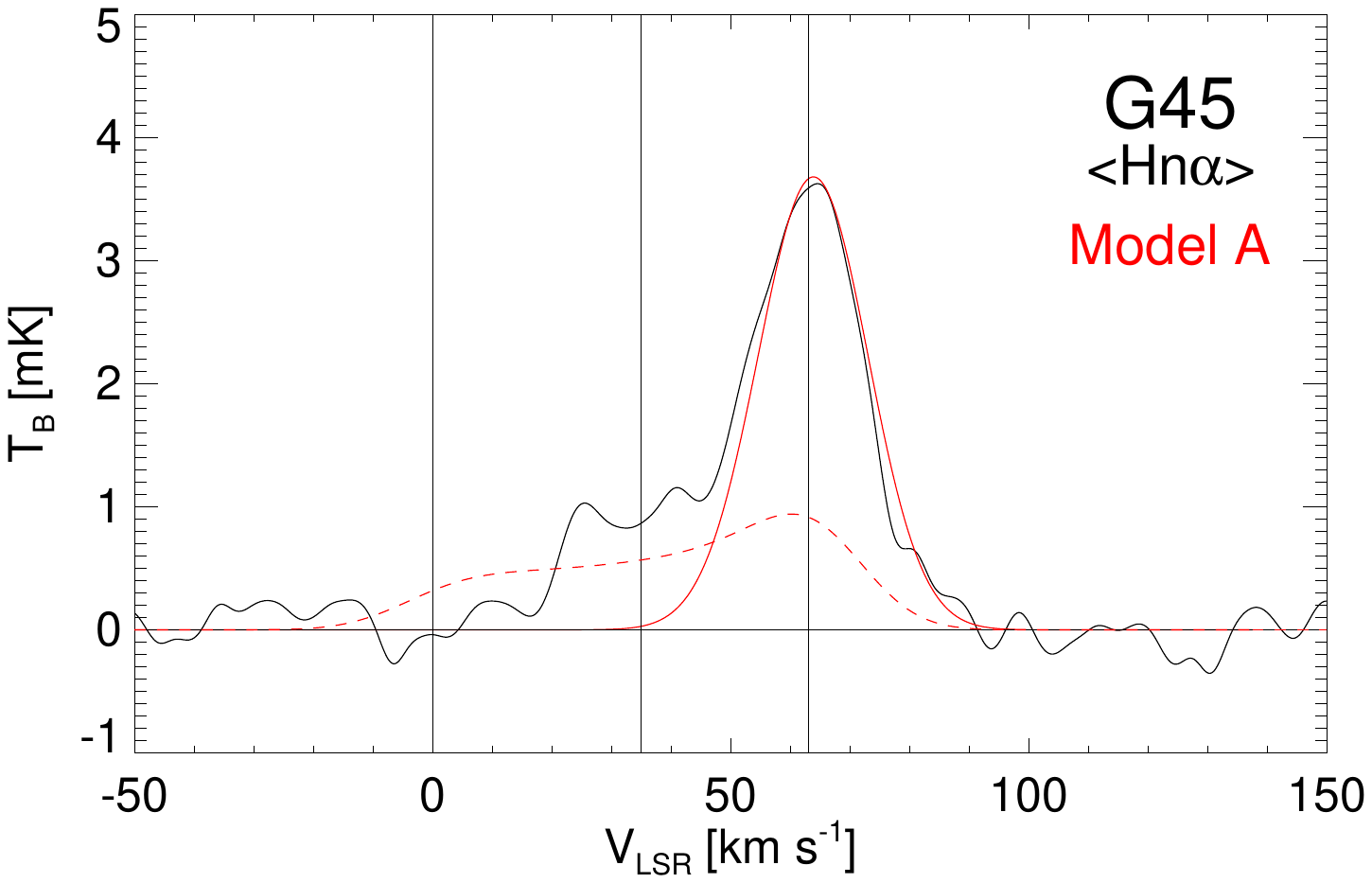}
\vskip -0.5cm
\caption{
LTE models for RRL emission (solid red curves) from the G20 and G45 high
velocity components compared with the observed \hna\ brightness
temperature spectrum (black curves).  Properties of the models are
summarized in Table~\ref{tab:finalModels}.  Vertical lines flag LSR
velocities at 0\kms\ and the observed component velocities listed in
Table~\ref{tab:WIMfits}. Model B is not shown because it is
indistinguishable from Model A.  The dashed red curve models have plasma
filling the entire LOS inside the solar orbit with the rms density from
Equation~\ref{eqn:Ne}.
}
\label{fig:finalModels}
\end{figure}

\subsection{LOS Tangent Point Models}

Models for RRL emitting gas located at the LOS tangent point distance,
however, can provide useful constraints for the plasma physical
properties. Gas is unambiguously located at the tangent point if it is
emitting at the LSR terminal velocity produced by Galactic rotation for
a particular first quadrant sight line.  The high velocity components
seen toward G20 and G45 emit at their terminal velocities and are hence
located at their respective tangent point distances from the Sun.

We model the RRL emission from these components using a single cloud
each for G20 and G45.  Each plasma cloud is specified by its LOS
distance, \dsun, cloud size, electron density, \Ne, electron
temperature, \te, and the velocity dispersion of the emission component,
$\sigma$.  Each cloud is located at the tangent point distance (see
Table~\ref{tab:LOSinfo}). The velocity dispersion is set by the observed
line width (see Table~\ref{tab:WIMfits} where $\sigma$ = \fwhm/2.355 for
a Gaussian line shape).

We use the observed RRL line width, \fwhm, to set the model plasma
electron temperature. This line width stems from a combination of
mechanisms including thermal, \dvth, and non-thermal broadening, \dvnt.
In addition, velocity shear along the line of sight produced by Galactic
rotation, $d\,\vlsr/d\,\dsun$, can also broaden the line emission,
\dvlos. (Quantum mechanical natural broadening is insignificant compared
with these mechanisms.)  The broadening processes add in quadrature to
produce the observed line width:
$\fwhm^2 = \Delta V^2_{\rm th} + \Delta V^2_{\rm nt} +
\Delta V^2_{\rm los}$. 
By definition the line of sight LSR velocity gradient at any Galactic
tangent point distance, $d_{\rm TP}$, is $d\,\vlsr/d\,d_{\rm TP} =
0$. Thus, for plasma located at the tangent point, spectral broadening
due to Galactic rotation, \dvlos, is negligible.
The WIM plasma is turbulent, however, so even at the tangent point the
observed RRL line width is an unknown combination of thermal and
non-thermal, supersonic turbulent broadening:
$\fwhm^2 = \Delta V^2_{\rm th} + \Delta V^2_{\rm nt}$.

We estimate \te\ in two ways, leading to two models each for G20 and
G45. Model A sets \te\ to be the upper limit derived using
Equation~\ref{eqn:telimit} and assuming that \dvobs\ is entirely due to
thermal broadening: \fwhm=\dvth. Because the plasma {\it is} turbulent,
this is a robust upper limit for \te. For Model A, \te\ is
4,000\,\K\ and 9,000\,\K\ for G20 and G45, respectively.  Model B
explores the effect of turbulent broadening by assuming that the thermal
and turbulent contributions to the line width are equal:
\dvth=\dvnt. The thermal contribution to the observed line width is then
\dvth =\fwhm/$\sqrt{2}$. For Model B, \te\ is 2,800\,\K\ and
6,400\,\K\ for G20, and G45, respectively.

Here, we seek to reproduce both the intensity and line shape of the
tangent point emission components. The model cloud parameters that are
yet to be determined are the electron density and cloud size.  For an
isothermal plasma the RRL intensity scales linearly with emission
measure so $\tb \propto \Ne^2$.  The model spectral intensities are very
sensitive to \Ne.  We find that changing \Ne\ by
$\pm$\,0.01\percc\ produces significant differences in the model peak
\tb\ intensity.  The cloud size is constrained by the line shape: if the
cloud is too large, Galactic rotation produces a distinctly non-Gaussian
line profile.

To craft the final model parameters we explored a range of choices for
\Ne\ and \Dcloud. 
The models are summarized in Table~\ref{tab:finalModels} which lists for
each cloud the minimum LOS Galactocentric distance, \rgal,
location along the LOS, \dsun,
LOS path length size (i.e., the cloud diameter), \Dcloud,
electron density, \Ne,
electron temperature, \te,
velocity dispersion, $\sigma$,
emission measure, EM, and
the plasma pressure, $p/k$. 
The Model A spectra for G20 and G45 are compared in
Figure~\ref{fig:finalModels} with the observed \hna\ emission.  The
Model B spectra are not shown because their \Ne\ has been adjusted to
fit the observed \hna\ and so they are indistinguishable from Model A.

The dashed red curve models in Figure~\ref{fig:finalModels} have plasma
filling the entire LOS inside the solar orbit with the rms electron
density from Equation~\ref{eqn:Ne} (see model ``RMS'' in
Table~\ref{tab:finalModels}). As expected these models are poor fits to
the observed \hna\ spectra because the plasma is surely not homogeneous
and isothermal throughout the \Dlos\ path length.  The models do,
however, show the LSR velocity span produced by plasma that fills the
entire \Dlos. For both sight lines these models produce larger
intensities for \vlsr's between $\sim$\,0 and $\sim$\,20\kms\ than what is
observed.  This may indicate that the plasma density is rather low near
the Sun for both directions.

\section {Discussion}

Here, we compare our results for G20 and G45 with other studies of the
WIM.  This is challenging because there is no universally accepted
definition of the WIM.  The plasma has been investigated using a variety
of spectral observations over a range of frequencies.  The WIM has been
studied in the optical, far infrared (FIR), and radio.  Interpreting
these spectra requires many different techniques and assumptions before
one derives physical properties such as density and temperature.
Furthermore, because of dust extinction and differing angular
resolutions, these three spectral regimes probe different volumes of
the Milky Way.

Historically, the distribution and physical properties of the WIM in
the Milky Way have primarily been characterized by observations of
optical spectral lines.  H$\alpha$ emission studies conclude that the
WIM accounts for $\sim 90$\% of the ionized gas in the ISM with a scale
height of $\sim 1-2$\kpc\ \citep[e.g.,][]{reynolds89, hill08}.
Dispersion measures from pulsars with known distances, together with
H$\alpha$ emission measures, estimate the range of the average electron
density to be $\sim$\, 0.03-0.10\percc\ with a filling factor, $f \sim $
0.4-0.2 \citep[e.g.,][]{reynolds91, taylor93, berkhuijsen06,
  gaensler08}.  The ratio of [\nii] $\lambda$5755 and [\nii]
$\lambda$6583 provides a direct measure of the electron temperature and
indicates the WIM is about 2,000\kel\ warmer than \hii\ regions
\citep{reynolds01}.  For a typical \hii\ region with $T_{\rm e} =
8,000$\kel, this corresponds to an electron temperature of
$\expo{4}$\kel.

Most of the ionized gas in the Milky Way, however, is located in the
inner Galaxy which is not probed by optical emission lines because of
extinction by dust.  Far-infrared collisionally excited lines are less
affected by dust and offer an alternative method to sample the WIM.  In
particular, [\nii] 122\,\micron\ and [\nii] 205\,\micron\ are excellent
diagnostics of the WIM since nitrogen has an ionization potential of 
14.5\ev, similar to hydrogen (13.6\ev), and will thus be associated with
\hii.

Using the {\tt HIFI} instrument on {\em Herschel}, \citet{persson14}
observed [\nii] 205\,\micron\ toward four \hii\ regions.  They detected
absorption of [\nii] from foreground gas toward W31C and W49N. This
absorbing gas must be located within the Solar orbit due to the velocity
range of the spectral components. Assuming a N/H abundance ratio and
filling factor they estimate mean electron densities of $\sim
0.1-0.3$\percc. These results, together with modeling of N$^{+}$ with
{\tt RADEX}, are consistent with the optically derived WIM properties,
albeit with somewhat higher electron densities.

Observations of RRLs offer an extinction-free probe of the WIM.  Early
RRL observations made with single-dish telescopes toward directions away
from OB-type star excited \hii\ regions detected weak RRL emission
\citep[e.g.,][]{gottesman70, lockman76, mezger78, heiles96, roshi00}.
This emission was interpreted by \citet{mezger78} as being associated
with the envelopes of \hii\ regions and called the ``extended low
density'' (ELD) \hii.  Because of their poor angular resolution,
however, all these observations suffer from confusion with the known
population of \hii\ regions \citep{anderson14} so any quantitative
interpretation of these data is problematic.

The more recent RRL surveys with the GBT and FAST, however, have
sufficient angular resolution to avoid emission from \hii\ regions and
in principle produce WIM-only images \citep{2021GDIGS}.  But there are
several reasons why these WIM-only images may still contain ELD
\hii\ emission and thus not be truly representative of WIM gas: 
(1) GDIGS is only sensitive to plasmas with mean electron densities
$\gsim 30$\percc\ (for a path length of 1\pc) and this is $\sim$\,100
times larger than the density expected for the WIM (see the
Figure~\ref{fig:GDIGS} sensitivity comparison);
(2) since H$\alpha$ studies imply that the WIM fills $\sim$70\% of the
ISM's volume, this semi-ubiquitously distributed plasma ought to emit
over the entire range of velocities that are allowed by Galactic
rotation yet for typical sight lines the GDIGS detected RRL spectra have
gaps with no emission at allowed LSR velocities; and
(3) the RRL emission appears to be correlated with the location of
\hii\ regions \citep{luisi20} and therefore may be part of the ELD \hii.

For the high sensitivity RRL observations toward G20 and G45, we
estimate rms electron densities of $(n_{\rm e}^{2})^{1/2} \sim
0.15-0.18$\percc\ over a wide range of velocities.  For a filling factor
of $f = 0.3$ this corresponds to mean electron densities of $\sim
0.3$\percc.  These electron densities are somewhat higher than expected
from the WIM based on optical data.  These higher values may not be
surprising since Figure~\ref{fig:wimHIIcomposite} suggests that some of
the emission detected toward both G20 and G25 might be associated with
OB-type star \hii\ regions. But the figure also shows that the G20 RRL
emission above 100\kms\ and the G45 emission below 50\kms\ have no
association with any nearby \hii\ regions and are probably tracing the
WIM.

\subsection{G20}\label{sec:discussG20}

The narrow, 13.5\kms, line width from plasma located at the G20 tangent
point distance sets a strong upper limit of $\lsim$\,4,000 K on the
electron temperature. At this \te, Model A with a density of
\Ne=0.38\,\percc\ replicates the observed intensity and line shape of
this emission component. The Model A $p/k=\Ne\te$ pressure is
$\lsim$\,1,520\,\percc\K. Models A and B together span \Ne\ between
$\sim$\,0.38--0.29\,\percc\ and \te\ between
$\sim$\,4,000--2,800\,\K\ resulting in WIM pressures ranging from $p/k$
$\sim$\,1,500 to 800\,\nT.

Measuring the pressure in the WIM is challenging because there do not
seem to be any observables from which one can derive both the density
and temperature without bias.  Combining the pulsar/\halpha\ density
determinations with the FIR derived electron temperature yields a
nominal pressure, $p/k$, of $\sim$\,500 to 1,000\nT\ for the WIM
plasma. But combining these \Ne\ and \te\ determinations may be
problematic. For example, these estimates for the WIM plasma's physical
parameters are not probing the same gas volumes.
In lieu of robust observational constraints for the WIM $p/k$, we can
seek insight from simulations. The {\tt TIGRESS-NCR} simulation of the
Galactic ISM \citep{2023TIGRESS} predicts a WIM pressure $p/k$ ranging
between $\sim$\expo{2} and $\sim$\expo{5}, with a mean value close to
$\sim$\expo{4}\nT.

Here, we show that a combination of RRL observations with models for the
emission together provide an alternative approach to deriving the WIM
pressure. Using the observed RRL line width to constrain \te\ and models
to estimate \Ne, we find that the $p/k$ pressure derived here for the
G20 tangent point emission is generally consistent with other
determinations of the WIM pressure that are based on observations. The
low \te\ limit, however, is not. Too, the electron density is at the
high end for values typically derived for WIM plasma. Altogether, the
G20 model for the tangent point emission challenges our understanding of
the WIM as a canonically pervasive, low density, $\lsim$\,0.1\,\percc,
$\sim$\,10,000\,\K\ plasma.

\subsection{G45}

The line width, 20.3\kms, for the G45 tangent point emission component
also sets a strong upper limit for the plasma electron temperature of
\te\ $\lsim$\,9,000\,\K. Model A adopts this \te\ limit and with a
density of \Ne=0.40\,\percc\ replicates the observed intensity and line
shape of this emission component.
Models A and B together span \Ne\ between
$\sim$\,0.40--0.31\,\percc\ and \te\ between
$\sim$\,9,000--6,400\,\K\ resulting in WIM pressures ranging from $p/k$
$\sim$\,3,600 to 2,000\,\nT.

These model $p/k$ pressures are also consistent with other
determinations of the WIM pressure and are more in line with the {\tt
  TIGRESS-NCR} simulation results.  Again, the density is at the high
end of values reported for the WIM, but overall this G45 tangent point
emission component's physical properties are consistent with WIM plasma.

In sum, the result of these efforts is that the WIM spans a range of
densities. Almost any density up to a $\sim$ few \percc\ may be
consistent with the WIM.
Optical studies derive \Ne\ $\sim$\,0.03--0.1\percc\ \citep{2009WIMreview}.
FIR observations of fine structure transitions infer
$\sim$\,0.1--0.3\percc\ \citep{persson14}.
Our G20 and G45 analysis gives $\sim$\,0.3--0.4\percc. 
Finally, there is evidence from low frequency RRL observations for a
more or less continuous distribution of plasma densities. Several
studies have found volume filling factors inversely proportional to
density, so the higher densities subsume an increasingly smaller
fraction of the ISM volume \citep[e.g.,][]{1985anantha,1986anantha,
  2001roshi}. 

%\clearpage
\section{Summary}

Studies of the Milky Way's WIM that use H$\alpha$ emission can only
probe the Galactic disk out to distances limited to a few \kpc\ from the
Sun due to extinction. Because the ISM is optically thin at cm
wavelengths radio recombination lines probe the WIM emission
at transgalactic path lengths.  Here, we use RRL emission to study two
Galactic sight lines located in the plane that are devoid of any OB-star
produced \hii\ regions and show no indication of nearby star formation.

\begin{itemize}

\item GBT observations of the WIM made toward Galactic sight lines G20
  and G45 show four Gaussian-shaped emission features in the
  \hna\ stacked RRL spectra. The emission measures of these spectral
  components range between $\sim$100 and $\sim$300\emeas.

\item For both sight lines, RRL emission can be seen at nearly all LSR
  velocities between 0\kms\ and the terminal velocity produced by
  Galactic rotation. The lack of any significant RRL emission at LSR
  velocities below 0\kms\ means that in these directions RRL
  emission from any WIM plasma located beyond the Solar orbit is below
  the EM $\gsim$\,10\emeas\ (3$\sigma$ limit for 5\kms\ resolution)
  sensitivity limit of our GBT observations. For G20 and G45, the line
  of sight path lengths with RRL emitting WIM plasma, \Dlos, must thus
  be no larger than 12 and 16 \kpc, respectively.

\item The observed \hnb/\hna\ intensity\ ratios are consistent with LTE
  excitation for the stronger, higher velocity components seen toward
  both sight lines at LSR velocities of $\sim$113\,\kms\ and
  $\sim$63\,\kms\ for G20 and G45, respectively.  The weaker, lower
  velocity components at $\sim$48\,\kms\ and $\sim$35\,\kms, however,
  show some evidence for non-LTE excitation.
  
\item Although our sight lines show no \hii\ regions in their fields,
  the WIM emission we observe might originate in extended WIM halos
  around \hii\ regions. Some of the emission detected toward both G20
  and G45 occurs at LSR velocities shared with nearby
  \hii\ regions. Empirical models for WIM emission near \hii\ regions
  predict comparable emission to what is seen (see
  Figure~\ref{fig:wimHIIcomposite}), although the model uncertainties
  preclude us from making any definitive conclusions. Despite these
  uncertainties, it seems unlikely that the G20 RRL emission above
  100\kms\ is associated with nearby known \hii\ regions.
  
\item Cloud models with plasma located at the G20 and G45 tangent points
  reproduce the observed RRL emission components at the tangent point
  velocities for these sight lines. These models have densities
  \Ne\,=\,$\sim$\,0.29--$\sim$\,0.40\,\percc, temperatures
  \te\,=\,$\sim$\,2,800--$\sim$\,9,000\,\K, and $p/k$ pressures
  $\sim$\,800--$\sim$\,3,600\,\percc\K.
  
\item The $p/k$ pressure derived for the G20 tangent point emission is
  consistent with other determinations of the WIM pressure. The low,
  $\lsim$\,4,000\K\ \te\ limit, however, is not. The G20 model for the
  tangent point emission challenges our understanding of the WIM as a
  canonically pervasive, low density, $\lsim$\,0.1\,\percc,
  $\sim$\,10,000\,\K\ plasma.

\end{itemize}

\begin{acknowledgments}

We thank the referee for a timely and close reading of our manuscript. 
We thank the GBT telescope operators whose diligence and expertise in
executing our observing scripts for project GBT/20A--483 were
exceptional.  The Green Bank Telescope is operated by the Green Bank
Observatory (GBO). The GBO and the National Radio Astronomy Observatory
(NRAO) are facilities of the National Science Foundation (NSF) operated
under cooperative agreements by Associated Universities, Inc.
This publication makes use of data products from the Wide-field Infrared
Survey Explorer, which is a joint project of the University of
California, Los Angeles, and the Jet Propulsion Laboratory/California
Institute of Technology, funded by the National Aeronautics and Space
Administration.
This research was supported by NSF awards AST-1714688 to
T.M.B. and AST-1516021 to L.D.A. T.V.W. is supported by the AST-2202340
NSF Astronomy and Astrophysics Postdoctoral Fellowship.

\end{acknowledgments}

\vspace{5mm}
\facilities{GBT}
\software{{\tt TMBIDL} \citep{2016tmbidl} }

%\clearpage
\vspace{0.5cm}
\restartappendixnumbering
\appendix
\vspace{-0.5cm}
\section{RRL Transitions used for Stacked Spectra}

Instrumental effects caused by the C-band receiver, the GBT IF system,
and VEGAS spectrometer compromise some of the transitions we observe.
These effects include unacceptable frequency structure in the spectral
baselines and loss of sensitivity due to high system temperatures. High
system temperatures mostly occur when the C-band receiver performance
deteriorates at both the high and low frequency extremes of the 
$\sim$4\ghz\ bandwidth. We inspected each RRL tuning to evaluate 
overall quality.  The usable RRL transitions are summarized in
Table~\ref{tab:bands}. Listed for the G20 and G45 targets are the
transition, the rest frequency, and the average system temperatures
for both linear polarizations: $\langle T_{XX} \rangle$ and
$\langle T_{YY} \rangle$. For these acceptable tunings system
temperatures typically range between $\sim$20\K\ and $\sim$40\K.

\vspace{-1cm}
\begin{deluxetable}{lcccccc}
  \tabletypesize{\scriptsize}
%  \tabletypesize{\footnotesize}
\tablecaption{RRL Transitions Used for Stacked Spectra\label{tab:bands}}
\tablehead{
  \colhead{Line} & \colhead{Transition} & \colhead{Frequency} &
  \colhead{$\langle$\emph{T}\textsubscript{XX}$\rangle_{\rm G20}$} &
  \colhead{$\langle$\emph{T}\textsubscript{YY}$\rangle_{\rm G20}$} &
  \colhead{$\langle$\emph{T}\textsubscript{XX}$\rangle_{\rm G45}$} &
  \colhead{$\langle$\emph{T}\textsubscript{YY}$\rangle_{\rm G45}$} \\
\colhead{} & \colhead{$n+\Delta n \rightarrow n$} & \colhead{(GHz)} &
\colhead{(K)} & \colhead{(K)} & \colhead{(K)} & \colhead{(K)}
}
\startdata
H97$\alpha$    & 98  $\rightarrow$ 97  & 7.09541 & 38.6 & 38.3 & 38.0 & 37.4 \\
H98$\alpha$    & 99  $\rightarrow$ 98  & 6.88149 & 35.4 & 33.4 & 34.6 & 31.9 \\
H99$\alpha$    & 100 $\rightarrow$ 99  & 6.67607 & 35.8 & 27.6 & 35.1 & 26.5 \\
H100$\alpha$   & 101 $\rightarrow$ 100 & 6.47876 & 24.6 & 25.1 & 23.9 & 24.1 \\
H101$\alpha$   & 102 $\rightarrow$ 101 & 6.28914 & 29.8 & 21.1 & 28.5 & 20.2 \\
H102$\alpha$   & 103 $\rightarrow$ 102 & 6.10685 & 28.5 & 21.7 & 27.6 & 20.8 \\
H103$\alpha$   & 104 $\rightarrow$ 103 & 5.93154 & 26.9 & 22.8 & 25.8 & 21.8 \\
H104$\alpha$   & 105 $\rightarrow$ 104 & 5.76288 & 24.2 & 23.4 & 23.0 & 22.4 \\
H105$\alpha$   & 106 $\rightarrow$ 105 & 5.60055 & 23.8 & 24.0 & 22.5 & 22.8 \\
H106$\alpha$   & 107 $\rightarrow$ 106 & 5.44426 & 21.2 & 24.2 & 20.1 & 23.1 \\
H107$\alpha$   & 108 $\rightarrow$ 107 & 5.29373 & 27.1 & 23.7 & 25.6 & 22.5 \\
H108$\alpha$   & 109 $\rightarrow$ 108 & 5.14870 & 24.6 & 26.4 & 23.1 & 25.1 \\
H109$\alpha$   & 110 $\rightarrow$ 109 & 5.00892 & 25.8 & 24.1 & 24.0 & 22.8 \\
H110$\alpha$   & 111 $\rightarrow$ 110 & 4.87416 & 26.0 & 24.0 & 24.2 & 22.8 \\
H111$\alpha$   & 112 $\rightarrow$ 111 & 4.74418 & 27.0 & 23.7 & 24.9 & 22.5 \\
H112$\alpha$   & 113 $\rightarrow$ 112 & 4.61879 & 27.1 & 25.2 & 25.2 & 24.0 \\
H114$\alpha$   & 115 $\rightarrow$ 114 & 4.38095 & 30.2 & 32.4 & 27.6 & 30.7 \\
H115$\alpha$   & 116 $\rightarrow$ 115 & 4.26814 & 38.9 & 39.1 & 36.0 & 36.0 \\
\hline
H122$\beta$    & 124 $\rightarrow$ 122 & 7.06872 & 34.1 & 35.0 & 33.3 & 34.8 \\
H123$\beta$    & 125 $\rightarrow$ 123 & 6.89905 & 34.6 & 32.9 & 33.7 & 31.8 \\
H124$\beta$    & 126 $\rightarrow$ 124 & 6.73479 & 34.0 & 31.3 & 33.1 & 29.9 \\
H125$\beta$    & 127 $\rightarrow$ 125 & 6.57570 & 27.7 & 26.4 & 26.7 & 25.1 \\
H126$\beta$    & 128 $\rightarrow$ 126 & 6.42158 & 24.5 & 29.8 & 23.7 & 28.6 \\
H127$\beta$    & 129 $\rightarrow$ 127 & 6.27223 & 30.3 & 22.4 & 29.2 & 21.4 \\
H128$\beta$    & 130 $\rightarrow$ 128 & 6.12748 & 31.0 & 22.2 & 29.4 & 21.2 \\
H129$\beta$    & 131 $\rightarrow$ 129 & 5.98714 & 28.8 & 22.5 & 27.6 & 21.5 \\
H130$\beta$    & 132 $\rightarrow$ 130 & 5.85107 & 25.5 & 22.6 & 24.3 & 21.5 \\
H131$\beta$    & 133 $\rightarrow$ 131 & 5.71909 & 24.3 & 23.1 & 23.1 & 21.9 \\
H132$\beta$    & 134 $\rightarrow$ 132 & 5.59105 & 24.3 & 23.9 & 23.2 & 22.8 \\
H133$\beta$    & 135 $\rightarrow$ 133 & 5.46680 & 22.5 & 23.3 & 21.2 & 22.0 \\
H134$\beta$    & 136 $\rightarrow$ 134 & 5.34619 & 21.7 & 25.7 & 20.5 & 24.3 \\
H135$\beta$    & 137 $\rightarrow$ 135 & 5.22913 & 25.6 & 17.3 & 23.8 & 16.3 \\
H136$\beta$    & 138 $\rightarrow$ 136 & 5.11544 & 26.4 & 22.9 & 24.8 & 21.8 \\
H137$\beta$    & 139 $\rightarrow$ 137 & 5.00502 & 25.4 & 24.1 & 23.8 & 22.8 \\
H138$\beta$    & 140 $\rightarrow$ 138 & 4.89778 & 25.9 & 24.2 & 24.0 & 23.0 \\
H139$\beta$    & 141 $\rightarrow$ 139 & 4.79357 & 26.3 & 23.7 & 24.4 & 22.5 \\
H140$\beta$    & 142 $\rightarrow$ 140 & 4.69229 & 28.2 & 24.8 & 25.9 & 23.3 \\
H141$\beta$    & 143 $\rightarrow$ 141 & 4.59384 & 30.0 & 23.5 & 27.5 & 22.2 \\
H143$\beta$    & 145 $\rightarrow$ 143 & 4.40508 & 29.9 & 30.5 & 27.4 & 28.7 \\
H145$\beta$    & 147 $\rightarrow$ 145 & 4.22650 & 42.4 & 32.8 & 38.3 & 30.8 \\
\hline
H139$\gamma$   & 142 $\rightarrow$ 139 & 7.11476 & 40.8 & 42.4 & 39.9 & 41.4 \\
H140$\gamma$   & 143 $\rightarrow$ 140 & 6.96495 & 34.1 & 32.3 & 33.5 & 31.5 \\
H141$\gamma$   & 144 $\rightarrow$ 141 & 6.81933 & 38.5 & 32.8 & 37.5 & 31.6 \\
H143$\gamma$   & 146 $\rightarrow$ 143 & 6.54004 & 26.5 & 26.2 & 25.6 & 25.1 \\
H144$\gamma$   & 147 $\rightarrow$ 144 & 6.40609 & 23.2 & 28.7 & 22.5 & 27.6 \\
H146$\gamma$   & 149 $\rightarrow$ 146 & 6.14898 & 23.9 & 23.5 & 22.9 & 22.4 \\
H147$\gamma$   & 150 $\rightarrow$ 147 & 6.02558 & 27.6 & 23.2 & 26.4 & 22.2 \\
\enddata
%\tablecomments{} 
%\tablenotetext{a}{}
\end{deluxetable}

\clearpage
\section{Radio Recombination Line Excitation Analysis}
\subsection{LTE Excitation}

Are the \hna, \hnb, and \hng\ stacked spectra consistent with LTE
excitation? 
To make this assessment we need to know the expected LTE ratio
between a specific $n+\Delta\,n \rightarrow\,n$ transition and a
fiducial $n+1 \rightarrow\,n$ transition.  Here, we use \hna\ as
the fiducial transition and calculate the LTE ratios expected for
\hnb\ and \hng.
First we calculate the statistical weights, $g_n$, and oscillator
strengths, $f_{n+\Delta\,n,n}$, for each RRL transition listed in
Appendix Table~\ref{tab:bands}. Since these transitions were used to
craft the stacked spectra, we use the average statistical weight,
$\langle g_n\rangle$, and average oscillator strength, $\langle
f_{n+\Delta\,n,n}\rangle$, for \hna, \hnb, and \hng\ to derive the
expected LTE ratios.

%{\tt\string \begin\{deluxetable\}[htb!]}{lcccc}
%\tabletypesize{\normalsize}
\begin{deluxetable}{lcccc}
\tablecolumns{5} \tablewidth{0pt}
\tablecaption{Oscillator Strengths\label{tab:oscillator}}
\tablehead{
  \colhead{Transition}            & \colhead{$n$}   &
  \colhead{$\Delta\,n$}             & \colhead{$g_{\rm n}$} &
  \colhead{$f_{\rm n+\Delta\,n,n}$} \\
}
%\vspace{-10pt}
\startdata
H97$\alpha$  &\phn 97 & 1 & 18818 &    18.79132765 \\
H98$\alpha$  &\phn 98 & 1 & 19208 &    18.98210255 \\
H99$\alpha$  &\phn 99 & 1 & 19602 &    19.17287745 \\
H100$\alpha$ &    100 & 1 & 20000 &    19.36365235 \\
H101$\alpha$ &    101 & 1 & 20402 &    19.55442725 \\
H102$\alpha$ &    102 & 1 & 20808 &    19.74520215 \\
H103$\alpha$ &    103 & 1 & 21208 &    19.93597705 \\
H104$\alpha$ &    104 & 1 & 21632 &    20.12675195 \\
H105$\alpha$ &    105 & 1 & 22050 &    20.31752685 \\
H106$\alpha$ &    106 & 1 & 22472 &    20.50830175 \\
H107$\alpha$ &    107 & 1 & 22898 &    20.69907665 \\
H108$\alpha$ &    108 & 1 & 23328 &    20.88985155 \\
H109$\alpha$ &    109 & 1 & 23762 &    21.08062645 \\
H110$\alpha$ &    110 & 1 & 24200 &    21.27140135 \\
H111$\alpha$ &    111 & 1 & 24642 &    21.46217625 \\
H112$\alpha$ &    112 & 1 & 25088 &    21.65295115 \\
H114$\alpha$ &    114 & 1 & 25992 &    22.03450095 \\
H115$\alpha$ &    115 & 1 & 26450 &    22.22527585 \\
\hline
H122$\beta$  &    122 & 2 & 29768 &\phn 3.29151250 \\
H123$\beta$  &    123 & 2 & 30258 &\phn 3.31784460 \\
H124$\beta$  &    124 & 2 & 30752 &\phn 3.34417670 \\
H125$\beta$  &    125 & 2 & 31250 &\phn 3.37050880 \\
H126$\beta$  &    126 & 2 & 31752 &\phn 3.39684090 \\
H127$\beta$  &    127 & 2 & 32258 &\phn 3.42317300 \\
H128$\beta$  &    128 & 2 & 32768 &\phn 3.44950510 \\
H129$\beta$  &    129 & 2 & 33282 &\phn 3.47583720 \\
H130$\beta$  &    130 & 2 & 33800 &\phn 3.50216930 \\
H131$\beta$  &    131 & 2 & 34322 &\phn 3.52850140 \\
H132$\beta$  &    132 & 2 & 34848 &\phn 3.55483350 \\
H133$\beta$  &    133 & 2 & 35378 &\phn 3.58116560 \\
H134$\beta$  &    134 & 2 & 35912 &\phn 3.60749770 \\
H135$\beta$  &    135 & 2 & 36450 &\phn 3.63382980 \\
H136$\beta$  &    136 & 2 & 36992 &\phn 3.66016190 \\
H137$\beta$  &    137 & 2 & 37538 &\phn 3.68649400 \\
H138$\beta$  &    138 & 2 & 38088 &\phn 3.71282610 \\
H139$\beta$  &    139 & 2 & 38642 &\phn 3.73915820 \\
H140$\beta$  &    140 & 2 & 39200 &\phn 3.76549030 \\
H141$\beta$  &    141 & 2 & 39762 &\phn 3.79182240 \\
H143$\beta$  &    143 & 2 & 40898 &\phn 3.84448660 \\
H145$\beta$  &    145 & 2 & 42050 &\phn 3.89715080 \\ 
\hline
H139$\gamma$ &    139 & 3 & 38642 &\phn 1.16315647 \\
H140$\gamma$ &    140 & 3 & 39200 &\phn 1.17126209 \\
H141$\gamma$ &    141 & 3 & 39762 &\phn 1.17936771 \\
H143$\gamma$ &    143 & 3 & 40898 &\phn 1.19557895 \\
H144$\gamma$ &    144 & 3 & 41472 &\phn 1.20368457 \\
H146$\gamma$ &    146 & 3 & 42632 &\phn 1.21989581 \\
H147$\gamma$ &    147 & 3 & 43218 &\phn 1.22800143 \\
\enddata
\end{deluxetable}

%\tablecomments{} 
%\tablenotetext{a}{}

The quantum properties of the Table~\ref{tab:bands} transitions are
compiled in Table~\ref{tab:oscillator}. Listed are the transition, the 
principle quantum number, $n$, the order of the transition, $\Delta\,n$,
the statistical weight, $g_n$, and oscillator strength,
$f_{n+\Delta\,n,n}$.  The statistical weight is $g_n=2\times n^2$ and
we use the prescription in \citet{1968Menzel} to derive the oscillator
strengths, $f_{n+\Delta\,n,n}$, for recombination transitions between
principle quantum numbers $ n+\Delta\,n\rightarrow\,n$.

The LTE RRL intensity ratio between two optically thin transitions is
given by the ratio of their oscillator strengths, $f_{n+\Delta\,n,n} / 
f^{\prime}_{n+\Delta\,n,n}$, and their statistical weights, $g_n /g^{\prime}_n $:  

\begin{equation}
  {LTE~Intensity~Ratio}~=~ {{\langle g_n\rangle \times \langle f_{n+\Delta\,n,n}\rangle}
    \over {\langle g^{\prime}_n\rangle \times \langle f^{\prime}_{n+\Delta\,n,n}\rangle}}.
\label{eq:LTEratio}
\end{equation}

\noindent Here, $\langle f^{\prime}_{n+\Delta\,n,n}\rangle $ and
$\langle g^{\prime}_n\rangle $ refer to the fiducial transition, \hna,
and the angle brackets denote the average of the quantum properties of
the transitions used to derive the stacked spectra.
The average quantum properties for the \hna, \hnb, and \hng\ stacked
spectra are summarized in Table~\ref{tab:LTEinfo} which lists $\langle
g_n\rangle $, $\langle f_{n+\Delta\,n,n}\rangle $, and $\langle
g_n\rangle \times \langle f_{n+\Delta\,n,n}\rangle $. The last column of
Table~\ref{tab:LTEinfo} gives the LTE intensity ratios for \hnb\ and
\hng\ relative to \hna.
 
\begin{deluxetable}{lcccc}[h]
\tabletypesize{\normalsize}
\tablecolumns{5} \tablewidth{0pt}
\tablecaption{Average Quantum Properties and LTE Ratio\label{tab:LTEinfo}}
\tablehead{
  \colhead{Transition}                      &
  \colhead{$\langle g_n\rangle $}                         &
  \colhead{$\langle f_{n+\Delta\,n,n}\rangle $}              &
  \colhead{$\langle g_n\rangle  \times \langle f_{n+\Delta\,n,n}\rangle $} &
  \colhead{LTE Ratio} \\
}
% \vspace{-10pt}
\startdata
\hna & 22365.000 &    20.43411151 & 4.5700890395e+05 & 1.00000 \\
\hnb & 35271.273 &\phn 3.57159029 & 1.2597453522e+05 & 0.27565 \\
\hng & 40832.000 &\phn 1.19442100 & 4.8770598447e+04 & 0.10672 \\
\enddata
\tablecomments{Average quantum properties from
  Table~\ref{tab:oscillator}. LTE intensity ratio relative to \hna\ from
  Equation~\ref{eq:LTEratio}.}
\end{deluxetable}

The ``LTE?'' column in Table~\ref{tab:LTEfits} gives the ratio between
the observed RRL transition ratios and the expected LTE ratio. In LTE
this ratio of ratios would be unity.  For the strongest WIM components
--- 113.4\kms\ for G20 and 63.3\kms\ for G45 --- this ratio is
$1.11\pm0.11$ and $1.08\pm0.11$, respectively.  We thus find that,
within the errors, our \hnb/\hna\ ratios are consistent with LTE
excitation for both sight lines.

The lower velocity components in these directions, however, give values
for the ``LTE?'' parameter that are significantly larger than one:
$1.34\pm0.13$ and $1.95\pm0.34$ for G20 and G45, respectively.
Moreover, all ratios involving the \hng\ spectra are compromised by the
much poorer sensitivity of these spectra due to their comparatively
small integration times compared with the \hna\ and \hnb\ data. These
\hng\ spectra can only provide upper limits and these limits are not
significant.

\subsection{Non-LTE Excitation}
We thus find that the \hnb\ lower velocity components seen in the G20
and G45 spectra show some evidence for non-LTE excitation in the WIM
gas.  These non-LTE effects can be described by the use of departure
coefficients, $b_{\rm n}$. Departure coefficients relate the true level
population, $N_{\rm n}$, to the population level under LTE, $N_{\rm n}^{*}$,
where $b_{\rm n} = (N_{\rm n} / N^*_{\rm n})$.
Non-LTE effects can alter the RRL intensities in the following way
\citep[see
  Equation 14.52 in][] {2009TOOLSwilson}:
\begin{equation}
  {T_{\rm L} \over T_{\rm L}^*} = b_{\rm n} \bigl( 1 - \frac{1}{2}~ \tau_{\rm c} ~\beta_{\rm n} \bigr),
\label{eq:nonLTE}
\end{equation}
where $T_{\rm L}$ is the observed line intensity, $T_{\rm L}^*$ is the
LTE intensity, and $\tau_{\rm c}$ is the continuum optical depth.  Here, 
$\beta_{\rm n}$ is a measure of the gradient of $b_{\rm n}$ with respect
to $ n$ \citep[see Equation 14.40 in][] {2009TOOLSwilson}:

\begin{equation}
\beta_{\rm n} = 1 - 20.836\, \Bigl(\frac{\te}{\K}\Bigr)
\Bigl(\frac{\nu}{\ghz}\Bigr)^{-1}
\Bigl(\frac{{\rm d\,ln}\,b_{\rm n}}{{\rm d\,n}}\Bigr)\, \dnn 
\label{eq:nonLTEbeta}
\end{equation}

\noindent
The first term in Equation~\ref{eq:nonLTE} accounts for the effect
of non-LTE line formation whereas the second describes non-LTE line
transfer effects which can include maser amplification of the line
radiation.  When $\beta$ becomes negative, $\beta_{\rm n} < 0$, maser
amplification occurs and the resulting RRL intensities will depend on
radiative transfer details.

For the low densities expected in the WIM, however, the continuum
opacity should be small at cm-wavelengths and thus $\tau_{\rm c} \ll 1$.
The main non-LTE effect will therefore be departures in the level
populations and so the non-LTE line intensity is the LTE intensity times
$b_{\rm n}$:\ $T_{\rm L} = b_{\rm n}\,T_{\rm L}^*$.

As the principle quantum number $n \rightarrow \infty$, $b_{\rm n}$
approaches unity.  Since $n$ increases with $\Delta n$ for RRLs at the
same frequency ---e.g., H102$\alpha$, H129$\beta$, and H147$\gamma$ have
nearly identical rest frequencies --- we expect the Hn$\beta$
transitions to be closer to LTE than the Hn$\alpha$ transitions.  The
observed \hnb\ intensities should thus be larger than expected when in
LTE, consistent with the results in Figure~\ref{fig:LTEspectra} and
Table~\ref{tab:LTEfits}.

In sum, we find that for the strongest WIM components seen in G20 and
G45 the \hnb/\hna\ ratios are consistent with LTE excitation. The weaker
components, however, show some evidence for non-LTE excitation. 
The \hna\ intensity can be weaker than the LTE value because
collisions are less effective due the size of the H atom compared with
the larger size of the nearby (in frequency) \hnb\ transition. For the
larger \hnb\ atom collisions are more effective in establishing LTE
level populations and the deviation from LTE is smaller.  As a
consequence the \hnb /\hna\ ratio can exceed the LTE value.

\clearpage
\section{\ion{H}{2} Regions Located Near the G20 and G45 Sight Lines}

Properties of {\em WISE} Catalog \hii\ regions located within
100\arcmin\ of the G20 and G45 sight lines are compiled in
Table~\ref{tab:HIIregions}.  Listed for each nebula are the galactic
co-ordinates, \lb, and angular separation from the sight line, together
with the IR radius, $R_{\rm IR}$, RRL intensity, $T_L$, LSR velocity,
\vlsr, FWHM line width, \fwhm, and their measurement errors. Some of the
entries in the {\em WISE} Catalog refer to nebulae that reside within
the same telescope beam. In those cases the Catalog lists identical
\vlsr\ and \fwhm\ values \citep[see][]{anderson14}. When such confusion
within the beam occurs we only list here (and use in
Figure~\ref{fig:wimHIIcomposite}) a single nebula, choosing the one with
the smallest separation from its fiducial sight line. Some {\em WISE}
\hii\ region spectra have emission at multiple velocities.  Because we
have no reason to choose otherwise, in these cases we plot all the
velocities in Figure~\ref{fig:wimHIIcomposite}.

\startlongtable
\begin{deluxetable}{lrcrrrcrccc}
\tabletypesize{\footnotesize}
\tablecaption{\hii\ Regions Near G20 and G45 \label{tab:HIIregions}}
\tablehead{    
\colhead{\hii\ Region} & \colhead{Separation}         &
\colhead{$R_{\rm IR}$}         &
\colhead{\gl}          & \colhead{\gb}                &
\colhead{$T_L$}          & \colhead{$\sigma_{T}$}           &
\colhead{\vlsr}        & \colhead{$\sigma_{\rm LSR}$}      &
\colhead{\fwhm}        & \colhead{$\sigma_{\rm \Delta V}$}\\
\colhead{}             & \colhead{(arcmin)}           &
\colhead{(arcmin)}           &
\colhead{(deg)}        & \colhead{(deg)}              &
\colhead{(mK)}        & \colhead{(mK)}              &
\colhead{(\kms)}       & \colhead{(\kms)}             &
\colhead{(\kms)}       & \colhead{(\kms)}             
}
\startdata
G020.098$-$00.123 &     9.44 &      0.7 &    20.10 &   $-$0.123 &     32.3 &      0.3 &     43.4 &      0.1 &     20.3 &      0.2 \\ 
G020.083$-$00.135 &     9.48 &      0.5 &    20.08 &   $-$0.134 &     31.0 &      2.8 &     42.2 &      1.4 &     31.7 &      3.3 \\ 
G019.818+00.010 &    10.92 &      1.9 &    19.82 &   +0.010 &     26.1 &      0.4 &     60.4 &      0.1 &     18.2 &      0.3 \\ 
G020.227+00.110 &    15.13 &      1.2 &    20.23 &   +0.110 &     10.5 &      0.2 &     22.1 &      0.2 &     14.9 &      0.4 \\ 
G019.728$-$00.113 &    17.67 &      0.7 &    19.73 &   $-$0.113 &     15.4 &      0.4 &     57.3 &      0.3 &     21.7 &      0.7 \\ 
G019.677$-$00.134 &    20.96 &      1.0 &    19.68 &   $-$0.133 &     48.0 &      5.8 &     55.0 &      1.2 &     20.0 &      2.8 \\ 
G019.780+00.286 &    21.65 &      1.9 &    19.78 &   +0.287 &     21.3 &      0.3 &     $-$9.3 &      0.2 &     21.7 &      0.4 \\ 
G020.363$-$00.014 &    21.80 &      0.7 &    20.36 &   $-$0.014 &     21.8 &      0.5 &     54.1 &      0.2 &     16.6 &      0.4 \\ 
G020.150$-$00.335 &    22.03 &      9.4 &    20.15 &   $-$0.335 &     19.0 &      2.9 &     67.3 &      1.5 &     20.3 &      3.6 \\ 
G019.741+00.280 &    22.89 &      0.7 &    19.74 &   +0.280 &     32.2 &      0.4 &     16.3 &      0.1 &     25.5 &      0.3 \\ 
G019.629$-$00.095 &    22.96 &      5.5 &    19.63 &   $-$0.094 &     54.0 &      3.7 &     58.6 &      0.8 &     22.8 &      1.8 \\ 
G019.716$-$00.261 &    23.11 &      1.0 &    19.72 &   $-$0.261 &     29.6 &      0.5 &     40.1 &      0.1 &     15.2 &      0.3 \\ 
G019.675$-$00.226 &    23.69 &      3.5 &    19.68 &   $-$0.225 &    168.5 &      1.7 &     42.5 &      0.1 &     26.5 &      0.4 \\ 
G019.594+00.024 &    24.40 &      0.7 &    19.59 &   +0.024 &     12.8 &      0.3 &     35.3 &      0.4 &     29.0 &      1.0 \\ 
G019.666$-$00.309 &    27.25 &      1.2 &    19.67 &   $-$0.308 &     33.9 &      0.8 &     44.3 &      0.2 &     19.2 &      0.5 \\ 
G019.609$-$00.239 &    27.46 &      1.4 &    19.61 &   $-$0.238 &    168.5 &      1.7 &     42.5 &      0.1 &     26.5 &      0.4 \\ 
G020.457+00.021 &    27.48 &      1.7 &    20.46 &   +0.022 &     66.7 &      1.1 &     73.8 &      0.1 &     15.0 &      0.3 \\ 
G019.554$-$00.248 &    30.60 &      3.2 &    19.55 &   $-$0.248 &    131.0 &      7.4 &     41.0 &      0.8 &     27.7 &      1.8 \\ 
G020.481+00.168 &    30.63 &      2.6 &    20.48 &   +0.169 &     24.0 &      2.3 &     24.1 &      1.0 &     20.5 &      2.2 \\ 
G019.494$-$00.150 &    31.65 &      1.9 &    19.49 &   $-$0.149 &     34.3 &      0.4 &     32.1 &      0.1 &     16.8 &      0.3 \\ 
G019.494$-$00.150 &    31.65 &      1.9 &    19.49 &   $-$0.149 &     20.5 &      0.4 &     54.6 &      0.2 &     15.1 &      0.5 \\ 
G019.489+00.135 &    31.68 &      0.9 &    19.49 &   +0.135 &     56.0 &      3.4 &     19.8 &      0.5 &     17.5 &      1.2 \\ 
G019.504$-$00.193 &    31.88 &      0.9 &    19.50 &   $-$0.193 &     45.4 &      0.4 &     37.8 &      0.1 &     19.9 &      0.2 \\ 
G019.466+00.168 &    33.58 &      4.8 &    19.47 &   +0.168 &     45.0 &      0.5 &    109.3 &      0.1 &     23.2 &      0.3 \\ 
G019.466+00.168 &    33.58 &      4.8 &    19.47 &   +0.168 &    145.0 &      0.6 &     19.9 &      0.0 &     14.9 &      0.1 \\ 
G019.466+00.168 &    33.58 &      4.8 &    19.47 &   +0.168 &      8.2 &      0.6 &     70.8 &      0.6 &     16.3 &      1.4 \\ 
G020.542$-$00.179 &    34.24 &      3.0 &    20.54 &   $-$0.179 &     40.5 &      0.6 &     48.6 &      0.1 &     16.2 &      0.3 \\ 
G020.576+00.103 &    35.13 &      1.6 &    20.58 &   +0.104 &     12.8 &      0.3 &     28.8 &      0.3 &     25.2 &      0.9 \\ 
G020.728$-$00.105 &    44.18 &      6.6 &    20.73 &   $-$0.104 &    141.0 &      0.5 &     56.0 &      0.0 &     26.5 &      0.1 \\ 
G020.727$-$00.259 &    46.34 &      4.1 &    20.73 &   $-$0.259 &     97.8 &      0.3 &     55.2 &      0.0 &     20.4 &      0.1 \\ 
G019.122$-$00.263 &    54.94 &      1.1 &    19.12 &   $-$0.262 &     32.3 &      0.4 &     56.0 &      0.2 &     35.3 &      0.5 \\ 
G019.122$-$00.263 &    54.94 &      1.1 &    19.12 &   $-$0.262 &     27.2 &      0.5 &    100.4 &      0.2 &     18.1 &      0.4 \\ 
G019.064$-$00.282 &    58.63 &      1.3 &    19.06 &   $-$0.282 &    164.8 &      1.8 &     64.4 &      0.1 &     25.2 &      0.3 \\ 
G019.604$-$00.905 &    59.24 &      1.0 &    19.60 &   $-$0.905 & \dots & \dots &     38.7 &      1.2 &     36.7 &      2.8 \\ 
G020.988+00.092 &    59.54 &      2.2 &    20.99 &   +0.092 &     47.0 &      5.6 &     18.6 &      1.3 &     21.4 &      2.9 \\ 
G021.004$-$00.056 &    60.33 &      0.7 &    21.00 &   $-$0.056 &     18.0 &      0.2 &     28.8 &      0.1 &     19.6 &      0.3 \\ 
G018.978+00.030 &    61.35 &      4.6 &    18.98 &   +0.031 &     26.0 &      2.5 &     52.3 &      1.7 &     36.1 &      3.9 \\ 
G019.030+00.423 &    63.47 &      1.3 &    19.03 &   +0.424 &      8.1 &      0.8 &     25.4 &      0.7 &     14.8 &      1.6 \\ 
G020.966$-$00.455 &    64.10 &     15.4 &    20.97 &   $-$0.455 &     10.0 &      1.1 &     38.5 &      2.5 &     45.5 &      6.0 \\ 
G019.045$-$00.588 &    67.27 &      2.0 &    19.05 &   $-$0.588 &     36.0 &      3.7 &     68.2 &      1.0 &     18.6 &      2.2 \\ 
G018.914$-$00.329 &    68.08 &     12.9 &    18.91 &   $-$0.329 &    217.9 &      0.8 &     68.0 &      0.1 &     23.8 &      0.1 \\ 
G018.914$-$00.329 &    68.08 &     12.9 &    18.91 &   $-$0.329 &     12.5 &      1.1 &     36.2 &      0.6 &     14.3 &      1.5 \\ 
G018.832$-$00.300 &    72.35 &      0.7 &    18.83 &   $-$0.300 &     46.1 &      0.4 &     46.4 &      0.1 &     30.0 &      0.3 \\ 
G018.881$-$00.493 &    73.37 &      1.8 &    18.88 &   $-$0.493 &     92.0 &      5.3 &     65.5 &      0.8 &     27.1 &      1.8 \\ 
G018.725$-$00.046 &    76.52 &     12.5 &    18.73 &   $-$0.045 &     22.0 &      2.4 &     63.5 &      2.4 &     45.4 &      6.0 \\ 
G018.741+00.250 &    77.01 &      1.1 &    18.74 &   +0.251 &     29.1 &      0.5 &     19.1 &      0.2 &     22.1 &      0.5 \\ 
G018.710+00.000 &    77.36 &      0.8 &    18.71 &   +0.000 &     16.6 &      0.3 &     31.8 &      0.3 &     44.5 &      0.8 \\ 
G018.677$-$00.236 &    80.63 &      1.2 &    18.68 &   $-$0.236 &     57.0 &     11.3 &     42.6 &      0.6 &     23.2 &      1.3 \\ 
G018.657$-$00.057 &    80.64 &      1.4 &    18.66 &   $-$0.056 & \dots & \dots &     44.1 &      0.9 &     32.4 &      2.2 \\ 
G018.750$-$00.535 &    81.57 &      1.1 &    18.75 &   $-$0.535 &     18.5 &      0.6 &     66.5 &      0.4 &     25.9 &      1.0 \\ 
G018.632+00.256 &    83.49 &      1.1 &    18.63 &   +0.256 &     22.8 &      0.7 &     18.4 &      0.2 &     12.9 &      0.4 \\ 
G018.630+00.309 &    84.26 &      0.7 &    18.63 &   +0.309 &     25.3 &      0.4 &     14.0 &      0.1 &     17.5 &      0.3 \\ 
G021.386$-$00.255 &    84.54 &      1.0 &    21.39 &   $-$0.254 & \dots & \dots &     92.1 &      1.1 &     32.3 &      2.7 \\ 
G018.594+00.321 &    86.51 &      3.1 &    18.59 &   +0.322 &     44.3 &      0.2 &     13.4 &      0.0 &     17.1 &      0.1 \\ 
G018.631$-$00.492 &    87.23 &      3.2 &    18.63 &   $-$0.492 &     31.2 &      0.2 &     61.6 &      0.1 &     30.3 &      0.2 \\ 
G018.584+00.344 &    87.43 &      0.7 &    18.58 &   +0.344 &     28.0 &      0.6 &     10.8 &      0.2 &     16.7 &      0.4 \\ 
G021.426$-$00.546 &    91.64 &      1.0 &    21.43 &   $-$0.546 & \dots & \dots &     70.2 &      0.6 &     25.3 &      1.8 \\ 
G018.461$-$00.003 &    92.33 &      0.5 &    18.46 &   $-$0.003 & \dots & \dots &     56.5 &      0.4 &     30.3 &      1.0 \\ 
G021.560$-$00.108 &    93.84 &      2.9 &    21.56 &   $-$0.108 &     24.9 &      0.3 &    115.1 &      0.2 &     23.0 &      0.5 \\ 
G021.450$-$00.590 &    93.95 &      2.0 &    21.45 &   $-$0.590 &     38.3 &      0.3 &     72.7 &      0.1 &     20.5 &      0.2 \\ 
G021.603$-$00.169 &    96.72 &      0.5 &    21.60 &   $-$0.169 &      8.2 &      0.4 &     $-$4.7 &      0.6 &     23.0 &      1.5 \\ 
G021.634$-$00.003 &    98.05 &      1.9 &    21.63 &   $-$0.002 &     24.5 &      0.3 &     20.2 &      0.1 &     19.8 &      0.3 \\ 
\hline
G045.070+00.132 &     8.99 &      1.0 &    45.07 &   +0.132 & \dots & \dots &     59.2 &      0.8 &     24.8 &      2.1 \\ 
G045.121+00.133 &    10.83 &      1.7 &    45.12 &   +0.133 & \dots & \dots &     56.7 &      0.5 &     44.5 &      1.6 \\ 
G045.453+00.044 &    27.31 &      4.1 &    45.45 &   +0.045 &    742.9 &      1.6 &     54.2 &      0.0 &     27.6 &      0.1 \\ 
G045.195$-$00.439 &    28.82 &      1.3 &    45.20 &   $-$0.439 &     18.9 &      0.4 &     72.9 &      0.2 &     25.1 &      0.6 \\ 
G045.475+00.130 &    29.55 &      2.4 &    45.48 &   +0.130 & \dots & \dots &     55.6 &      0.1 &     29.5 &      0.3 \\ 
G044.552$-$00.239 &    30.42 &      3.8 &    44.55 &   $-$0.239 &     12.8 &      0.4 &     58.4 &      0.2 &     13.6 &      0.5 \\ 
G044.811$-$00.492 &    31.59 &      4.6 &    44.81 &   $-$0.492 &     17.0 &      2.5 &     44.8 &      4.0 &     30.0 &      5.5 \\ 
G045.542$-$00.006 &    32.52 &      0.7 &    45.54 &   $-$0.006 &     42.3 &      0.3 &     55.0 &      0.1 &     26.2 &      0.2 \\ 
G044.501+00.332 &    35.97 &      0.8 &    44.50 &   +0.332 &     48.5 &      0.2 &    $-$43.0 &      0.1 &     22.0 &      0.1 \\ 
G045.002$-$00.611 &    36.62 &      5.8 &    45.00 &   $-$0.610 &      8.7 &      0.3 &     61.3 &      0.4 &     24.6 &      0.9 \\ 
G044.521+00.385 &    36.87 &      0.7 &    44.52 &   +0.385 &     18.1 &      0.2 &    $-$49.7 &      0.1 &     25.9 &      0.3 \\ 
G044.375$-$00.076 &    37.76 &      4.2 &    44.38 &   $-$0.076 &     11.4 &      0.1 &     56.7 &      0.1 &     33.5 &      0.3 \\ 
G045.634$-$00.016 &    38.08 &      1.8 &    45.63 &   $-$0.016 &      3.6 &      0.2 &      9.2 &      0.6 &     19.6 &      1.6 \\ 
G044.689$-$00.579 &    39.42 &      2.1 &    44.69 &   $-$0.579 &      6.8 &      0.2 &     41.9 &      0.4 &     19.6 &      0.9 \\ 
G044.379$-$00.327 &    42.11 &      5.0 &    44.38 &   $-$0.327 &     22.8 &      0.5 &     61.1 &      0.2 &     18.6 &      0.5 \\ 
G045.503+00.495 &    42.38 &      1.9 &    45.50 &   +0.496 &      7.4 &      0.4 &    $-$35.9 &      0.7 &     23.8 &      1.9 \\ 
G045.689$-$00.235 &    43.70 &      4.0 &    45.69 &   $-$0.235 &      8.4 &      0.3 &     19.1 &      0.5 &     34.0 &      1.3 \\ 
G044.904$-$00.733 &    44.30 &      2.2 &    44.90 &   $-$0.732 &     15.7 &      0.3 &     65.6 &      0.2 &     17.2 &      0.4 \\ 
G045.197+00.740 &    45.96 &      1.3 &    45.20 &   +0.740 &     18.7 &      0.2 &    $-$35.0 &      0.2 &     31.4 &      0.4 \\ 
G044.224+00.085 &    46.79 &      5.3 &    44.22 &   +0.085 &     21.0 &      2.8 &     59.6 &      3.4 &     30.4 &      4.7 \\ 
G044.418+00.535 &    47.43 &      1.4 &    44.42 &   +0.536 &     13.1 &      0.2 &    $-$55.1 &      0.2 &     25.0 &      0.4 \\ 
G045.391$-$00.725 &    49.39 &      3.2 &    45.39 &   $-$0.724 &     53.0 &      0.5 &     52.5 &      0.1 &     20.6 &      0.2 \\ 
G045.773$-$00.378 &    51.62 &      1.4 &    45.77 &   $-$0.377 &     11.7 &      0.6 &     51.0 &      0.4 &     15.1 &      0.9 \\ 
G045.825$-$00.291 &    52.47 &      1.2 &    45.83 &   $-$0.290 & \dots & \dots &     61.2 &      0.2 &     26.3 &      0.5 \\ 
G045.882$-$00.088 &    53.23 &      4.8 &    45.88 &   $-$0.087 &     13.2 &      0.1 &     62.4 &      0.1 &     19.1 &      0.2 \\ 
G045.882$-$00.088 &    53.23 &      4.8 &    45.88 &   $-$0.087 &      2.2 &      0.1 &     18.2 &      0.5 &     28.9 &      1.3 \\ 
G044.094$-$00.015 &    54.31 &      1.8 &    44.09 &   $-$0.014 &     12.6 &      0.3 &     66.6 &      0.3 &     23.6 &      0.8 \\ 
G045.933$-$00.403 &    60.98 &      1.0 &    45.93 &   $-$0.402 & \dots & \dots &     63.9 &      0.5 &     21.1 &      1.3 \\ 
G046.033$-$00.097 &    62.28 &      3.1 &    46.03 &   $-$0.096 &      8.0 &      0.1 &     63.5 &      0.2 &     28.5 &      0.6 \\ 
G044.331$-$00.837 &    64.26 &      2.8 &    44.33 &   $-$0.837 &     13.9 &      0.4 &     62.5 &      0.2 &     17.0 &      0.6 \\ 
G046.069+00.216 &    65.47 &      1.1 &    46.07 &   +0.216 &      6.2 &      0.3 &     12.4 &      0.4 &     17.6 &      0.9 \\ 
G046.088+00.254 &    67.05 &      0.4 &    46.09 &   +0.255 &      7.6 &      0.3 &     13.4 &      0.4 &     17.0 &      0.9 \\ 
G043.894+00.197 &    67.37 &      2.8 &    43.89 &   +0.198 &      5.7 &      0.2 &    $-$38.1 &      0.4 &     21.3 &      1.0 \\ 
G043.794$-$00.129 &    72.74 &      0.6 &    43.79 &   $-$0.129 & \dots & \dots &     43.3 &      1.1 &     31.9 &      3.0 \\ 
G043.774+00.057 &    73.64 &      2.5 &    43.77 &   +0.058 &     33.1 &      0.3 &     70.5 &      0.1 &     20.8 &      0.2 \\ 
G043.818+00.395 &    74.78 &      1.8 &    43.82 &   +0.395 &     29.6 &      0.2 &    $-$10.5 &      0.1 &     27.6 &      0.2 \\ 
G043.730+00.114 &    76.49 &      2.1 &    43.73 &   +0.115 &      8.3 &      0.3 &     73.1 &      0.4 &     22.4 &      0.9 \\ 
G046.173+00.533 &    77.36 &      1.0 &    46.17 &   +0.533 &     10.2 &      0.2 &      6.3 &      0.2 &     24.8 &      0.6 \\ 
G046.203+00.532 &    78.95 &      0.7 &    46.20 &   +0.532 &      5.5 &      0.2 &      4.6 &      0.6 &     29.9 &      1.4 \\ 
G046.213+00.547 &    79.88 &      0.6 &    46.21 &   +0.548 &      5.5 &      0.2 &      5.8 &      0.6 &     26.3 &      1.3 \\ 
G043.890$-$00.780 &    81.40 &      1.0 &    43.89 &   $-$0.780 & \dots & \dots &     53.4 &      0.4 &     28.4 &      1.0 \\ 
G043.999+00.978 &    83.97 &      2.3 &    44.00 &   +0.979 &      4.4 &      0.2 &    $-$17.1 &      0.6 &     25.4 &      1.5 \\ 
G043.968+00.993 &    85.92 &      0.8 &    43.97 &   +0.993 &     11.1 &      0.5 &    $-$21.6 &      0.6 &     23.8 &      1.5 \\ 
G046.495$-$00.241 &    90.85 &      5.0 &    46.49 &   $-$0.240 &    257.8 &      1.1 &     57.7 &      0.0 &     18.3 &      0.1 \\ 
G043.523$-$00.648 &    96.75 &      1.5 &    43.52 &   $-$0.648 &      4.4 &      0.4 &     55.4 &      1.1 &     26.8 &      2.7 \\ 
G043.432+00.516 &    99.04 &      1.2 &    43.43 &   +0.517 &     22.5 &      0.3 &    $-$12.8 &      0.2 &     22.7 &      0.3 \\ 
\enddata
\end{deluxetable}

\clearpage
\section{Numerical Modeling of RRL Emission from WIM Plasmas}

Models for WIM RRL emission from LOS plasmas must specify the plasma
electron density, \Ne, temperature, \te, and velocity dispersion,
$\sigma$, at every point along the \Dlos.  We use a numerical code to
craft synthetic RRL emission spectra for G20 and G45.  {\tt TMBIDL}
already includes code, {\tt MODEL\_HI}, that calculates
21\cm\ \hi\ spectra for any Galactic LOS direction and distribution of
physical properties.  The radiative transfer calculation is
one-dimensional: the LOS is modeled as a series of slabs of thickness,
$dx$, with some total LOS path length.  The LOS can have any arbitrary
distribution of density, excitation temperature, and velocity dispersion.

We modified the {\tt TMBIDL} {\tt MODEL\_HI} code to calculate instead
the LOS radiative transfer for a pure hydrogen plasma in LTE. The models
compute spectra for the H109$\alpha$ transition whose 5.00\ghz\ rest
frequency is the average for the \hna\ spectra.
We follow the \cite{balser21} technique.  Assuming the Rayleigh-Jeans
limit, $h\nu \ll kT_{\rm e}$, the brightness temperature as a function
of frequency, $\nu$, is given by
\begin{equation}
  T_{\rm B}(\nu) = T_{\rm e}(1 - {\rm e}^{-\tau_{\rm L} (\nu)}),
\end{equation}
where $\tau_{\rm L} (\nu)$ is the optical depth, $\tau_{\rm L} (\nu) = \int \kappa_{\rm
  L} (\nu) \,d\ell$, and $T_e$ is the plasma electron temperature.  Here
$\kappa_{\rm L} (\nu)$ is the absorption coefficient and $d\ell$ is the
path length through the ionized gas.

The RRL absorption coefficient \citep{condon16} is:
\begin{equation}
\kappa_{\rm L}(\nu) = \frac{c^2}{8\pi \nu_{\rm L}^2}\frac{g_{\rm
    u}}{g_{\rm l}}n_{\rm l}A_{\rm ul}\left[ 1 - {\rm exp}\left(-\frac{h\nu_{\rm L}}{kT_{\rm e}}\right) \right] \phi(\nu),
\end{equation}
where $c$ is the speed of light, $\nu_{\rm L}$ is the frequency of the
RRL transition, ($g_{\rm u}/g_{\rm l}$) is the ratio of statistical
weights for the upper and lower levels, $n_{\rm l}$ is the number
density in the lower state, $A_{\rm ul}$ is the spontaneous emission
rate from the upper to lower state, $h$ is Planck's constant, $k$ is
Boltzmann's constant, and
$\phi(\nu)$ is the normalized line shape.
The statistical weights for hydrogen are $g_{\rm n} = 2 n^{\rm 2}$.
The spontaneous emission rate from the upper to the lower state,
$A_{\rm ul}$, can be approximated \citep{condon16} as:
\begin{equation}
A_{\rm n+1,n} = \left(\frac{64\pi^{6} m_{\rm e} e^{10}}{3 c^{3} h^{6}}\right) \frac{1}{n^{5}},
\end{equation}
where $m_{\rm e}$ is the electron mass, $e$ is the electric charge,
and $n$ is the principal quantum number.   We assume a Gaussian line
profile, $\phi(\nu)$: 
\begin{equation}
  \phi(\nu) = \frac{1}{\sqrt{2 \pi \sigma^{2}}}\,{\rm exp} \left[-\frac{(\nu - \nu_{\rm L})^{2}}{2\sigma^{2}}\right].
\end{equation}

In order to compare model $T_{\rm B}(\nu)$ spectra with the
\hna\ observations we must calculate a \tbv\ model spectrum. To do this
each slab must be assigned an LSR velocity derived from the Galactic LOS
direction, \gl, location along the LOS, \dsun, and an assumed Galactic
rotation curve model.
Once this \vlsr\ vs \dsun\ relation is established, we use the Doppler
equation at the H109$\alpha$ rest frequency to transform $T_{\rm
  B}(\nu)$ to \tbv.
{\tt TMBIDL} has several Galactic rotation curve choices available.  The
conclusions we reach here about the WIM do not depend the choice of any
particular rotation curve. Here, for all our models we adopt the
\citet{1985Clemens} rotation curve scaled to \Ro=8.5\kpc.

Each model is comprised of one or more plasma ``clouds'' distributed
along the LOS. All clouds are homogeneous, isothermal and in LTE. Each
cloud's properties are specified at input.  A cloud is defined by:
location along the LOS, \dsun, LOS path length size (aka the cloud
diameter), \Dcloud, electron density, \Ne, plasma electron temperature,
\te, and velocity dispersion, $\sigma$.

\clearpage
\vspace{-2.0cm}
\bibliography{wim}

\end{document}